\documentclass[12pt,english]{book}
\usepackage{psfig,epsfig,babel,epsf,here,
amsthm,amsfonts,amssymb,wrapfig}
\usepackage{subeqn}
\usepackage{psfrag}
\usepackage{latexsym}

\def \bq{\begin{quote}}
\def \eq{\end{quote}}

\def\be{\begin{equation}}
\def\ee{\end{equation}}
\def\bc{\begin{center}}
\def\ec{\end{center}}
\def\bea{\begin{eqnarray}}
\def\eea{\end{eqnarray}}
\def\dd{\displaystyle}
\def\nn{\nonumber}

\def\ov{\overline}

\def\cL{{\cal L}}

\author{ {\Huge Carla Biggio} \vspace{8 cm}}

\title{\Huge\bf
Symmetry Breaking in Extra Dimensions
 \vspace{2 cm}}
\date{data}

\newcommand{\clearemptydoublepage}{\newpage{\pagestyle{empty}\cleardoublepage}}

\begin{document}


\pagestyle{empty}

\begin{titlepage}

\hfill{DFPD-03/TH/49}
\vspace{1cm}

\begin{center}
{\LARGE \bf Symmetry Breaking in Extra Dimensions}\\[1cm]
{\Large Carla Biggio}\\[0.5cm] 
{\large Dipartimento di Fisica `G.~Galilei', Universit\`a di Padova
\&}\\[0.1cm] 
{\large INFN, Sezione di Padova, Via Marzolo~8, I-35131 Padova,
Italy~\footnote{Now at IFAE, Universitat Aut\`onoma de Barcelona,
E-08193 Bellaterra (Barcelona), Spain}\\[1cm]}
{\bf Ph.D. thesis}\\[0.1cm]
Universit\`a degli Studi di Padova, Dipartimento di Fisica `G.~Galilei'\\[0.1cm]
Dottorato di Ricerca in Fisica, Ciclo XVI\\[0.1cm]
Advisor: Prof. F. Feruglio\\[1.5cm]
{\bf Abstract}
\end{center}
\noindent
In this thesis we analyze the problem of symmetry breaking in models
with extra dimensions compactified on orbifolds. In the first chapter
we briefly review the main symmetry breaking mechanisms peculiar of
extra dimensions such as the Scherk-Schwarz mechanism, the Hosotani
mechanism and the orbifold projection. In the second chapter we study
the most general boundary conditions for fields on the orbifold
$S^1/Z_2$ and we apply them to gauge and SUSY breaking. In the third
chapter we focus on flavour symmetry and we present a six dimensional
toy model for two generations that can solve the fermion hierarchy
problem.

\vspace{1cm}
\begin{center}
October 2003
\end{center}

\end{titlepage}

\pagebreak
\mbox{}
\clearemptydoublepage


\pagenumbering{roman}
\pagestyle{headings}

\tableofcontents
\clearemptydoublepage


\addcontentsline{toc}{chapter}{Introduction}
\pagestyle{plain} 

\chapter*{Introduction} 

\pagenumbering{arabic}

The introduction of extra dimensions in theoretical high energy
physics is mainly due to the quest for unification of particle
interactions. This duality -unification of forces on one side and
introduction of new coordinates on the other- was born already at the
end of the $19^{th}$ century, after the unification of electricity and
magnetism carried out by Maxwell. Indeed once the special relativistic
invariance of Maxwell's theory was recognized, it became clear that a
unified description of electricity and magnetism implied a unified
description of space and time, which for the first time began to be
considered as different coordinates of a continuum
space-time. Inspired by this idea, in the following years many
physicists attempted to unify gravitation and electromagnetism
starting from a theory defined on a five-dimensional space-time, which
was obtained from the usual one by adding a spatial coordinate.  The
first was the Finnish physicist Gunnar Nordstrom who, in 1914, built a
model starting from Maxwell equations for a five-dimensional vector
boson of an abelian group~\cite{nordstrom}. Then, after the
publication of Einstein's general relativity, the mathematician
Theodor Kaluza proposed another, more complex, unified theory,
originating from a five-dimensional gravitational Einstein
action~\cite{kaluza}. Some years later the same theory was
independently rediscovered by Oskar Klein~\cite{klein}.

After these first attempts at unification~\cite{KKtheories}, extra
dimensions were forgotten for many years, obscured by the successes of
four-dimensional quantum field theory, which culminated in the
discovery of the Standard Model of electroweak
interactions~\cite{standardmodel}. Following experimental
confirmation, most research was focused on building a unified theory
of strong and electroweak interactions, the so-called Grand Unified
Theory, still in four dimensions. However each time gravitation was
incorporated, extra dimensions naturally appeared. In particular
string theories~\cite{strings}, which arguably offer the only
consistent quantum description of gravitation and other fundamental
forces, are defined in ten (heterotic, type I and type II strings) or
in eleven (M-theory) dimensions. Motivated by this observation, many
physicists returned to the original Kaluza-Klein theories and began to
study quantum field theory in higher dimensions. And a new world
revealed to their eyes. Indeed extra dimensions offer a new
perspective for the interpretation of data, for the description of
physical phenomena, for overcoming problems and, in particular, they
offer new mechanisms for symmetry breaking.

Symmetry breaking is one of the most important and interesting aspects
of theoretical particle physics, since symmetries provide the basis of
our current description of nature. In the usual four-dimensional
theories we know that symmetries can be broken explicitly or
spontaneously; in this latter case if the broken symmetry is global
and continuous the Goldstone theorem~\cite{goldstone} applies, whereas
if the symmetry is local we have an Higgs mechanism~\cite{higgs}. In
theories defined in extra dimensions new ways of symmetry breaking
appear, associated with the different compactifications of the extra
dimensions.

If we start from a scenario with infinite extra dimensions, the
simplest way to compactify is to impose a periodicity condition on
each extra coordinate, in such a way as to obtain a multi-dimensional
torus. Already at this stage we have one first symmetry breaking since
the higher-dimensional Lorentz invariance is spoiled. When we
introduce fields and write down a lagrangian, we require that physics
depends only on points of the compact space, so a kind of periodicity
condition on the extra-dimensional action must be imposed. Obviously
if fields are already periodic this condition is immediately satisfied
and it does not imply any new interesting features. However if the
lagrangian is invariant under the transformations of some symmetry
group, we can use this symmetry to say that fields are periodic up to
such a transformation. This is known as the Scherk-Schwarz
mechanism~\cite{SS1,SS2} and is used to break four-dimensional
symmetries. In theories in extra dimensions four-dimensional fields
are recovered as the modes of a Fourier expansion along the extra
coordinates (Kaluza-Klein modes). For every mode there is a
corresponding four-dimensional mass (Kaluza-Klein levels) and in
general a higher-dimensional field possesses one zero mode
corresponding to a four-dimensional field of zero mass. If fields are
no longer periodic (if they are ``twisted'') the conventional Fourier
expansion is modified and this leads to a constant shift of every
Kaluza-Klein level which also includes the zero mode that no longer
corresponds to a massless four-dimensional field. Now if we consider a
multiplet of some symmetry group and we assign different periodicity
conditions to different components of the multiplet, we discover that
from the four-dimensional perspective the symmetry is broken, since
some fields maintain the usual Kaluza-Klein levels while others are
shifted.

In addition to the toroidal compactification, there exists an
alternative type of compactification which is called an
orbifold~\cite{orb-vafa-witten-ecc}. This is obtained by imposing a
discrete symmetry on a compact space that leads to fixed points
invariant under the discrete symmetry transformations. This
``orbifolding'' breaks translational invariance and, as the
Scherk-Schwarz mechanism, can be used to break other symmetries. Also
in this case we must require that the action only depends on the
points of the orbifold and this translates into particular
transformation properties for fields. In the simplest cases the
orbifold transformation corresponds to a parity operation where fields
can be even or odd assigned parity. The number of Kaluza-Klein modes
for each field of definite parity is now reduced and in particular odd
fields lose their zero modes. If once again we consider a multiplet of
some symmetry group and assign different parities to different
components of the multiplet, we find that the symmetry is broken from
the four-dimensional point of view.

Finally another symmetry breaking mechanism typical of
extra-dimensional frameworks exists for gauge theories, defined on
non-simply connected manifolds independently of the compactification
considered. Indeed if the extra-dimensional component of a gauge field
acquires a constant background or vev, the gauge symmetry can be
broken by a Wilson loop. This is known as the Hosotani
mechanism~\cite{hosotani} and it has been shown to be equivalent to
the Scherk-Schwarz mechanism. In fact it is possible to perform a
gauge transformation that reabsorbs the vev, but now leads to non
periodic boundary condition for the gauge field considered. Since the
Hosotani mechanism is a spontaneous symmetry breaking, we can exploit
this equivalence to state that Scherk-Schwarz breaking is also
spontaneous.

These novel mechanisms of symmetry breaking can be applied to various
types of symmetries, such as gauge symmetries, supersymmetry or
flavour symmetries. For every different symmetry considered, the
compactification scale must take a particular value. For example when
extra dimensions are introduced to explain the weakness of gravity
with respect to the other forces, they have to be very large, of order
$TeV^{-1}$~\cite{ADD}. In contrast if we want to break a Grand Unified
symmetry the radius of the extra dimension must be very small, of the
order of the inverse of the grand unification scale. In this thesis we
are not concerned with the size of the compactification radius as we
investigate the mechanisms of symmetry breaking on orbifolds,
independently of the compactification scale. However, since we apply
them to gauge symmetries, supersymmetry and flavour symmetry, it will
become apparent that we are always dealing with small extra
dimensions.\\

\noindent The rest of the thesis is organized as follows.

In chapter 1 we describe in more detail the symmetry breaking
mechanisms typical of extra-dimensional frameworks that we have
outlined above and apply them to gauge symmetry and supersymmetry. To
do this we choose to describe in detail some realistic models which
exploit these mechanisms to break their symmetries. Of course the
literature on the subject is too large, so this chapter is far from
being a complete phenomenological review on models in extra
dimensions. We simply choose some models as an example to give an idea
of the importance of these new methods for symmetry breaking in model
building.

Chapters 2 and 3 contain the original parts of this thesis. In chapter
2 we study the particular features the Scherk-Schwarz mechanism shows
when implemented on orbifolds. It is well known that when this
mechanism is applied to orbifolds, there are various consistency
conditions that must hold between the operators defining the twist and
those defining the orbifold itself. However there is also a more
interesting feature. At variance with manifolds where fields must be
smooth everywhere, on orbifolds they can have discontinuities at the
fixed points, provided the physical properties of the system remain
well defined. So the most general boundary conditions for fields are
specified not only by parity and periodicity, but also by possible
jumps at the fixed points. In sections~\ref{GBCferm} and~\ref{GBCbos}
we discuss the most general boundary conditions respectively for
fermions and bosons on the orbifold $S^1/\mathbb{Z}_2$ and we
calculate the spectra and eigenfunctions in various cases, discussing
the relationship between these boundary conditions and the
Scherk-Schwarz mechanism.

We find that the most general boundary conditions for fields on
orbifolds are identical for fermions and bosons, but in the bosonic
case identical conditions are required also for the $y$-derivative of
fields (where $y$ is the extra coordinate). This is due to the
requirement of self-adjointness for the differential operator which
determines the spectrum. These generalized boundary conditions include
twist and jumps at the fixed points and the matrices defining them are
required to be unitary and must satisfy certain consistency
conditions.

Once we have assigned periodicity, parity and jumps to fields, we can
calculate the corresponding spectra and eigenfunctions by solving the
equations of motion. In this thesis this is done in the case of one
fermion field and one or more scalar fields. In every case we find
that the spectrum is a Scherk-Schwarz-like spectrum, since all the
Kaluza-Klein levels are always shifted by a universal amount. However
the shift is no longer determined by the twist parameter alone, as in
the conventional Scherk-Schwarz mechanism, but now also depends on
parameters that define the behaviour of fields at the fixed points. As
required from boundary conditions, eigenfunctions are either
discontinuous or have cusps at the fixed points and they can be
periodic or not, depending on the twist.

As the shift in the Kaluza-Klein levels corresponding to a given set
of generalized boundary conditions is similar to the shift induced by
usual twisted boundary conditions, we can try to relate the two
systems. This can be achieved by choosing an appropriate twist for the
``smooth'' system, so that eigenfunctions associated to this twist are
now continuous, i.\,e. different from the previous ones, whereas the
mass spectra remain the same. We can move from one system to the other
by using a local field redefinition. As the physical properties of a
quantum mechanical system are invariant under a local field
redefinition, we can therefore state that the two systems - the one
characterized by generalized boundary conditions involving twist and
jumps and the other characterized only by twist - are equivalent.

Therefore we conclude that there is an entire class of different
boundary conditions that correspond to the same spectrum, i.\,e. to
the same physical properties, with eigenfunctions that are related by
field redefinitions. By performing this redefinition at the level of
the action, we observe, both for the fermionic and the bosonic cases,
that the generalized boundary conditions lead to $y$-dependent
five-dimensional mass terms that can be even localized at the fixed
points. Although sometimes these terms are singular and an appropriate
regularization is required, they are necessary for the consistency of
the theory, as they encode the behaviour of the fields at the
boundaries.

We previously stated that different mass terms, corresponding to
different sets of boundary conditions, can give rise to the same
four-dimensional spectrum. Then it is useful to determine the most
general set of five-dimensional mass terms that correspond to a given
mass spectrum, i.\,e. to a given Scherk-Schwarz twist parameter. In
section~\ref{bfwz} we discuss the conditions that a five-dimensional
mass term must satisfy in order to be associated to a Scherk-Schwarz
twist and we find a relationship between the twist parameter and the
Wilson loop obtained by integrating over the mass terms. We also
discuss some examples of equivalent mass terms.

The work done in this first part is quite formal. In
sections~\ref{gauge} and~\ref{susy} we discuss some phenomenological
applications of our generalized boundary conditions to gauge symmetry
breaking and to supersymmetry breaking respectively. In the first case
we study the breaking of the symmetry of a toy model based on the
gauge group $SU(2)$ and then discuss a realistic model based on
$SU(5)$. For the second example we consider pure five-dimensional
supergravity.

In chapter~\ref{chap3} we focus on the problem of flavour symmetry
breaking and we construct a six-dimensional toy model for flavour
where the number of generations arises dynamically as a consequence of
the presence of extra dimensions. It is already known in the
literature that extra-dimensional frameworks offer new mechanisms to
obtain four-dimensional chiral fermions. For instance they can
originate as zero modes of higher-dimensional fermions coupled to a
solitonic background or, alternatively, to a scalar field with non
constant profile in orbifold models, where the same scalar field
forces the fermion to be localized.  This can be used to explain the
fermion mass hierarchy, since Yukawa constants are given by the
overlap of fermionic and Higgs wave functions. If the overlap among
these functions is different due to their position in the extra space,
we obtain different values for the Yukawa constants and thus large
hierarchies can be generated.  There are several variants of this
idea, such as the case of constant Higgs vev and fermions localized in
$ad$ $hoc$ regions and also a scenario with varying Higgs vev and
fermion families localized in three different places. Note that in all
of these simple models the number of fermion replica is introduced by
hand. However it is known that it is possible to obtain an arbitrary
number of four-dimensional chiral fermions by coupling the
higher-dimensional fermions to a topological defect. This fact has
been exploited in six-dimensional models and, by requiring that the
winding number of the defect is three, the authors have built a
semi-realistic model for flavour in which three families dynamically
arise.

In our toy model we exploit another fact to simultaneously address
both the flavour problem and the question of fermion replica. It is
well known that a spinor in higher dimensions consists of many
four-dimensional spinors. For instance a six-dimensional Dirac spinor
contains two left-handed and two right-handed four-dimensional
spinors. After projecting out the unwanted chirality, for example by
orbifolding, we are left with two four-dimensional spinors with the
same chirality and the same quantum numbers, i.\,e. with two replica
of the same fermion. Although this is insufficient to build a
realistic model of flavour since in this scheme we can obtain only two
families, we feel that the study of such a toy model is essential and
it has revealed extremely interesting features that may also apply to
a more realistic theory.

In section~\ref{3.2.1} we illustrate the basis of our construction and
describe the localization of families in the extra dimensions. We work
on the orbifold $T^2/\mathbb{Z}_2$ and start from six vector-like
fermions with the Standard Model quantum numbers. With appropriate
parity assignments, after orbifolding we obtain two four-dimensional
chiral zero modes for every spinor, which we can identify with
($q_{1L},u_R,d_R,l_{1L},e_R,\nu_{eR}$) and
($q_{2L},c_R,s_R,l_{2L},\mu_R,\nu_{\mu R}$). In the absence of other
interactions these zero modes have a constant profile along the extra
dimensions, which would suggest that it is impossible to reproduce the
hierarchical fermion spectrum.  However the picture drastically
changes if we localize the two families of fermions in different
regions of the extra space, in such a way that the introduction of a
non constant Higgs profile can reproduce the measured fermion mass
spectrum. We achieve this aim by adding a Dirac mass term for every
fermion to the lagrangian, where parity assignments to fields require
that the mass should have an odd profile. We thus choose a mass
proportional to the periodic sign function along one of the two extra
coordinates (and constant along the other), with the proportionality
constant different for each fermion. By solving the new equations of
motion we obtain the shape of zero modes that are now localized around
the lines where the six-dimensional mass changes sign, where the
amount of localization depends on the absolute value of this mass.

We have described how we obtain two sets of identical four-dimensional
fermions localized in two different regions of the extra space and if
we introduce a non constant Higgs vev we can attempt to derive the
fermionic mass spectrum. For the sake of simplicity we adopt a Higgs
vev that is completely localized on the brane around which the second
generation lives. In section~\ref{masses} we compute the mass spectrum
and we observe that these masses are naturally hierarchical, or rather
from order one parameters of the fundamental theory we obtain a
hierarchical pattern of masses. Moreover, as Majorana masses are
allowed in six dimensions, the smallness of neutrino masses could be
potentially explained through a higher-dimensional see-saw
mechanism. These results are of course very interesting, but
unfortunately our toy model contains many parameters and therefore its
predictability is weak.

In section~\ref{3gen} we discuss how to extend our toy model to a more
realistic scenario with three generations. As the most promising
framework, we suggest a ten-dimensional space-time, where Majorana
masses are allowed and fermions contains enough four-dimensional
components. We have just begun to investigate this proposal and a lot
of work is still required. However the toy model we present is
certainly an important step towards a realistic construction that
simultaneously addresses both the flavour and fermion replica problem.

\clearemptydoublepage
\pagestyle{headings}

\chapter[Symmetry Breaking in Extra Dimensions]{Symmetry Breaking in Extra Dimensions}
\label{chap1}
%
%


In this chapter we briefly review the main issues on symmetry breaking
in extra dimensions already present in the literature. In
section~\ref{SYMbreak} we describe the main features of the
Scherk-Schwarz (SS) mechanism, of the Hosotani mechanism and of
orbifold compactification and we discuss how they can break
symmetries. In particular in section~\ref{SUSY-breaking} we consider
supersymmetry (SUSY), while in section~\ref{gauge-breaking} we analyze
gauge symmetry, discussing in details some examples. In these sections
we also briefly outline some realistic models that exploit these
symmetry breaking mechanisms. Of course this is far from being a
complete phenomenological review on theories in extra dimensions, but
we simply discuss some examples in order to give an idea of the
importance of these new methods for symmetry breaking.

Before analyzing in detail the problem of symmetry breaking, we would
like to discuss some general features of theories in extra dimensions.
We consider a $D$-dimensional space-time ($D=4+d$) of coordinates
$(x^\mu,y^\alpha)$ with $\mu=0,1,2,3$ and
$\alpha=1,...,d$\,\footnote{Alternatively we can use the notation
$x^M$ with $M=0,1,2,3,5,...,D$.}.  The extra dimensions can be
factorizable or non-factorizable. If they are factorizable the
space-time is given by the product of the Minkowsky space $M_4$ times
a compact space $C$ and the line element is $ds^2\, =\,
\eta_{\mu\nu}\, dx^\mu\, dx^\nu\, +\, dy_\alpha^2$. On the contrary,
if the space-time is non-factorizable, the line element is $ds^2\,
=\,a(y^\alpha)\, [\eta_{\mu\nu}\, dx^\mu\, dx^\nu]\, +\, dy_\alpha^2$
and we cannot isolate $M_4$. From here on we forget this last case
(for references see~\cite{warped}) and we deal only with factorizable
geometry. Before entering into calculations we have to define the
metric. As we shall see, we will adopt different conventions about the
metric in chapters~\ref{chap2} and~\ref{chap3}, but the important
thing is that all the spatial coordinates have the same sign.

We suppose to work on the space-time $M_4\times C$, where $C$ is a
$d$-dimensional torus $T^d$ of radii $R_1,...,R_d$. The action in $D$
dimensions is defined by:
\be
\label{intro-actionD}
S_D=\int d^4x\,d^dy\, \cL_D(\phi,\partial\phi) 
\ee
and the four-dimensional (4D) lagrangian is obtained after integration
of the compact coordinates $y^\alpha$ as
\be
\label{intro-lagr4}
\cL_4=\int d^dy\, \cL_D(\phi,\partial\phi)~. 
\ee
The field $\phi(x^\mu,y^\alpha)$ represents a generic field depending
on the whole set of coordinates. Since the extra coordinates are
compact we can develop $\phi$ in Fourier series along $y^\alpha$:
\be
\label{intro-Fourier}
\phi(x^\mu,y^\alpha)=\sum_{n_1 ... n_d=-\infty}^\infty
e^{\dd \,i\, \bigg(\frac{n_1}{R_1}y^1+...+\frac{n_d}{R_d}y^d\bigg)} 
\phi_{n_1 ... n_d}(x^\mu)~,~~~~~~n_i\in\mathbb{Z}~.
\ee
Each $\phi_{n_1 ... n_d}(x^\mu)$ is a 4D field called Kaluza-Klein (KK)
mode. From a 4D point of view it corresponds to a field of mass square
\be
\label{intro-kkmass}
m^2_{n_1 ... n_d}=\bigg(\frac{n_1}{R_1}\bigg)^2
+ ... + \bigg(\frac{n_d}{R_d}\bigg)^2~.
\ee
We can show this with a simple example, supposing $\phi$ to be a real
massless D-dimensional scalar field. The lagrangian reads:
\be
\label{intro-lagrphi}
\cL_D = \frac{1}{2} \partial_M \phi\, \partial^M \phi
\ee
and the corresponding equation of motion\footnote{Here we choose to
work with the metric $\eta_{MN}=\textrm{diag}(-1,+1,...,+1)$.} is:
\be
\label{intro-eomphi} 
\partial_\mu\partial^\mu\phi + \partial_{y^1}\partial^{y^1}\phi + ... +
\partial_{y^d}\partial^{y^d}\phi = 0~.
\ee
Substituting eq.~(\ref{intro-Fourier}) into eq.~(\ref{intro-eomphi})
we obtain:
\bea
\label{intro-eomkk}
\sum_{n_1 ... n_d=-\infty}^\infty 
e^{\dd \,i\, \bigg(\frac{n_1}{R_1}y^1+...+\frac{n_d}{R_d}y^d\bigg)}\times
~~~~~~~~~~~~~~~~~~~~~~~~~~~~~~~~~~~~~~~~~~~~~~~~~~~~~\\
\times \left[
\partial_\mu\partial^\mu\phi_{n_1 ... n_d}(x^\mu) - 
\bigg(\frac{n_1}{R_1}\bigg)^2 \phi_{n_1 ... n_d}(x^\mu) - ... -
\bigg(\frac{n_d}{R_d}\bigg)^2 \phi_{n_1 ... n_d}(x^\mu) \right] = 0\nn
\eea
which is equivalent to
\be
\label{intro-eom}
\partial_\mu\partial^\mu\phi_{n_1 ... n_d}(x^\mu) - 
\left[ \bigg(\frac{n_1}{R_1}\bigg)^2 + ... + \bigg(\frac{n_d}{R_d}\bigg)^2
\right]  \phi_{n_1 ... n_d}(x^\mu) = 0 ~~~~~~~~\forall~~~n_1 ... n_d~.
\ee
Eq.~(\ref{intro-eom}) is precisely the equation of motion of a 4D
massive scalar field with mass given by eq.~(\ref{intro-kkmass}).
Masses $m_{n_1 ... n_d}$ are called KK levels.

After this brief reminder on field theory in extra dimensions we can
proceed with the analysis of symmetry breaking, following the lines of
ref.~\cite{Quiros}.


\section[Mechanisms of Symmetry Breaking]{Mechanisms of Symmetry Breaking}
\label{SYMbreak}

\subsection{Compactification}
\label{compactification}

We consider the space-time $M_4\times C$, where $M_4$ is the usual
Minkowsky space while $C$ is a compact $d$-dimensional space. In
general we can write $C=M/G$, where $M$ is a (non-compact) manifold
and $G$ is a discrete group acting freely on $M$ by operators
$\tau_g:\ M\to M$ for $g\in G$. $M$ is defined the covering space of
$C$. That $G$ is acting freely on $M$ means that only $\tau_{\imath}$
has fixed points in $M$, where $\imath$ is the identity in $G$. In our
case we have $\tau_{\imath} (y)=y,\ \forall y\in M$. The operators
$\tau_g$ constitute a representation of $G$, which means that
$\tau_{g_1 g_2}=\tau_{g_1}\cdot\tau_{g_2}$. Finally $C$ is constructed
by the identification
\be
\label{comp-id}
y \equiv \tau_g(y)~.
\ee

To be more concrete we focus on a simple example with one extra
dimension. We take $M=\mathbb{R}$, $G=\mathbb{Z}$ and $C=S^1$ (the
circle). The $n$-th element of the group $\mathbb{Z}$ can be
represented by $\tau_n$ with
\be
\label{comp-taun}
\tau_n(y)=y+2\pi n R~,~~~ y\in \mathbb{R}~,~~~ n\in \mathbb{Z}
\ee
where $R$ is the radius of the circle $S^1$. The identification
(\ref{comp-id}) leads to the fundamental domain of length $2\pi R$, the
circle, as $[y,y+2\pi R)$ or $(y,y+2\pi R]$. The interval must be
opened at one end because $y$ and $y+2\pi R$ describe the same point
in $S^1$ and should not be counted twice. Any choice for $y$ leads to
an equivalent fundamental domain in the covering space $\mathbb R$. A
convenient choice is $y=-\pi R$ which leads to the fundamental domain
$(-\pi R,\pi R]$.

After the identification (\ref{comp-id}) the physics should not depend
on individual points in $M$ but only on points in $C$. This means:
\be
\label{comp-idlag}
S_D[\phi(x,y)]=S_D[\phi(x,\tau_g(y))] ~.
\ee
A sufficient condition to fulfill eq.~(\ref{comp-idlag}) is
\be
\label{comp-ord}
\phi(x,\tau_g(y))=\phi(x,y)
\ee
which is known as ordinary compactification. However condition
(\ref{comp-ord}) is sufficient but not necessary. In fact a more
general condition to satisfy eq.~(\ref{comp-idlag}) is provided by
\be
\label{comp-SS}
\phi(x,\tau_g(y))=T_g\phi(x,y)
\ee
where $T_g$ are the elements of a symmetry group of the
theory. Condition (\ref{comp-SS}) is known as SS compactification and
and will be the subject of the next section.

\subsection{The Scherk-Schwarz Mechanism}
\label{scherk-schwarz}

The Scherk-Schwarz mechanism was introduced in 1979 first for
``external'' symmetries, i.\,e. that do not involve the
space-time~\cite{SS1}, and then for ``internal'' symmetries involving
space-time transformations~\cite{SS2}. It applies to theories which
are invariant under the transformations of some symmetry group and it
occurs when the operator $T_g$ ($g\in G$) of eq.~(\ref{comp-SS}) is
different from the identity. We say that in this case we have a
twist. The operators $T_g$ are a representation of the group $G$
acting on field space, i.\,e. they satisfy the property: $T_{g_1
g_2}=T_{g_1} T_{g_2},\ g_1,g_2\in G$. The SS compactification reduces
to ordinary compactification when $T_g=1,\ \forall g\in G$. Both for
ordinary and SS compactifications fields are functions on the covering
space $M$, but while for ordinary compactification fields are also
functions on the compact space $C$, in the twisted case fields are not
single-valued on $C$.

In order to give a simple explanation of how this mechanism works, we
consider the example of the previous section. We work on the circle
$S^1$ and we have $G=\mathbb{Z}$.  This group has infinitely many
elements but all of them can be obtained from just one generator, the
translation $2\pi R$. Then only one independent twist can exist acting
on the fields, as
\be
\label{SStwist}
\phi(x,y+2\pi R)=T\, \phi(x,y)~,
\end{equation}
while twists corresponding to the other elements of $\mathbb{Z}$ are
just given by $T_n=T^n$.  For simplicity we consider one complex
scalar field $\phi$ and we assume that the theory is invariant under
$U(1)$ transformations on this field.  Then $T$ can be written as:
\be
\label{SSphase}
T=e^{\dd 2\pi i \beta}~.
\ee
With this twist $\phi$ is no more periodic and the development in
Fourier series becomes:
\be
\label{SSfourier}
\phi(x^\mu,y)= e^{\dd i\,\frac{\beta}{R}\,y}
\sum_{n=-\infty}^\infty e^{\dd \,i \frac{n}{R}y} \phi_{n}(x^\mu)~.
\ee
If we calculate the KK levels we observe that they are shifted by a
constant amount and precisely they are:
\be
\label{SSmass}
m_n=\frac{n+\beta}{R}~.
\ee
If instead of a single field we consider a multiplet of some symmetry
group $F$ and we assign different periodicity conditions to different
members of the multiplet, we will obtain a breaking of $F$ in 4D,
since after compactification some components of the multiplet will
have the usual KK levels, while others will have levels shifted by a
constant amount. In particular if we look at the zero modes, only
periodic fields maintain them and the symmetry $F$, at the level of
the zero modes, is spoiled.

All what discussed above can be easily generalized to $p$-extra
dimensions, with $M=\mathbb{R}^p$, $G=\mathbb{Z}^p$ and $C=T^p$ is the
$p$-torus. In that case the torus periodicity is defined by a lattice
vector $\vec{v}=(v_1,\dots,v_p)$, where $v_i=2\pi R_i$ and $R_i$ are
the different radii of $T^p$.  Twisted boundary conditions are defined
by $p$ independent twists given by $T_i$
\be
\label{SStwistgen}
\phi(x,y^i+ 2\pi R_i)=T_{i}\,\phi(x,y^i)~.
\ee
%

\subsection{Orbifold}
\label{orbifold}

Firstly introduced in string theory, orbifolding is a technique used
to obtain chiral fermions from a (higher-dimensional) vector-like
theory~\cite{orb-vafa-witten-ecc}. Orbifold compactification can be
defined in a similar way to ordinary or SS compactification. Let $C$
be a compact manifold and $H$ a discrete group represented by
operators $\zeta_h:\ C\to C$ for $h\in H$ acting non freely on $C$. We
mod out $C$ by $H$ by identifying points in $C$ which differ by
$\zeta_h$ for some $h\in H$ and require that fields defined at these
two points differ by some transformation $Z_h$, a global or local
symmetry of the theory:
\bea
\label{orb-comp}
y &\equiv & \zeta_h(y)\nonumber\\
\phi(x,\zeta_h(y))&= & Z_h \phi(x,y)~.
\eea
The fact that $H$ acts non-freely on $C$ means that some
transformations $\zeta_h$ have fixed points in $C$. The resulting
space $O=C/H$ is not a smooth manifold but it has singularities at the
fixed points: it is called orbifold.

To illustrate this with a simple example we continue with the case
analyzed in the previous section with $d=1$ and $C=S^1$. Now we can
take $H=\mathbb{Z}_2$ and the resulting orbifold is
$O=S^1/\mathbb{Z}_2$. The action of the only non-trivial element of
$\mathbb{Z}_2$ (the inversion) is represented by $\zeta$ where
\be
\label{orb-z2}
\zeta(y)=-y
\ee
that obviously satisfies the condition
$\zeta^2(y)=\zeta(-y)=y\Rightarrow\zeta^2=1$. For fields we can write
as in (\ref{orb-comp})
\be
\label{orb-z2field}
\phi(x,-y)=Z\,\phi(x,y)
\ee
where using (\ref{orb-z2}) and (\ref{orb-z2field}) one can easily
prove that $Z^2=1$. This means that in field space $Z$ is a matrix
that can be diagonalized with eigenvalues $\pm 1$. The orbifold
$S^1/\mathbb{Z}_2$ is a manifold with boundaries and the boundaries
are the fixed points. Its fundamental domain is a segment of length
$\pi R$ and can be chosen to be the interval $[0,\pi R]$; the fixed
points are precisely $0$ and $\pi R$.  While all orbifolds possess
fixed points, not all possess boundaries: for example in $d=2$,
$T^2/\mathbb{Z}_2$ is a ``pillow'' with four fixed points and no
boundaries.

How this orbifold can break a symmetry? Suppose that $\phi$ is a
collection of $N$ fields which form a multiplet of some symmetry group
$F$ of the lagrangian and suppose that the matrix $Z$ does not
coincide with the identity. We can choose $Z$ with the first $p$
entries equal to $+1$ and the remaining ($N-p$) equal to $-1$. This
means that the first $p$ fields are even, while the others are
odd. When we develop them in KK modes we obtain for the even fields
\be
\label{orb-evenkkmodes}
\phi^+ (x^\mu,y) = \sum_{n=0}^\infty 
\cos\bigg(\frac{n}{R}\,y\bigg)\, \phi_n^+(x^\mu)
\ee
and for odd fields
\be
\label{orb-oddkkmodes}
\phi^- (x^\mu,y) = \sum_{n=1}^\infty 
\sin\bigg(\frac{n}{R}\,y\bigg)\, \phi_n^-(x^\mu)~.
\ee
We observe that the $\mathbb{Z}_2$-symmetry projects out half of the
tower of the KK modes of eq.~(\ref{intro-Fourier}). Moreover, after
orbifolding, only even fields maintain a zero mode. From the 4D point
of view the symmetry $F$ is broken down to a symmetry group $H\subset
F$.

\subsection{The Scherk-Schwarz Mechanism on Orbifold}
\label{ss-orb}

In this section we analyze the behaviour of the SS mechanism on
orbifold. In order to discuss the conditions holding among the
operators defining the orbifold parity and those defining the twist,
we begin by remembering how these operators were introduced. We
started from a non-compact space $M$ with a discrete group $G$ acting
freely on the covering space $M$ by operators $\tau_g$ ($g\in G$) and
defining the compact space $C=M/G$. The elements $g\in G$ are
represented on field space by operators $T_g$, eq.~(\ref{comp-SS}).
Subsequently we introduced another discrete group $H$ acting
non-freely on $C$ by operators $\zeta_h$ ($h\in H$) and represented on
field space by operators $Z_h$, eq.~(\ref{orb-comp}). We can always
consider the group $H$ as acting on elements $y\in M$ and then
considering both $G$ and $H$ as subgroups of a larger discrete group
$J$. Since in general $\tau_g\cdot
\zeta_h(y)\neq\zeta_h\cdot\tau_g(y)$, this means that $g\cdot h\neq
h\cdot g$ so we can conclude that $J$ is not the direct product
$G\otimes H$. Furthermore the twists $T_g$ have to satisfy some
consistency conditions. In fact from eqs.~(\ref{comp-SS}) and
(\ref{orb-comp}) one can easily deduce a set of identities as
\bea
\label{ssorb-comp}
T_g Z_h\,\phi(x,y)&=&\phi(x,\tau_g\cdot\zeta_h(y))\equiv
Z_{gh}\,\phi(x,y) \nonumber\\
Z_hT_g\,\phi(x,y)&=&\phi(x,\zeta_h\cdot\tau_g(y))\equiv
Z_{hg}\,\phi(x,y)\nonumber\\ 
T_{g_1}\, Z_h\,
T_{g_2}\,\phi(x,y)&=&\phi(x,\tau_{g_1}\cdot\zeta_h\cdot\tau_{g_2}(y))\equiv
Z_{g_1hg_2}\,\phi(x,y)
\eea
where $g_1,g_2,h$ are considered as elements in the larger group
$J$. The conditions (\ref{ssorb-comp}) impose compatibility constraints in
particular orbifold constructions with twisted boundary conditions as
we will explicitly illustrate in the following example.

We continue by analyzing the simple case of the orbifold
$S^1/\mathbb{Z}_2$ with twisted boundary conditions. In this case
there is only one independent group element for $G=\mathbb{Z}$ which
is the translation $\tau(y)=y+2\pi R$ while the orbifold group
$H=\mathbb{Z}_2$ contains only the inversion $\zeta(y)=-y$.  First of
all, notice that the translation and the inversion do not commute to
each other. In fact $\zeta\cdot\tau(y)=-y-2\pi R$ while
$\tau\cdot\zeta(y)=-y+2\pi R$. It follows then that
$\zeta\cdot\tau\cdot\zeta=\tau^{-1}$ and
$\tau\cdot\zeta\cdot\tau=\zeta$, which imply the consistency condition
on the possible twist operators
\be
\label{ssorb-cons}
Z\,T\,Z=T^{-1} ~\Leftrightarrow~ T\,Z\,T=Z~,
\ee  
as can be easily deduced from
eq.~(\ref{ssorb-comp})~\cite{UZU=Z,pq98}.

We now give an explicit example of how this condition constraints the
twist $T$. We consider a theory invariant under $SU(2)$ transformations
and we choose $\Phi$ to be a doublet of fields. Now both the parity
$Z$ and the twist $T$ are $2\times 2$ matrices. There are two
possibilities for $Z$: $Z=\sigma^3$ or $Z=\pm \mathbf{1}$. If we
require that condition~(\ref{ssorb-cons}) is satisfied, we obtain:
\be
\label{ssorb-final}
\Bigg\{
\begin{array}{lll}
Z=\sigma^3\; & ~\Rightarrow~ &
T=e^{\dd 2\pi i (\beta_1\sigma^1+\beta_2\sigma^2)}\\ 
Z=\pm \mathbf{1} & ~\Rightarrow~ & T=\pm \mathbf{1}~,
\end{array}
\ee
where $\beta_{1,2}$ are real parameters.  In the case of $Z=\sigma^3$,
 using a global residual invariance, we can rotate
$(\beta_1,\beta_2)\to(0,\omega)$ and consider twists given by
\be
\label{ssorb-twistfin}
T=e^{2\pi i \omega\sigma^2}=\left(
\begin{array}{rr}
\cos 2\pi \omega & \sin 2\pi\omega\\
-\sin 2\pi\omega &\cos 2\pi \omega
\end{array}
\right)~.
\ee
The twist (\ref{ssorb-twistfin}) is a continuous function of $\omega$
and so it is continuously connected with the identity that corresponds
to the trivial no-twist solution (i.\,e. $\omega=0$). In this way
eq.~(\ref{ssorb-twistfin}) describes a continuous family of solutions
to the consistency condition (\ref{ssorb-cons}).  In the case of
$Z=\pm\mathbf{1}$ eq.~(\ref{ssorb-cons}) implies that boundary
conditions can be either periodic or anti-periodic,
i.\,e. $T=\pm\mathbf{1}$.

\subsection{The Hosotani Mechanism}
\label{hosotani}

In this section we illustrate another symmetry breaking mechanism that
applies to local symmetries and it has been introduced by Hosotani in
1983~\cite{hosotani}.  It is based on the fact that the
extra-dimensional components of a gauge field can acquire a vev,
breaking the gauge symmetry itself. We show the main features of this
mechanism in a simple example in 5D, without discussing the problem of
the origin of the vev.

We consider the space-time $M^4\times S^1$ and a gauge theory based on
$SU(2)$. We write down the 5D lagrangian and we focus on the
quadri-linear part which contains the terms
\be
\label{hoso1}
A^{5i}\,A^{\nu j}\,(A_5^i\,A_\nu^j - A_5^j\,A_\nu^i)~,
\ee
where $i=1,2,3$ are the $SU(2)$ indices. We now assume that the fifth
component of $A_M^3$ acquires a vev: $\langle A_5^3\rangle\ne0$. This
may have a dynamical origin, coming from the minimization of the
1-loop effective potential in the presence of matter fields, as
discussed in~\cite{hosotani}. If we substitute this into
eq.~(\ref{hoso1}) we obtain:
\be
\label{hoso2}
\langle A_5^3\rangle^2\,(A^{\nu 1}\,A_\nu^1 + A^{\nu 2}\,A_\nu^2)~.
\ee
This is a mass term for fields $A_\nu^{1,2}$. If we calculated the KK
spectrum for gauge fields, we would obtain the usual KK spectrum for
$A_\nu^3$, while the KK levels for $A_\nu^{1,2}$ would be shifted by a
constant amount proportional to $\langle A_5^3\rangle$.  From a 4D
point of view the symmetry $SU(2)$ is broken down to $U(1)$.  The vev
$\langle A_5^3\rangle$ is a continuous parameter so it can be made as
small as we want and we can move continuously from a phase of broken
symmetry to another in which it is restored.

\subsection{Scherk-Schwarz vs Hosotani}
\label{ss-hoso}

In this section we will show that if the symmetry exploited by the SS
mechanism is local, this symmetry breaking mechanism is equivalent to
a Hosotani breaking, where the extra-dimensional components of the
corresponding gauge fields acquire a vev.

For simplicity we consider the case studied before of a 5D gauge
theory based on $SU(2)$. Now we do not postulate anything on the fifth
component of the gauge fields, but we assign the following twist to
the fields:
\be
\label{SShoso1}
\left (
\begin{array}{c}
A_M^1\\
A_M^2\\
A_M^3
\end{array}
\right )
(y+2\pi R)
=\left ( 
\matrix{ \cos\beta & \sin\beta & 0 \cr
        -\sin\beta & \cos\beta & 0 \cr
             0     &     0     & 1 }
\right )
\left (
\begin{array}{c}
A_M^1\\
A_M^2\\
A_M^3
\end{array}
\right )
(y)
\ee
The corresponding spectrum is:
\be
\label{SShoso2}
m_{1,2}=\frac{n}{R}-\frac{\beta}{2\pi R} ~~~~~~~~ m_3=\frac{n}{R}~.
\ee
Also in this case we observe that from a 4D point of view $SU(2)$ is
broken down to $U(1)$. Moreover if we make the identification
\be
\label{SShoso3}
\beta = - 2\pi R \langle A_5^3\rangle~,
\ee
the spectrum~(\ref{SShoso2}) is identical to the one calculated in the
Hosotani scheme. Are these two systems really equivalent? The answer
is yes, and it was given by Hosotani himself in one of his papers. He
showed that by performing a gauge transformation with non periodic
parameters one can switch from one picture to the other. In particular
if we start from the Hosotani scheme, in which fields are periodic and
$\langle A_5^3\rangle\ne0$, and we perform the gauge transformation
\be
\label{SShoso4}
A^\prime_M = \Omega^\dagger A_M \Omega - \Omega^\dagger \partial_M \Omega
\ee
with
\be
\label{SShoso5}
\Omega = e^{\dd i \langle A_5^3\rangle \frac{\sigma^3}{2} y}~,
\ee
we observe that $\langle A^{3\prime}_5\rangle = 0$. But this is not
the only effect of the gauge transformation. Indeed now fields
$A^\prime_M$ are twisted and the twist is precisely the one of
eq.~(\ref{SShoso1}) with $\beta$ given by~(\ref{SShoso3}).

In both pictures the symmetry is broken and we would like to have an
order parameter for symmetry breaking which is gauge independent. Such
a parameter exists and it is the Wilson line defined in the following
way:
\be
\label{SShoso   }
W=P\, exp \left\{ i g \int_0^{2\pi R} dy A_5(y) \right\}~,
\ee
where $P$ is a path-ordered integral.


\section[Phenomenological Implications]{Phenomenological Implications}
\label{phen}

In this section we apply the methods described above to gauge symmetry
and SUSY. The literature offers many realistic models in which these
symmetries are broken by means of the previously discussed
mechanisms. Of course we cannot discuss here all the existing
theories, thereby we choose two models which we consider particularly
suitable to our scopes. The first one, which we adopt to describe SUSY
breaking, was studied in ref.~\cite{pq98} and developed
in~\cite{after-pq}. It uses the orbifold projection and the SS
compactification to obtain the 4D Standard Model (SM) starting from a
SUSY theory in 5D. The second one, which we adopt to describe gauge
symmetry breaking, was introduced in ref.~\cite{kawa}, extended to
include fermions in ref.~\cite{AF} and completed in a realistic theory
in refs.~\cite{hn1,hn2}. It is a 5D SUSY Grand Unified Theory (GUT)
based on the gauge group $SU(5)$ which reduces to the SM through
orbifold and SS compactification.

\subsection{Supersymmetry Breaking}
\label{SUSY-breaking}

The model we consider is a 5D theory defined on the space-time
$M_4\times S^1/\mathbb{Z}_2$. It possesses $N=1$ SUSY in 5D, which
corresponds to $N=2$ in 4D. Non-chiral matter, as the gauge and Higgs
sectors of the theory, lives in the bulk of the fifth dimension, while
chiral matter, i.\,e. the three generations of quarks and leptons and
their superpartners, lives on the 4D boundary, i.\,e. at the fixed
points of the orbifold.

In 5D, the vector supermultiplet $(V_M,\lambda^i_L,\Sigma)$ of an
$SU(N)$ gauge theory consists of a vector boson $V_M$
($M=0,\dots,3,5$), a real scalar $\Sigma$ and two bispinors
$\lambda^i_L$ ($i=1,2$) (which are symplectic Majorana spinors), all
in the adjoint representation of $SU(N)$.  The 5D lagrangian is given
by
\be
\label{lagr1-pq}
{\cal L}=\frac{1}{g^2}{\rm Tr}\Big[-\frac{1}{2}{F_{MN}^2+|D_{M}\Sigma|^2+
i\overline\lambda_i\gamma^M D_M\lambda^i-
\overline\lambda_i}[\Sigma,\lambda^i]\Big]~.
\ee
The 5D matter supermultiplet, ($H_i,\Psi$), consists of two scalar
fields, $H_i$ ($i=1,2$), and a Dirac fermion $\Psi=(\Psi_L,\Psi_R)^T$.
We consider two matter supermultiplets, ($H^a_i$, $\Psi^a$) ($a=1,2$),
associated with the two Higgs doublets of the Minimal Supersymmetric
Standard Model~(MSSM).  The lagrangian for the matter supermultiplet
interacting with the vector supermultiplet is:
\bea
\label{lagr2-pq}
{\cal L} & = & |D_{M}H^a_i|^2+ i \overline\Psi_a\gamma^M D_M\Psi^a-
(i\sqrt{2}H^{\dagger i}_a \overline \lambda_i \Psi^a+h.c.)-
\overline\Psi_a\Sigma\Psi^a\nonumber\\ &-& H^{\dagger
i}_a\Sigma^2H^{a}_i- \frac{g^2}{2}\sum_{m,\alpha}[H^{\dagger i}_a
(\sigma^{m})^{j}_{i}T^\alpha H^{a}_j]^2 ~,
\eea
where $\sigma^m$ are the Pauli matrices.  The lagrangians of
eqs.~(\ref{lagr1-pq}) and (\ref{lagr2-pq}) have an $SU(2)_R\times
SU(2)_H$ global symmetry under which the fermionic fields transform as
doublets, $\lambda^i\in$({\bf 2},{\bf 1}), $\Psi^a\in$({\bf 1},{\bf
2}), while Higgs bosons transform as bidoublets $H^a_i\in$({\bf
2},{\bf 2}).  The rest of the fields in the vector multiplet are
singlets.

The model based on the SM gauge group is constructed from the above
lagrangians. It contains 5D vector multiplets in the adjoint
representation of $SU(3)\times SU(2)\times U(1)$ [({\bf 8},{\bf
1},0)+({\bf 1},{\bf 3},0)+({\bf 1},{\bf 1},0)] and two Higgs
hypermultiplets in the representation [({\bf 1},{\bf 2},1/2)+({\bf
1},{\bf 2},-1/2)]; the chiral matter is located on the boundary and
contains the usual chiral $N=1$ 4D multiplets\footnote{For details on
auxiliary fields, SUSY transformations, etc. see
ref.~\cite{Quiros}.}.

Since we work on an orbifold we have to impose consistent parity
assignments to fields. These are displayed in table 1.1, where we have
rearranged fields in components of $N=1$ $D=4$ SUSY multiplets that
are disposed along the same column.
\begin{table}
\label{1.1}
\begin{center}
\begin{tabular}{|c|c|c|c|c|c|} \hline
\multicolumn{3}{|c|}{Even fields} &\multicolumn{3}{|c|}{Odd fields}\\ 
\hline
$V_\mu$ & $H^2_2 $ & $H^1_1$ & $V_5,\Sigma$ & $H^2_1$ & $H^1_2$ \\
$\lambda^1_L$  & $\Psi^2_L$ & $\Psi^1_R$ & $\lambda^2_L$ & $\Psi^2_R$ &
$\Psi^1_L$ \\ 
\hline
\end{tabular}
\caption{Parity assignments for the vector multiplet and the
Higgs hypermultiplets.}
\end{center}
\end{table}   
We observe that, for $n\ne 0$, KK levels form massive $N=2$ 4D
multiplets.  The $V_5^{(n)}$ field is eaten by the vector
$V^{(n)}_\mu$ to become massive while $\lambda^{1\,(n)}_L$ and
$\lambda^{2\,(n)}_L$ become the components of a massive Dirac fermion.
These fields together with $\Sigma^{(n)}$ form an $N=2$ vector
multiplet.  $H^{a\,(n)}_i$ and $\Psi^{a\,(n)}$ form two $N=2$
hypermultiplets. These are displayed in table 1.2.
\begin{table}
\label{1.2}
\begin{center}
\begin{tabular}{|c|c|c|} 
\hline
Vector multiplet &\multicolumn{2}{|c|}{ Hypermultiplets}\\ 
\hline
$V_\mu^{(n)}$~~~$\Sigma^{(n)}$ & $H_1^{1\,(n)}$~~$H_2^{1\,(n)}$ &
$H_1^{2\,(n)}$~~$H_2^{2\,(n)}$ \\
$\lambda_L^{1\,(n)}$~~~$\lambda_L^{2\,(n)}$ & $\Psi^{1\,(n)}$ & 
$\Psi^{2\,(n)}$ \\  
\hline
\end{tabular}   
\caption{$N=2$ 4D multiplets for massive KK modes.}
\end{center}
\end{table}
On the contrary if we consider the zero modes we are left only with
even fields which form $N=1$ chiral multiplets, evidence that we still
have a SUSY theory. In particular we see that the massless spectrum of
this model coincide with the MSSM.  Finally we observe that at the
fixed points odd fields vanish $\forall \, n$. This means that at
boundaries we have $N=1$ SUSY, while $N=2$ holds only in the bulk for
massive modes.

Here we have shown how SUSY can be reduced by orbifold
projection. However we still have a residual $N=1$ SUSY which, in
order to obtain the SM, should be broken. To perform this further
breaking we use the SS mechanism and the symmetries we exploit for
twisting fields are $SU(2)_R$ and $SU(2)_H$. We impose the following
periodicity conditions:
\bea
\label{twist-pq}
\left (\matrix{\lambda^1\cr \lambda^2\cr }\right ) & = & 
e^{iq_R\sigma_2y/R}
\left (\matrix{\widetilde\lambda^1\cr \widetilde\lambda^2\cr }\right )~,\nn\\ 
\left (\matrix{\Psi^1\cr \Psi^2\cr}\right ) & = & 
e^{iq_H\sigma_2y/R}
\left (\matrix{\widetilde\Psi^1\cr \widetilde\Psi^2\cr }\right )~,\nn\\ 
\pmatrix{H^1_1 & H^1_2 \cr H^2_1 & H^2_2 \cr } & = & 
e^{iq_H\sigma_2y/R} \pmatrix{\widetilde H^1_1 & \widetilde H^1_2 \cr
\widetilde H^2_1 & \widetilde H^2_2 \cr }e^{-iq_R\sigma_2y/R}~.
\eea
Other fields are periodic, since they are singlets under
$SU(2)_R\times SU(2)_H$.  After the integration over the fifth
dimension we obtain the following mass spectrum for $n\ne 0$:
\bea
\label{massnotzero-pq}
{\cal L}_{m1}&=&
\frac{1}{R}\left\{\left (
\lambda^{1\,(n)}_L\, 
\lambda^{2\,(n)}_L
\right )
\pmatrix{
q_R & n \cr
n & q_R\cr
}\left (\matrix{
\lambda^{1\,(n)}_L\cr \lambda^{2\,(n)}_L
\cr
}\right ) + \right. \nonumber \\ 
&+&  \left.
\left (
\overline\Psi^{1\,(n)}_L\,
\overline\Psi^{2\,(n)}_L
\right )
\pmatrix{n&
-q_H  \cr
q_H&-n\cr
}\left (\matrix{
\Psi^{1\,(n)}_R\cr \Psi^{2\,(n)}_R\cr
}\right )+h.c.\right\} - \\ 
&-&{\dd \frac{1}{R^2}} \left (
H^{(n)\dagger}_0\, 
  H^{(n)\dagger}_2\, 
H^{(n)\dagger}_1\,  
H^{(n)\dagger}_3 \right )
M_H
\left (\matrix{
H^{(n)}_0\cr
H^{(n)}_2\cr
H^{(n)}_1\cr
H^{(n)}_3
}\right )\nonumber 
\eea
with
\be
\label{matrix}
M_H = \left (\matrix{
n^2+q_-^2&  -2inq_-&  &  \cr
2inq_- & n^2+q_-^2 &  &  \cr
 &  & n^2+q_+^2 & -2nq_+ \cr
 &  & -2nq_+ & n^2+q_+^2 \cr
}\right )
\ee
where we have redefined the scalar fields as
$H^a_i=H_\mu(\sigma^\mu)^a_i$, $\sigma^\mu\equiv(1,\vec{\sigma})$, and
$q_{\pm}=q_H\pm q_R$.  Therefore the massive KK modes are now given by
two Majorana fermions $(\lambda^{1\,(n)}_L\pm \lambda^{2\,(n)}_L)$
with masses $|n\pm q_R|$, two Dirac fermions, $(\Psi^{1\, (n)}
\pm\Psi^{2\, (n)})$ with masses $|n\pm q_H|$, and four scalars,
$(H^{(n)}_0\pm i\;H^{(n)}_2)$ and $(H^{(n)}_1\pm H^{(n)}_3)$ with
masses $|n\pm(q_R-q_H)|$ and $|n\pm(q_R+q_H)|$ respectively. The mass
spectrum of the fields $V^{(n)}_\mu$, $V_5^{(n)}$ and $\Sigma^{(n)}$
is not modified by the SS compactification, since they are periodic.

For $n=0$, we have
\bea
\label{masszero-pq}
{\cal L}_{m2}&=&\frac{1}{R}\left\{q_R\lambda^{1\;(0)}_L\lambda^{1\;(0)}_L+
q_H\overline\Psi^{2\;(0)}_L\Psi^{1\;(0)}_R+h.c.\right\}\nonumber\\
&-&\frac{1}{R^2}\left\{(q_R-q_H)^2\left|H_0^{(0)}\right|^2+
(q_R+q_H)^2\left|H_3^{(0)}\right|^2\right\}~.
\eea
The massless spectrum now consists of only the $n=0$ mode of the
vector fields $V^\mu$. Nevertheless a massless scalar Higgs can be
obtained if either $q_R-q_H=n$ or $q_R+q_H=n$ ($n\in \mathbb{Z}$) are
satisfied. Therefore, after SS compactification, the massless spectrum
of the model can be reduced to the SM, with one or two Higgs doublets.
From eq.~(\ref{masszero-pq}) we can see how the SS mechanism breaks
SUSY in the massless sector: it provides a mass for gauginos
$M_{1/2}=q_R/R$ and a mass for Higgsinos $\mu=q_H/R$, providing an
extra-dimensional solution to the MSSM $\mu$-problem.

As we said before, the fermionic sector is localized at $y=0$. Since
the theory is SUSY there are scalar partners for all fermions (squarks
and sleptons) which are massless at tree level.  How can we break this
SUSY? Sfermions (like fermions) are 4D, so we cannot break it through
compactification, since they are independent from the fifth
coordinate. But if we calculate the radiative
correction~\cite{after-pq} we find non vanishing contributions to the
sfermions masses both from the gauge and the Higgs sectors at one
loop. These contributions are finite and transmit SUSY breaking from
the bulk to the boundaries.

In the model just described, after compactification, we are left with
the SM, with its good qualities but also its defects. In particular we
have not improved our knowledge of electroweak symmetry breaking
(ESB). However this can be done with little modifications, as it is
shown in ref.~\cite{BHN} where a realistic and testable model have
been built. At variance with before, now also fermions live in the
bulk and only one Higgs field is introduced. The two SUSYs are broken
with two orbifolds, which is equivalent to the previous case with an
appropriate twist parameter. The novelty is that ESB is triggered by
the interactions of the top quark KK modes with the Higgs. These give
rise to an effective potential which contains a negative mass scale
depending on the compactification scale $R$ which induces spontaneous
breaking. The Fermi constant can thus be used to determine $R$ which
results to be $R^{-1}=352\pm 20~GeV$; consequently the Higgs mass
turns out to be $m_H=127\pm 8~GeV$. This predictions, and others which
are discussed in the paper, make this model testable at LHC.

\subsection{Gauge Symmetry Breaking}
\label{gauge-breaking}

We begin with the model introduced by Kawamura in ref.~\cite{kawa} and
then extended in ref.~\cite{AF} with the introduction of fermions. It
is a 5D SUSY theory defined on the orbifold
$S^1/\mathbb{Z}_2\times\mathbb{Z}^\prime_2$ based on the gauge group
$SU(5)$.  The bulk fields content is the following: there is a vector
multiplet $V = (A^\alpha_M, \lambda^{1\alpha}_L, \lambda^{2\alpha}_L,
\Sigma^\alpha )$, with $\alpha = 1,...,24$ and $M=\mu ,5$, which forms
an adjoint representation of $SU(5)$, and two hypermultiplets $H^1$
and $H^2$ equivalent to four chiral multiplets $H_5$,
$\widehat{H}_{\overline{5}}$, $\widehat{H}_{5}$ and $H_{\overline{5}}$
which form two fundamental representations of $SU(5)$.  The gauge
vector bosons are the fields $A^\alpha_M$: $A^a_M$ with $a=1,...,12$
are the SM vector bosons , while $A^{\hat{a}}_M$ with
$\hat{a}=13,...,24$ are those of the coset $SU(5)~/~SU(3)\times
SU(2)\times U(1)$. The bulk lagrangian is given by
eqs.~(\ref{lagr1-pq}) - (\ref{lagr2-pq}).

\begin{table}[!t]
\label{kawamura}
\centering{
\begin{tabular}{|c|c|c|}
\hline
{} &{} &\\[-1.5ex]
Fields & $(Z,Z^\prime)$  & $m$ \\
{} &{} &\\[-1.5ex]
\hline
{} &{} &\\[-1.5ex]
$A^{a}_{\mu}, \lambda_{L}^{2a}, H^{D}_{u}, H^{D}_{d}$ &
$(+,+)$   & $\frac{2n}{R}$ \\
{} &{} &\\[-1.5ex]
\hline
{} &{} &\\[-1.5ex]
$A^{\hat{a}}_{\mu}, \lambda_{L}^{2\hat{a}}, H^{T}_{u}, H^{T}_{d}$
&
$(+,-)$  & $\frac{2n+1}{R}$ \\
{} &{} &\\[-1.5ex]
\hline
{} &{} &\\[-1.5ex]
$A^{\hat{a}}_{5}, \Sigma^{\hat{a}},\lambda_{L}^{1\hat{a}},
\widehat{H}^{T}_{u}, \widehat{H}^{T}_{d}$ &
$(-,-)$  & $\frac{2n+1}{R}$ \\
{} &{} &\\[-1.5ex]
\hline
{} &{} &\\[-1.5ex]
$A^{a}_{5}, \Sigma^{a},\lambda_{L}^{1a}, \widehat{H}^{D}_{u},
\widehat{H}^{D}_{d}$ &
$(-,+)$  & $\frac{2n+2}{R}$ \\[1ex]
\hline
\end{tabular}}
\caption{Boundary conditions and spectra for fields in the Kawamura
model.}
\end{table}
Parity assignments, displayed in table 1.3, are given is such a way
that $\mathbb{Z}_2$ breaks one 4D SUSY, while $\mathbb{Z}^\prime_2$
breaks $SU(5)$ down to $SU(3)\times SU(2)\times U(1)$. After
orbifolding we thus obtain a massless spectrum which contains the
vector bosons of the SM, $A_\mu^a$, their SUSY partners,
$\lambda^{2a}_L$, and two Higgs doublets $H^{D}_{u}$ and $H^{D}_{d}$
which are precisely the fields of the MSSM. These boundary conditions
solve the doublet-triplet splitting problem, since while the doublet
is massless, the lightest triplet mode has mass of order
$R^{-1}$. Since the unification of gauge couplings is achieved at the
compactification scale, it is natural to identify $R^{-1}$ with
$M_{GUT}\sim 10^{16}~GeV$. We observe that the compactification scale
is here much bigger then in the model of the previous section.

What about fermions? Although they are already present, localized on
the brane, in the model of ref.~\cite{kawa}, what follows is based on
ref.~\cite{AF} which offers a more detailed description. Fermions
cannot live in the bulk, unless we double the number of matter
fields. Moreover constant Yukawa couplings respecting also $SU(5)$
lead to vanishing up-type quark mass. Then matter fields and their
SUSY partners are introduced as multiplets of the SM gauge
groups. Proton decay is analyzed and it is shown it cannot proceed via
tree-level Higgsino or gauge boson exchange for any parity assignment
compatible with non vanishing fermion masses. Appropriate parity
assignments can also forbid 4D and 5D B/L violating operators, playing
the r\^ole of $R$-parity symmetry. The problem of proton decay is thus
solved and the good properties of traditional GUTs are
maintained. Indeed fermion mass relations are preserved, Dirac and
Majorana neutrino masses can be included and the see-saw mechanism can
be realized.

Although it solves two of the problems of 4D GUTs, the model described
above is far from being a realistic theory. First of all SUSY has
still to be broken and then we do not have yet a realistic fermion
mass spectrum. A first step in the direction of building a realistic
model have been done in ref.~\cite{hn1}, while a complete 5D GUT is
constructed in ref.~\cite{hn2}. 

\begin{figure}[!t]
\begin{center}
\epsfig{figure=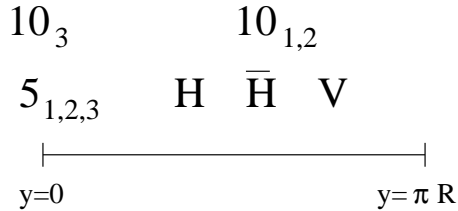}
\end{center}
\vspace{-0.5cm}
\caption{Locations of matter, Higgs and gauge multiplets in the model
of ref.~\cite{hn2}.}
\label{hn-fig}
\end{figure}

The model introduced by Hall and Nomura is an upgrade of previous
models, although it is based on the orbifold $S^1/\mathbb{Z}_2$ rather
than on $S^1/\mathbb{Z}_2\times\mathbb{Z}^\prime_2$. However we could
demonstrate there is an equivalence between the orbifold and the SS
mechanism, so in this case the r\^ole of the second parity is played
by a twist. Moreover, at variance with parity, the twist parameter can
be continuous and we will see that on this fact is based the breaking
of the residual SUSY. The gauge and Higgs fields content is identical
to the one of previously described models, with fields in the bulk and
boundary conditions breaking $SU(5)$ and one SUSY. The grand unified
group is broken on the brane at $y=\pi R$, while elsewhere it remains
unbroken. This structure allows the introduction of three types of
fields: 4D $N=1$ superfields localized on the $y=0$ brane in
representations of $SU(5)$, 4D $N=1$ superfields on the $y=\pi R$
brane in representations of the SM gauge group, and bulk fields
forming 5D $N=1$ supermultiplets in representations of $SU(5)$. In
order to preserve the understanding of matter quantum numbers given by
$SU(5)$, at variance with ref.~\cite{AF} where they were introduced in
representation of the SM, here quarks and leptons are put either in
the bulk or on the brane at $y=0$, in $SU(5)$ representations. In
principle one can choose where to put fermions for each $SU(5)$
representation in each generation but in practice the choice is unique if
we want a realistic theory. In agreement with ref.~\cite{AF}, the
authors show that Yukawa interactions are forbidden in the bulk by 5D
SUSY, but they also demonstrate that if all the matter fields were
localized on the brane, too rapid proton decay would be induced. In
order to avoid this, and considering also the size of the top Yukawa
coupling and $b/\tau$ unification, they find that the location of the
$\mathbf{10}$ of the first and third generations and of the
$\bar{\mathbf{5}}$ of the third is uniquely determined. The location
of other fermions is instead determined after the breaking of the
residual SUSY, which will be discussed in a while. The final location
of all the fields of the model is represented in fig.~\ref{hn-fig}.
SUSY is broken by a SS mechanism with twist parameter $\alpha$,
analogously to what done in the previous section. Since the breaking
parameter is continuous, at variance with the twist that breaks
$SU(5)$ which is discrete, this SUSY breaking can be reinterpreted \`a
la Hosotani, that is we have spontaneous symmetry breaking.

After the introduction of all the fields, and after compactification
with appropriate boundary conditions, we are left with the SM
fields. Then it is possible to work out the predictions of the
model. First of all the fermion mass spectrum is calculated and, even
if a part of the hierarchy must be introduced by hands with
appropriate Yukawa couplings, it is interesting to observe that we
have the typical $SU(5)$ mass relation ($m_b=m_\tau$) only for the
third generation, while there is a small mixing of the first two
generations with the third. Neutrino masses are generated via the
see-saw mechanism and large mixing is found in the neutrino
sector. Also the superpartners spectrum is calculated. In this model
SUSY breaking effects depend only on one parameter.  With a particular
choice of this parameter it is possible to give predictions for
$\alpha_s(M_Z)$ and $m_b(M_Z)$ which are in good agreement with
data. Moreover they calculate branching ratios for flavour violation
lepton decays which are found to be close to the present experimental
limits. A limit on proton decay is estimated and it results to be of
order of $10^{34}$ years. The first possible direct experimental
signal should be the observation of scalar fermions. Thus we can say
that also this model is realistic and testable, although it depends on more
parameters than the model of ref.~\cite{BHN}.

\clearemptydoublepage


\chapter[Generalized Scherk-Schwarz Mechanism]{Symmetry Breaking 
via Generalized Scherk-Schwarz Mechanism}
\label{chap2}
%
%
%

In section~\ref{SYMbreak} we have described in details the SS
mechanism. We have seen that on a circle $S^1$ one can twist the
periodicity conditions on the fields by a symmetry of the action. The
result is a shift in the KK levels of the spectrum and this can be
used to break symmetries. In this chapter we deal in more detail with
the orbifold $S^1/\mathbb{Z}_2$. Since an orbifold contains fixed
points, the boundary conditions are fully specified not only by the
periodicity of field variables, but also by the possible jumps of the
fields across the orbifold fixed points. These jumps are forbidden on
manifolds, where the fields are required to be smooth everywhere, but
are possible on orbifolds at the singular points, provided the
physical properties of the system remain well defined.

In section~\ref{GBCferm}, following the lines of ref.~\cite{bfz1}, we
study the general boundary conditions for fermions and we calculate
spectrum and eigenfunctions. Along similar lines, in
section~\ref{GBCbos}, based on ref.~\cite{bf}, we study the bosonic
case. In both cases KK levels are shifted precisely as in the SS
mechanism and a field redefinition exists mapping discontinuous fields
into continuous, twisted ones. Since the spectra are identical,
generalized boundary conditions can be used to break symmetries, as in
the case of twisted boundary conditions. In section~\ref{gauge} we
exploit them to break gauge invariance, as described in
refs.~\cite{bf,b}, while in section~\ref{susy} we apply our
considerations to SUSY breaking, as shown in
refs.~\cite{bfwz,bfz2}. From the point of view of the spectrum the two
mechanisms are identical; which are the differences? For free theories
the difference is only in the explicit form of the 5D mass
terms. Generalized boundary conditions lead to $y$-dependent mass
terms while twisted boundary conditions produce constant
ones. Different mass terms can give rise to the same 4D spectrum and
it is useful to determine the most general set of mass terms that
correspond to a given spectrum. All this is analyzed in
section~\ref{bfwz}, following the lines of ref.~\cite{bfwz}.

All along this chapter we will work in a 5D space-time with the extra
coordinate compactified on the orbifold $S^1/\mathbb{Z}_2$. The metric
$\eta_{MN}$ and the $\Gamma$ matrices we use are defined in the first
part of appendix~\ref{gamma-mat}.


\section[Generalized Boundary Conditions for Fermions]{Generalized 
Boundary Conditions for Fermions}
\label{GBCferm}

\subsection{Boundary Conditions for Fermionic Fields on $S^1/\mathbb{Z}_2$}
\label{GBCferm-bc}
We consider a generic 5D theory compactified on the orbifold
$S^1/\mathbb{Z}_2$ and we introduce a set of $n$ 5D fermionic fields
$\Psi(x,y)$, which we classify in representations of the 4D Lorentz
group. We define the $(T,\mathbb{Z}_2)$ transformations of the fields
by
\bea
\label{ferm-trasf}
\Psi(y+ 2 \pi R) &=& U_\beta~ \Psi(y)\nn\\ 
\Psi(-y) &=& Z~ \Psi(y)~,
\eea
where $U_\beta$ and $Z$ are constant unitary matrices and $Z$ has the
property $Z^2={\mathbf 1}$.  It is not restrictive for us to take a
basis in which $Z$ is diagonal,
with the first $p$ entries equal to $+1$ and the remaining $(n-p)$
equal to $-1$.

We look for boundary conditions on the fields $\Psi(x,y)$ in the
general class
\be 
\Psi(\gamma^+) = U_\gamma \ \Psi(\gamma^-)~,
\label{ferm-bc}
\ee
where $\gamma=(0, \pi,\beta)$, $\gamma^\pm=(0^\pm,\pi R^\pm, y^\pm)$
and $U_\gamma$ are constant $2n\times 2n$ matrices\footnote{In fact to
each fermion corresponds a $2\times 2$ matrix, since a 5D spinor is
composed by two 4D Weyl spinors.}.  We have defined
$0^\pm\equiv\pm\xi$, $\pi R^\pm\equiv\pi R\pm\xi$, $y^-\equiv y_0$ and
$y^+\equiv y_0+2 \pi R$. Here $\xi$ is a small positive parameter and
$y_0$ is a generic point of the $y$-axis, for convenience chosen
between $-\pi R+\xi$ and $-\xi$. $U_\beta$ is the operator associated
with the twist, while $U_{0,\pi}$ define the possible discontinuities
of fields at the fixed points.

We will now constrain $U_\gamma$ by imposing certain consistency
requirements on our theory.  The spectrum of the theory is determined
by the eigenmodes of the momentum along $y$, represented by the
differential operator $- i \partial_y$. In order to deal with a good
quantum mechanical system, we demand that this operator is
self-adjoint with respect to the scalar product:
\be
\langle\Psi \vert \Phi\rangle =
\int_{y^-}^{0^-} d y~ \Psi^{\dagger}(y) \Phi(y)+
\int_{0^+}^{\pi R^-} d y~ \Psi^{\dagger}(y) \Phi(y)+
\int_{\pi R^+}^{y^+} d y~ \Psi^{\dagger}(y) \Phi(y)~,
\label{ferm-sprod}
\ee
where the limit $\xi\to 0$ is understood. If we consider the matrix
element $\langle\Psi \vert \left(- i \partial_y \Phi \right)\rangle$
and we perform a partial integration we obtain:
\bea
\label{ferm-selfadj}
\langle\Psi \vert \left(-i\partial_y\Phi\right)\rangle &=&
\langle \left(-i\partial_y\Psi\right) \vert \Phi\rangle+\nn\\
&+ & i \left[ \Psi^{\dagger}(0^-)\Phi(0^-) - \Psi^{\dagger}(y^-)\Phi(y^-)
          + \Psi^{\dagger}(\pi R^-)\Phi(\pi R^-) - \right.\\
& & \left.-\ \Psi^{\dagger}(0^+)\Phi(0^+) + \Psi^{\dagger}(y^+)\Phi(y^+)
          - \Psi^{\dagger}(\pi R^+)\Phi(\pi R^+)\right]~.\nn
\eea
A necessary condition for the self-adjointness of the operator
$-i\partial_y$ is that the square bracket vanishes. However this is
not sufficient, in general. To guarantee self-adjointness, the domain
of the operator $-i\partial_y$ should coincide with the domain of its
adjoint. In other words, we should impose conditions on $\Phi(y)$ in
such a way that the vanishing of the unwanted contribution implies
precisely the same conditions on $\Psi(y)$. We observe that if
$U_{\gamma}$ are unitary all these requirements are satisfied and the
operator is self-adjoint.

Finally, we should take into account consistency conditions among the
twist, the jumps and the orbifold projection.  If we combine a
translation $T$ with a reflection $\mathbb{Z}_2$, we have seen that
the operators $U_\beta$ and $Z$ must satisfy the
relation~(\ref{ssorb-cons}): $U_\beta~Z~U_\beta=Z$.  An analogous
relation is also obtained if we combine a jump with a
reflection. Finally, each of the two possible jumps should commute
with the translation $T$. We thus have:
\bea
\label{cons-cond}
U_\gamma~ Z~ U_\gamma &=& Z~~~
~~~~~~~~~~~~~~\gamma\in (0, \pi , \beta)~,\nn\\
\left[U_0,U_\beta\right]&=& 0~,\\
\left[U_\pi,U_\beta\right]&=& 0~.\nn
\eea 
If $[Z,U_\gamma]=0$, then $U_\gamma^2={\mathbf 1}$ and twist and/or
jumps have eigenvalues $\pm 1$. In particular, if also $U_0$ and
$U_\pi$ commute, there is a basis where they are all diagonal with
elements $\pm 1$: in this special case the boundary conditions do not
involve any continuous parameter.  When $[Z,U_\gamma]\ne 0$ or when
$[U_0,U_\pi]\ne 0$ continuous parameters can appear in $U_\gamma$.

In summary the most general boundary conditions for a set of $n$ 5D
fermionic fields are
\be
\Psi(\gamma^+) = U_\gamma \ \Psi(\gamma^-)~~~~~~ 
\textrm{with}~~~~~~ U_{\gamma}^{\dagger}U_{\gamma}={\mathbf 1}
\label{ferm-finalbc}
\ee
and $U_{\gamma}$ satisfying conditions~(\ref{cons-cond}). Obviously,
in order to assign these generalized boundary conditions to fermions,
$U_{\gamma}$ must be a symmetry of the theory.

\subsection{One Fermion Field}
\label{GBCferm-one}
To illustrate how these general boundary conditions determine the
physics, we focus now on the case of a single 5D fermion. We start by
deriving the lagrangian and the equation of motion in terms of 4D
spinors and then we solve it with general boundary conditions.

A 5D spinor is composed by two 4D Weyl spinors and can be represented
with different notations:
\be
\begin{array}{cc}
\Phi = \left(\begin{array}{c}\psi_{1} \\
\overline{\psi}_{2}\end{array}\right)
&~~~~~~~~ (\textrm{A}) \\[0.5cm]
\Psi = \left(\begin{array}{c}\psi_{1} \\
\psi_{2}\end{array} \right)
&~~~~~~~~ (\textrm{B})\\
\end{array}~.
\label{ferm-notations}
\ee
With notations (A) we have $\overline{\Phi} = (\psi_{2}\
\overline{\psi}_{1})$ while with notations (B) we have
$\overline{\Psi} = (\overline{\psi}_{1}\ \overline{\psi}_{2})$. We
remember that in this case $\overline{\psi}$ is not the usual Dirac
$\overline{\Psi}=\Psi^{\dag} \gamma^0$, but it is defined by
$\overline{\psi}=\psi^{*}$.  Within formalism (A) we can write the 5D
lagrangian in the usual way:
\be
\cL (\Phi,\partial\Phi)=
i\overline{\Phi}\Gamma^M\partial_M\Phi~.
\label{ferm-lagr1}
\ee
By substituting the explicit expression for $\Phi$ and the
$\Gamma$-matrices we obtain the lagrangian in terms of 4D Weyl
spinors:
\be
\cL (\psi,\partial\psi)
=  i\, \overline{\psi}_1\, \bar{\sigma}^{\mu}\, \partial_{\mu} \psi_1 +%
i\, \overline{\psi}_2\, \bar{\sigma}^{\mu}\, \partial_{\mu} \psi_2 -%
\frac{1}{2}\, (\psi_1\, \partial_y \psi_2 - \psi_2\, \partial_y
\psi_1 + h.c.)~.
\label{ferm-lagrW}
\ee
For simplicity we can move to notation (B) and we rewrite the
lagrangian in the following way:
\be
\cL (\Psi, \partial\Psi) =
i\, \overline{\Psi}\, \bar{\sigma}^{\mu}\, \partial_{\mu} \Psi -
\frac{1}{2}\, (\Psi^T\, i\widehat{\sigma}^2\, \partial_y \Psi + h.c.)~.
\label{lagrBFWZ}
\ee
Here (and in the following) $\widehat{\sigma}^i$ are Pauli matrices
acting in the space $(\psi_1, \psi_2)$, while
$\bar{\sigma}^{\mu}$ are the usual matrices rotating the
components of the Weyl spinors $\psi_i$.

From eq.~(\ref{lagrBFWZ}) we can derive the equation of motion
for the fermion. Varying with respect to $\Psi$ we obtain:
\be
i\, \bar{\sigma}^{\mu}\, \partial_{\mu} \overline{\Psi}^T -
i\, \widehat{\sigma}^2\, \partial_y \Psi = 0~.
\label{ferm-eom1} 
\ee
Substituting the 4D equation of motion for Weyl spinors $i\,
\bar{\sigma}^{\mu}\, \partial_{\mu} \overline{\psi} = m\, \psi$ into
eq.~(\ref{ferm-eom1}), we finally obtain:
\be
i\, \widehat{\sigma}^2\, \partial_y \Psi = m\, \Psi~.
\label{ferm-eom2}
\ee

We solve this equation of motion first with the usual SS twisted
boundary conditions and then in a more general case including also
jumps.

With respect to the $\mathbb{Z}_2$ reflection that defines the
orbifold, we adopt the following parity assignment:
\be \Psi(-y) = Z~ \Psi (y) ~, ~~~~~~
Z=\widehat{\sigma}^3~. 
\label{ferm-parity} 
\ee
We observe that $\cL$ is not only invariant under $\mathbb{Z}_2$, as
required for the consistency of the orbifold construction, but also
under
\be 
\Psi'(y) = U \, \Psi(y) ~, 
\label{su2} 
\ee
where $U$ is a global $SU(2)$ transformation\footnote{Notation (B) is
useful to display this symmetry.}.  In this framework the field
$\Psi(y)$ does not need to be periodic and continuous, but
conditions~(\ref{ferm-finalbc}) can be adopted with $U_{\gamma}\in
SU(2)$. The most general twist we can assign to the fermion is the
following:
\be 
\Psi(y+2 \pi R)=U_{\vec{\beta}} \Psi(y) ~,~~~~~~
U_{\vec{\beta}} \equiv \displaystyle{ e^{i
\vec{\beta} \cdot \vec{\sigma} } } = \cos \beta \, {\mathbf 1} +
{\sin \beta \over \beta} \, \vec{\beta} \cdot \vec{\sigma} ~,
\label{twistBFWZ} 
\ee
where $\vec{\sigma} = (\widehat{\sigma}^1, \widehat{\sigma}^2,
\widehat{\sigma}^3)$, $\vec{\beta} = (\beta^1, \beta^2, \beta^3)$ is a
triplet of real parameters, $\beta$ is the absolute value of the
vector $\vec{\beta}$, $\beta \equiv \sqrt{\beta_1^2 + \beta_2^2 +
\beta_3^2}$, and it is not restrictive to assume $\beta \le \pi$. The
operators $Z$ and $U_{\vec{\beta}}$ acting on the fields should
satisfy consistency conditions~(\ref{cons-cond}) which, for
$Z=\widehat{\sigma}^3$, implies~\footnote{The choice
$\vec{\beta}=(0,0,\pi)$ gives $U_{\vec{\beta}} = - {\mathbf 1}$, as
any other choice with $\beta = \pi$.}
\be 
\vec{\beta}=(\beta_1,\beta_2,0)~. 
\label{betaBFWZ} 
\ee
The 4D modes have a spectrum characterized by a universal shift of KK
 levels, with respect to the mass eigenvalues $n/R$ of the periodic
 case, controlled by $\beta$:
\be 
m= \frac{n}{R} - \frac{\beta}{2\pi R} ~, ~~~~~~
(n \in {\mathbb Z}) ~. 
\label{ferm-massesBFWZ} 
\ee
The corresponding eigenfunctions are
\be 
\Psi^{(n)}(x,y)= \chi(x) \, e^{\dd{\, i \, \gamma  \,
\widehat{\sigma}^3}} \left(
\begin{array}{c}
\cos m y\\
\sin m y
\end{array}
\right) ~, 
\label{ferm-eigfunBFWZ} 
\ee
where $\chi(x)$ is $y$-independent 4D Weyl spinor satisfying the
equation $i \sigma^m \partial_m \, \ov{\chi} = m \, \chi$, and,
barring the trivial case $\vec{\beta} = 0$ in which the rotation
angle $\gamma$ is arbitrary: 
\be 
\gamma = {1 \over 2} \arctan
\left( \beta_1 \over \beta_2 \right) + \delta + \rho \pi ~, ~~~~~~
\delta = \left\{
\begin{array}{lcc}
0      & {\rm for} & \beta_2 \ge 0 \\
\pi/2  & {\rm for} & \beta_2<0
\end{array}
\right. ~, ~~~~~~
(\rho \in {\mathbb Z}) ~. 
\label{gammaBFWZ}
\ee
Here we have reproduced the usual SS mechanism: starting from twisted
fields we obtained a shift in the KK spectrum proportional to the
twist parameter itself. In the following we will show that the same
spectrum can be obtained also by means of jumps or a combination of
both.

For the sake of simplicity we do not consider the most general
boundary conditions but we focus on a simpler case with
\bea
\label{ferm-jumps}
U_{\beta} &\equiv~~ e^{\dd{i \beta \widehat{\sigma}^2}} &=~~ 
\left( \begin{array}{cc}
       \cos \beta & \sin \beta \\
     - \sin \beta & \cos \beta
       \end{array} \right) \nn                      \\
U_0 &\equiv~~ e^{\dd{i \delta_0 \widehat{\sigma}^2} } &=~~ 
\left( \begin{array}{cc}
       \cos \delta_0 & \sin \delta_0 \\
     - \sin \delta_0 & \cos \delta_0
       \end{array} \right)    ~.                    \\
U_{\pi} &\equiv~~ e^{\dd{i \delta_{\pi} \widehat{\sigma}^2} } &=~~
\left( \begin{array}{cc}
       \cos \delta_{\pi} & \sin \delta_{\pi} \\
     - \sin \delta_{\pi} & \cos \delta_{\pi}
       \end{array} \right) \nn 
\eea
The solution to eq.~(\ref{ferm-eom2}) with boundary
conditions~(\ref{ferm-jumps}) is the following:
\be
\Psi^{(n)}(x,y)= \chi(x) 
\left(
\begin{array}{c}
\cos \left[ m y -\alpha (y) \right]\\
\sin \left[ m y -\alpha (y) \right]
\end{array}
\right) ~, 
\label{ferm-eigfun} 
\ee
with
\be
m= \frac{n}{R} - \frac{\beta - \delta_0 - \delta_{\pi}}{2\pi R} ~, ~~~~~~
(n \in {\mathbb Z})~. 
\label{ferm-mass} 
\ee
Here $\alpha(y)$, depicted in fig.~\ref{aofy} for some representative
choices of $\delta_0$ and $\delta_\pi$, is given by:
\begin{figure}[!t]
\begin{center}
\epsfig{figure=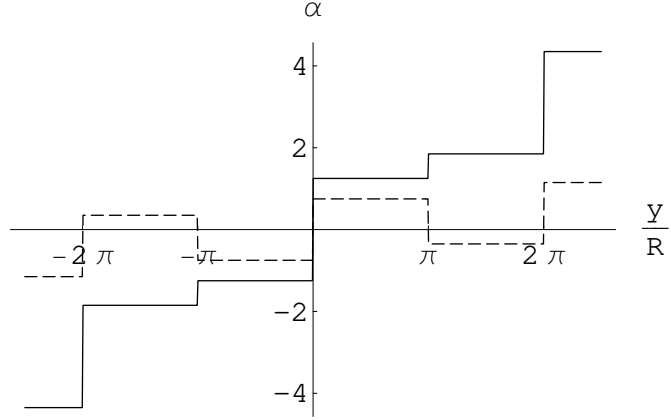,height=6cm}
\end{center}
\caption{The function $\alpha(y)$ for two representative parameter
choices: the solid line corresponds to $\delta_0 = 2.5$,
$\delta_\pi = 0.6$, the dashed one to $\delta_0 = 1.5$,
$\delta_\pi = -1.1$.}
\label{aofy}
\end{figure}
\be
\alpha(y)=\frac{\delta_0 - \delta_{\pi}}{4} \epsilon(y) + 
          \frac{\delta_0 + \delta_{\pi}}{4} \eta(y) ~,
\label{alpha}
\ee
where $\epsilon(y)$ is the periodic sign function defined on $S^1$ and 
\be
\eta(y)=2q+1 ~,~~~~~~ q\,\pi R< y < (q+1)\,\pi R~,~~~~~~(q\in {\mathbb Z})
\label{eta}
\ee
is the staircase function that steps by two units every $\pi R$ along
$y$. The function $\alpha(y)$ satisfies
\be
\alpha(y+2\pi R)=\alpha(y)+\delta_0+\delta_{\pi}~,
\label{cond-alpha}
\ee
so that the eigenfunctions have the correct twist.

\begin{figure}[!t]
\begin{center}
\epsfig{figure=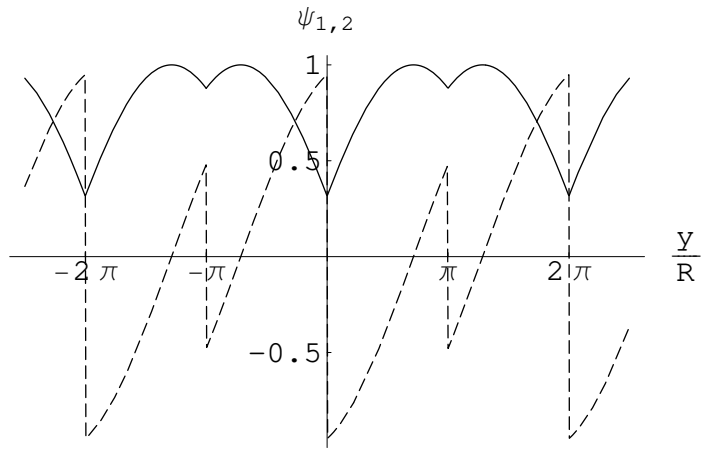,height=7cm}
\end{center}
\caption{The even (continuous) and odd (dashed) components of the zero
mode of a fermion field for $\beta=0$, $\delta_0=2.5$ and
$\delta_{\pi}=1$.}
\label{fermion}
\end{figure}

The spectrum~(\ref{ferm-mass}) is characterized by a uniform shift
with respect to the KK levels, as in the traditional SS picture (see
eq.~(\ref{ferm-massesBFWZ})). But while there the spectrum depends
only on the twist $\beta$, here it depends also on jumps $\delta_0$
and $\delta_{\pi}$. In particular it is possible to have a vanishing
shift for a non-vanishing twist. We observe that we recover the usual
SS spectrum in the limit $\delta_{\gamma} \to 0$. What about
eigenfunctions? They are discontinuous at the fixed points: the even
part has cusps while the odd one has jumps, as required by boundary
conditions (see fig.~\ref{fermion}).  If we take the limit
$\delta_{\gamma} \to 0$ in eq.~(\ref{ferm-eigfun}) we recover
precisely eq.~(\ref{ferm-eigfunBFWZ}) with $\beta_1=0$.

For any $\delta_{\gamma}$ the system is equivalent to a conventional
SS compactification with $\beta_c = \beta - \delta_0 -
\delta_{\pi}$. We can move to the new continuous eigenfunctions
$\Psi_c$ performing the following field redefinition:
\be
\Psi_c(x,y) = e^{\dd{-i\, \alpha(y)\, \widehat{\sigma}_2}}\ \Psi(x,y)~.
\label{ferm-redef}
\ee
Here the exponential factor removes the discontinuities from $\Psi$
and add a twist $-\delta_0-\delta_{\pi}$ to the wave function. Passing
from one system to the other we have only performed a local field
redefinition. It is a general statement that the physical properties
of a quantum mechanical system are invariant under a local field
redefinition. We can therefore say that the two systems are really
physically equivalent. This is interesting since it suggests that we
can move from a description in terms of discontinuous variables to
another in terms of smooth fields.

\subsection{Localized Mass Terms}
\label{GBCferm-mass}
In the previous section we discussed the equivalence between systems
characterized by discontinuous fields and `smooth' systems in which
fields are continuous but twisted, showing that a field redefinition
exists mapping the mass eigenfunctions of one description into those
of the other.  Here we would like to further explore the relation
between smooth and discontinuous frameworks by showing that the field
discontinuities are strictly related to lagrangian terms localized at
the fixed points.

We start from the lagrangian $\cL$ of eq.~(\ref{lagrBFWZ}) for the
continuous field $\Psi_{c}$ and we perform the
redefinition~(\ref{ferm-redef}). We obtain:
\bea
\label{ferm-brane1}
\cL (\Psi_c,\partial\Psi_c)
&=&
\cL (\Psi,\partial\Psi) + \cL_{brane} (\Psi,\partial\Psi) \ \ = \\
&=&
i\, \overline{\Psi}\, \bar{\sigma}^{\mu}\, \partial_{\mu} \Psi -
\frac{1}{2}\, (\Psi^T\, i\widehat{\sigma}^2\, \partial_y \Psi + h.c.)
-\frac{1}{2}\, \bigg[\alpha'(y) \, \Psi^T\Psi + h.c.\bigg] \nn
\eea
with 
\be
\alpha'(y) = \sum_{q=-\infty}^{+\infty} 
[\delta_0\,\delta(y-2q\,\pi R) + \delta_{\pi}\,\delta(y-(2q\,+1)\,\pi R)]~. 
\label{alphaprime}
\ee
We observe that jumps are related to mass terms localized at the
orbifold fixed points. If we want to explore this relation more deeply
we should derive the equation of motion, integrate it around the fixed
points and eventually we will recover the discontinuities of
fields. But we must pay attention in deriving the equation of motion!
Indeed the lagrangian~(\ref{ferm-brane1}) involves singular terms and
the naive use of the variational principle, which is tailored on
continuous functions and smooth functionals, would lead to
inconsistent results. In order to avoid these problems we regularize
$\alpha(y)$ by means of a smooth function $\alpha_{\lambda}(y)$
($\lambda>0$) which reproduces $\alpha(y)$ in the limit $\lambda\to
0$. Now we can derive the equation of motion which reads:
\be
i \widehat{\sigma}^2 \partial_y \Psi = [m -\alpha'_{\lambda}(y)]\, \Psi~.
\label{ferm-eombrane}
\ee
Since we are working with regularized functions also $\Psi$ is now
continuous, so we can divide eq.~(\ref{ferm-eombrane}) by
$\Psi$. Integrating the result around the fixed points and then taking
the limit $\lambda\to 0$ we obtain precisely the jumps of
eqs.~(\ref{ferm-finalbc})-(\ref{ferm-jumps}). If instead we take this
limit directly in eq.~(\ref{ferm-eombrane}), we immediately see that
this equation in the bulk coincides with eq.~(\ref{ferm-eom2}), as
expected.

To summarize, when going from a smooth to a discontinuous description
of the same physical system, singular terms encoding the informations
on the discontinuities of fields are generated in the lagrangian.
Conversely, when localized terms for bulk fields are present in the 5D
lagrangian, as for many phenomenological models currently discussed in
the literature, the field variables are affected by
discontinuities. These can be derived by analyzing the regularized
equation of motion and can be crucial to discuss important physical
properties of the system, such as its mass spectrum. In some case we
can find a field redefinition that eliminate the discontinuities and
provide a smooth description of the system. In section~\ref{bfwz} we
will explain in which cases it is possible to find a field
redefinition that completely reabsorbs the localized mass term or, in
other terms, which kind of masses can be ascribed to a generalized
SS mechanism.


\section[Generalized Boundary Conditions for Bosons]{Generalized 
Boundary Conditions for Bosons}
\label{GBCbos}

\subsection{Boundary Conditions for Bosonic Fields on $S^1/\mathbb{Z}_2$}
\label{GBCbos-bc}
In strict analogy with previous sections, we perform here the
discussion of the bosonic case, stressing its peculiarities with
respect to the fermionic case. As before, we begin by considering a
generic 5D theory compactified on the orbifold $S^1/\mathbb{Z}_2$. We
introduce a set of $n$ real 5D bosonic fields $\Phi(x,y)$, which we
classify in representations of the 4D Lorentz group. We define the
$(T,\mathbb{Z}_2)$ transformations of the fields by
\bea
\label{z2rep}
\Phi(y+ 2 \pi R) &=& U_\beta~ \Phi(y)\nn\\
\Phi(-y) &=& Z~ \Phi (y)~,
\eea
where $U_\beta$ and $Z$ are constant orthogonal matrices and $Z$ has
the property $Z^2={\mathbf 1}$.  Also in this case we can choose a
basis in which $Z$ is diagonal.

We look for boundary conditions on the fields $\Phi(x,y)$
and their derivatives, in the general class
\be
\left(
\begin{array}{c}
\Phi\\
\partial_y\Phi
\end{array}
\right)(\gamma^+) =
V_\gamma~\left(
\begin{array}{c}
\Phi\\
\partial_y\Phi
\end{array}
\right)(\gamma^-)~,
\label{gbc}
\ee 
where $V_\gamma$ are constant $2n\times2n$ matrices and $\gamma$ and
$\gamma^{\pm}$ are defined in section~\ref{GBCferm-bc}.  We observe
that eq.~(\ref{z2rep}) imply a specific form for the matrix $V_\beta$
in~(\ref{gbc}).  For the time being we keep a generic expression for
$V_\beta$, as well as for $V_{0,\pi}$. The reason for which we
consider also the derivatives of $\Phi$ will be clear in a while.

We will now constrain the matrices $V_\gamma$ by imposing certain
consistency requirements on our theory.  The spectrum of the theory is
determined by the eigenmodes of the differential operator
$-\partial^2_y$, which we require to be self-adjoint with respect to
the scalar product defined in eq.~(\ref{ferm-sprod}), where now $\Psi$
and $\Phi$ are real scalar fields and $\dagger$ must be converted
into $T$.  If we consider the matrix element $\langle\Psi \vert
\left(-\partial^2_y\Phi\right)\rangle$ and we perform a partial
integration we obtain:
\bea
\label{...}
\langle\Psi \vert \left(-\partial^2_y\Phi\right)\rangle &=&
\langle \left(-\partial^2_y\Psi\right) \vert \Phi\rangle+\nn\\
&+ &
\left[\Psi^T(y)\partial_y \Phi(y)-\partial_y \Psi^T(y) \Phi(y)
\right]^{0^+}_{0^-}+~~~\nn\\
&+ &
\left[\Psi^T(y)\partial_y \Phi(y)-\partial_y \Psi^T(y) \Phi(y)
\right]^{\pi R^+}_{\pi R^-}+~~~\nn\\
&+ &
\left[\Psi^T(y)\partial_y \Phi(y)-\partial_y \Psi^T(y) \Phi(y)
\right]^{y^-}_{y^+}~,
\eea
where $[f(y)]^a_b=f(a)-f(b)$. 
Necessary and sufficient conditions for the self-adjointness of the
operator $-\partial^2_y$ are that the three square brackets vanish and
the domain of the operator $-\partial^2_y$ coincide with the domain of
its adjoint.  In other words, we should impose conditions on $\Phi(y)$
and its first derivative in such a way that the vanishing of the
unwanted contributions implies precisely the same conditions on
$\Psi(y)$ and its first derivative.

It is easy to show that, in the class of boundary
conditions~(\ref{gbc}), this happens when
\be
V_\gamma J~ V_\gamma^T = J~~~~~
~~~~~~ \gamma\in (0, \pi , \beta)~,
\label{ug1}
\ee
where $J\equiv i \sigma^2$ is the symplectic form in the space
$(\Phi,\partial_y\Phi)$.  Eq.~(\ref{ug1}) restricts $V_\gamma$ in the
symplectic group $Sp(2 n)$. The simplest possibility is offered by
$V_\gamma={\mathbf 1}$, for $\gamma=(0, \pi , \beta)$. In this case
$U_\gamma={\mathbf 1}$, the fields are periodic and continuous across
the orbifold fixed points.  When $V_\beta\ne {\mathbf 1}$, the field
variables are not periodic and we have a twist. Such boundary
conditions are characteristic of the conventional SS
mechanism.  When $V_0$ or $V_\pi $ differs from unity, the
fields or their derivatives are not continuous across the fixed points
and we have jumps. Therefore, in close analogy with the fermionic
case, we find that the boundary conditions for bosons allow for both
twist and jumps.  At variance with the fermionic case, twist and jumps
can also affect the first derivative of the field variables.
Moreover, the self-adjointness alone does not forbid boundary
conditions that mix the fields and their $y$-derivatives.  For
instance, if we have a single real scalar field $\varphi(y)$, and we
parametrize the generic $2\times 2$ matrix $V_\gamma$ as:
\be
V_\gamma=
\left(
\begin{array}{cc}
A_\gamma& B_\gamma\\
C_\gamma& D_\gamma
\end{array}
\right)~,
\ee
the condition~(\ref{ug1}) reduces to $\det V_\gamma=1$,
as expected since $Sp(2)$ and $SL(2,R)$ are isomorphic. If $B_\gamma$ and
$C_\gamma$ are not both vanishing,
the boundary conditions will mix $\varphi$ and $\partial_y \varphi$.

While the field variables and their first derivatives may have twist
and jumps, we should require that physical quantities remain periodic
and continuous across the orbifold fixed points. This poses a further
restriction on the matrices $V_\gamma$. If the theory is invariant
under global transformations of a group G, we can satisfy this
requirement by asking that the matrices $V_\gamma$ are in a
$2n$-dimensional representation of G.  The choice of scalar product
made in~(\ref{ferm-sprod}) does not allow to consider symmetry
transformations of the 5D theory that mix fields and $y$-derivatives.
Moreover, it is not restrictive to consider orthogonal representations
of G on the space of real fields $\Phi$.  In this case, the solution to
eq.~(\ref{ug1}) reads
\be
V_\gamma=
\left(
\begin{array}{cc}
U_\gamma& 0\\
0& U_\gamma
\end{array}
\right)~~~
~~~~~~~~~~~~~U_\gamma\in {\rm G}~,
\label{ug2}
\ee
where $U_\gamma$ is in an orthogonal $n$-dimensional representation of
G.

Finally, we should take into account consistency conditions among the
twist, the jumps and the orbifold projection. These are identical to
the ones discussed for fermions and are reported here only for
completeness:
\bea 
\label{cons2}
U_\gamma~ Z~ U_\gamma &=& Z~~~ ~~~~~~~~~~~~~~\gamma\in (0, \pi ,
\beta)~,\nn\\ \left[U_0,U_\beta\right]&=& 0~,\\
\left[U_\pi,U_\beta\right]&=& 0~.  \nn
\eea 
%
Also in this case the parameters of $U_\gamma$ can be discrete or
continuous, depending on the commutators $[Z,U_\gamma]$ and
$[U_0,U_\pi]$.

In summary, the allowed boundary conditions on $\Phi(y)$ are specified
in eq.~(\ref{gbc}), with the matrices $V_\gamma$ satisfying the
requirements~(\ref{ug2})-(\ref{cons2}).

\subsection{One Scalar Field}
\label{GBCbos-one}
It is instructive to analyze in detail the case of a single massless
scalar field $\varphi(x,y)$, of definite parity, $Z=+ 1$ to begin
with.  We start by writing the equation of motion for $\varphi$
\be
-\partial^2_y \varphi = m^2 \varphi~,
\ee
in each region $y_q<y<y_{q+1}$ of the real line, where $y_q\equiv q
\pi R$ and $q\in {\mathbb Z}$. We have defined the mass $m$ through
the 4D equation $\partial^2\varphi= m^2 \varphi$.  The solutions of
these equations can be glued by exploiting the boundary conditions
$V_0$ and $V_\pi$, imposed at $y=y_{2q}$ and $y=y_{2q+1}$,
respectively. Finally, the spectrum and the eigenfunctions are
obtained by requiring that the solutions have the twist described by
$V_\beta$.

The equation of motion remains invariant if we multiply $\varphi$ by
$\pm 1$, so that the group G of global symmetries is a parity (to be
distinguished from the orbifold symmetry $\mathbb{Z}_2$ that acts also
on the $y$ coordinate). We have $V_\gamma=\pm{\mathbf 1}$. For
instance, we are allowed to consider either periodic or anti-periodic
fields.  We start by analyzing the case of no jumps,
$V_0=V_\pi=+{\mathbf 1}$.  The solutions of the equations of motion,
up to an arbitrary $x$-dependent factor, are
\be
\varphi_1(y)=\cos m y~~~~~~~~m~R=
\left\{
\begin{array}{ccc}
n& &V_\beta=+{\mathbf 1}\\
n+\dd\frac{1}{2}& &V_\beta=-{\mathbf 1}
\end{array}
\right.~,
\label{cos}
\ee
and $n$ is a non-negative integer.  It is interesting to compare the
result for $V_\beta=-{\mathbf 1}$ with that obtained by assuming a
jump in $y=0$: $(V_0,V_\pi,V_\beta)=(-{\mathbf 1},+{\mathbf
1},+{\mathbf 1})$.  We find
\be
\varphi_2(y)=\epsilon(y/2)~ \sin m y~~~~~~~m~R=
n+\frac{1}{2}~.
\label{epsin}
\ee
where $\epsilon(y)$ is the sign function on $S^1$.

%
%
\begin{figure}[!t]
\centerline{
\psfig{figure=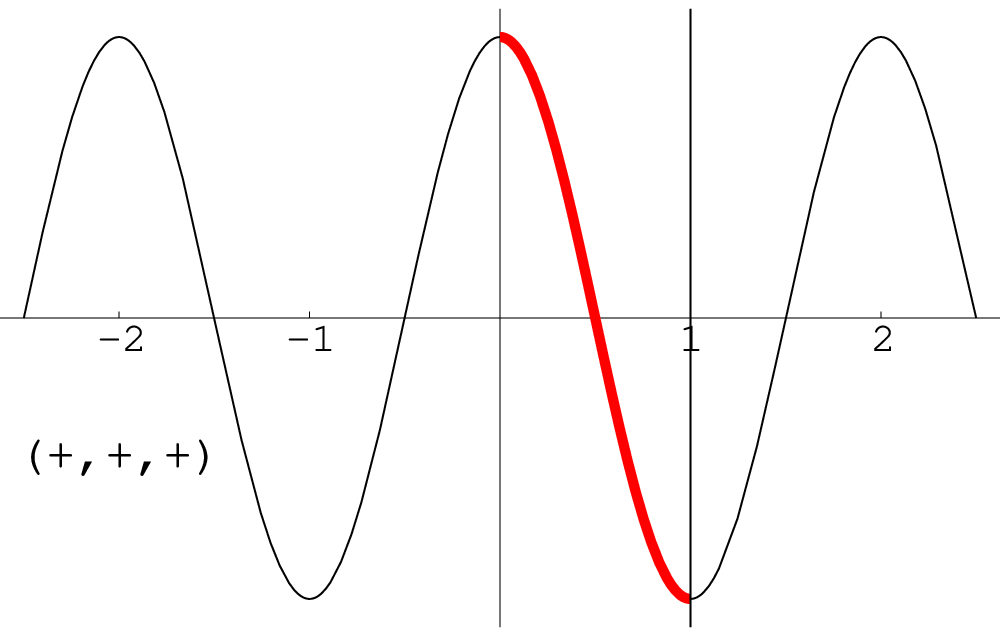,height=3cm,width=6cm}
\psfig{figure=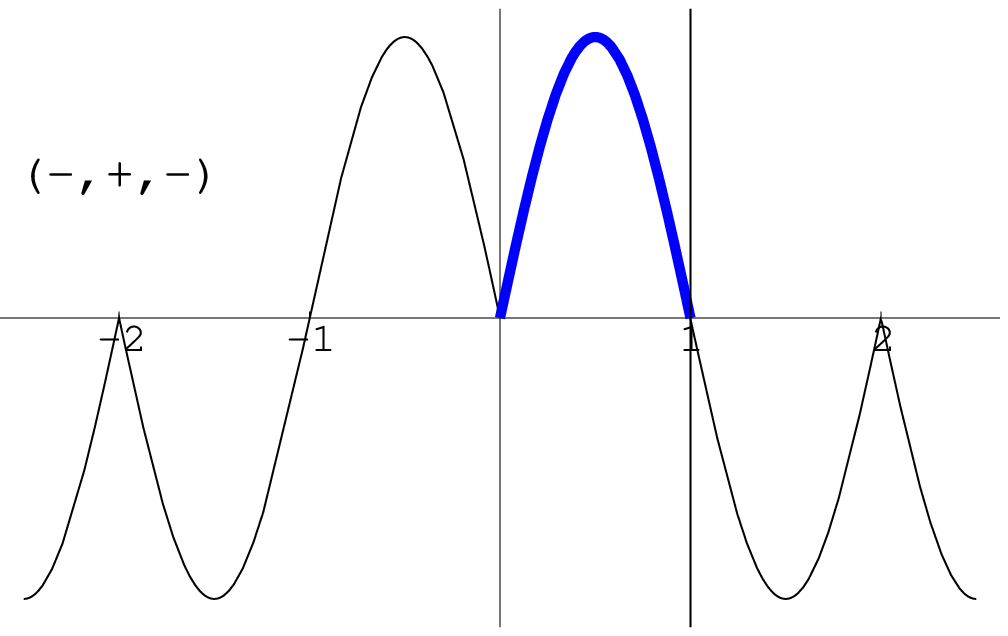,height=3cm,width=6cm}}
\vspace{0.3cm}
\centerline{
\psfig{figure=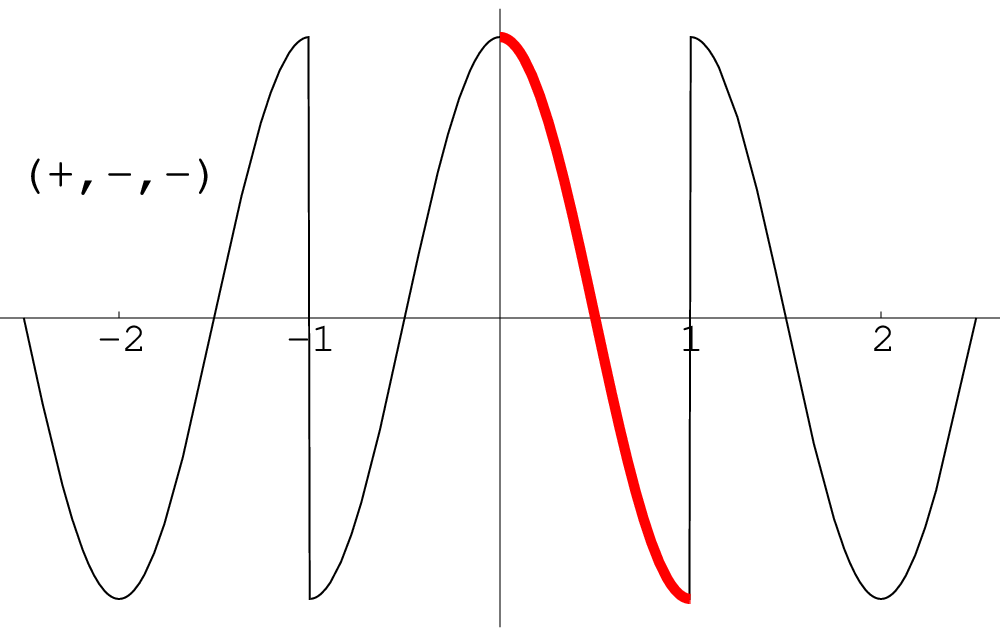,height=3cm,width=6cm}
\psfig{figure=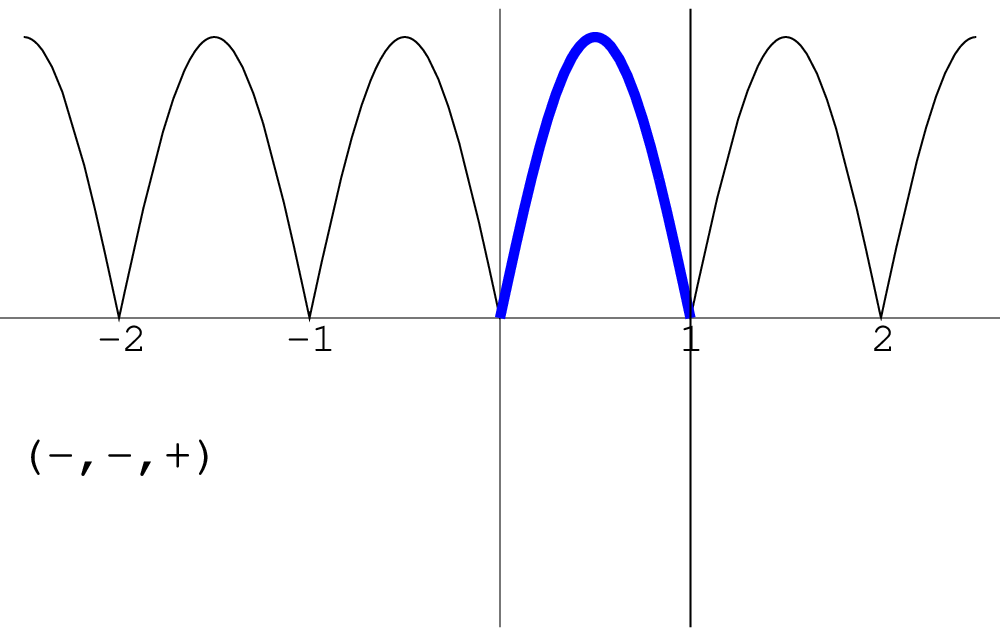,height=3cm,width=6cm}}
\vspace{0.3cm}
\centerline{
\psfig{figure=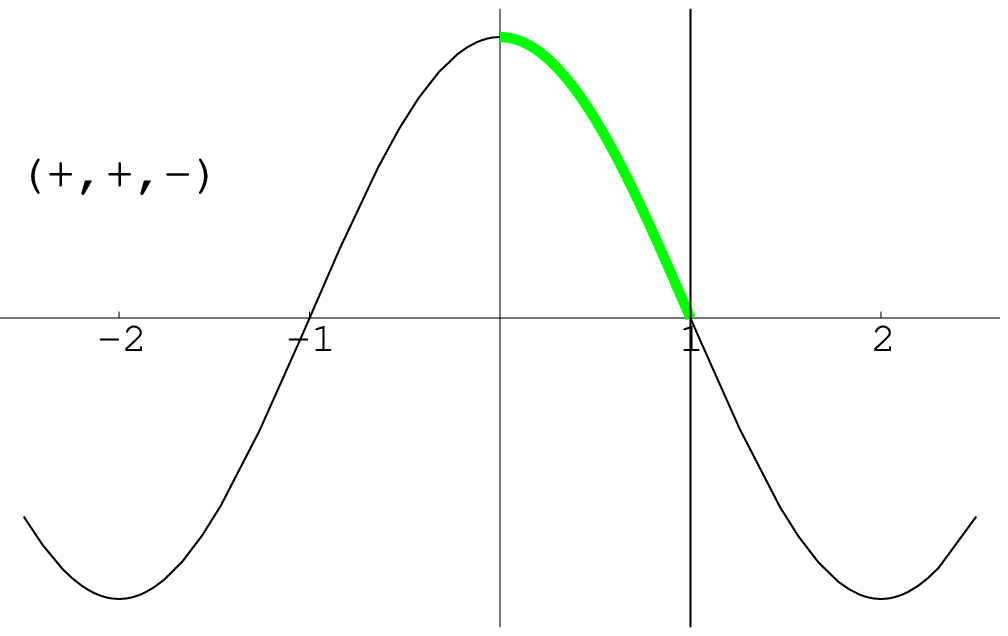,height=3cm,width=6cm}
\psfig{figure=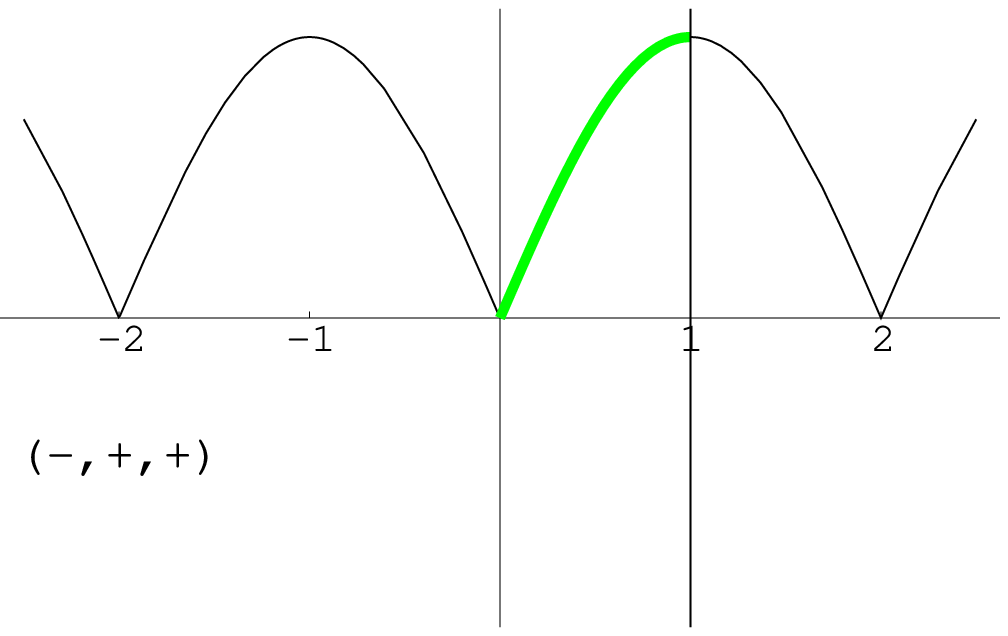,height=3cm,width=6cm}}
\vspace{0.3cm}
\centerline{
\psfig{figure=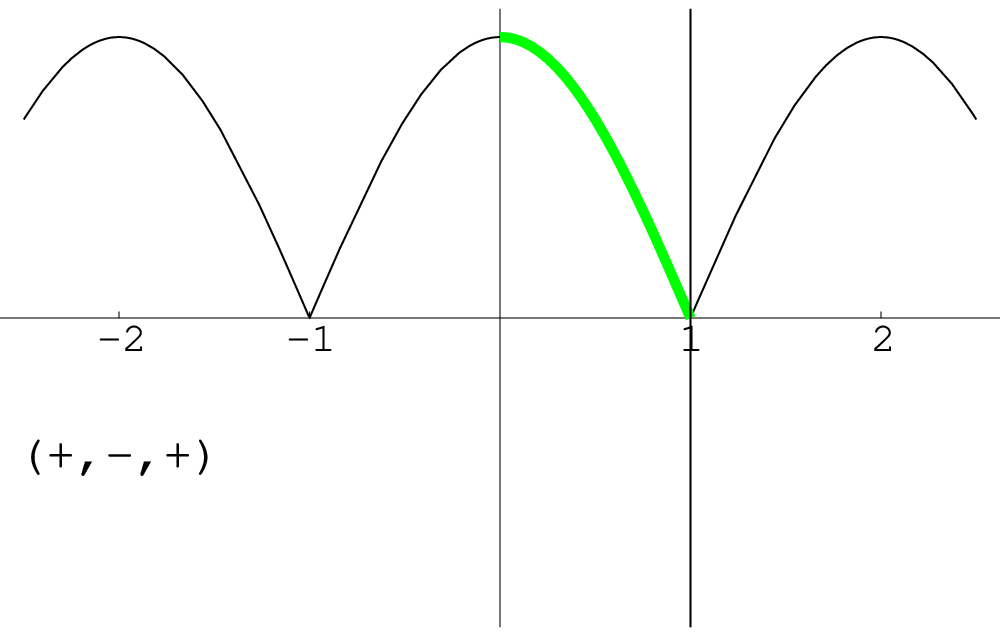,height=3cm,width=6cm}
\psfig{figure=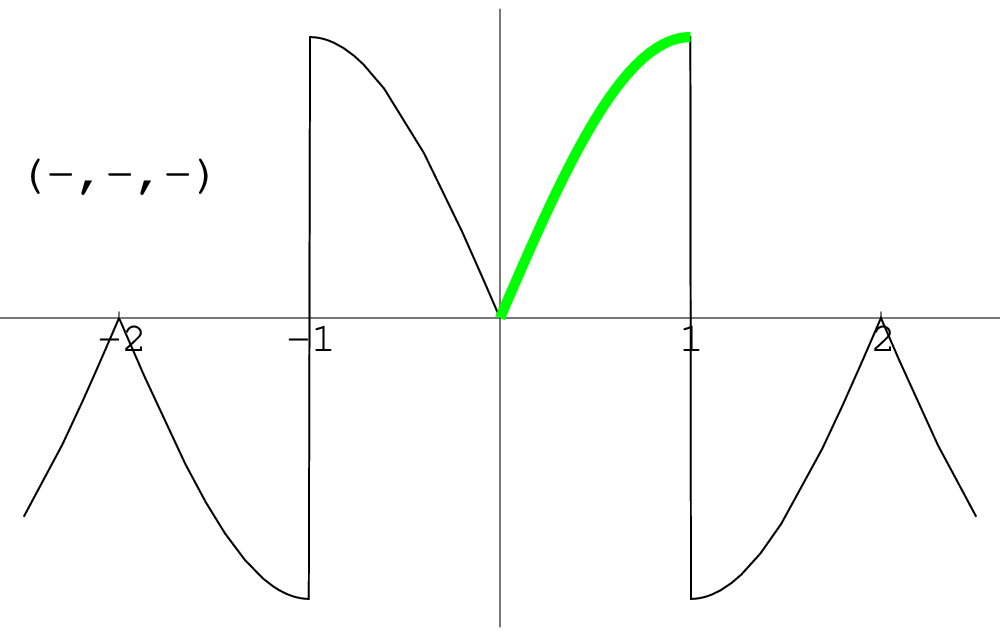,height=3cm,width=6cm}}
\caption{Eigenfunctions of $-\partial^2_y$, for a real even field $\varphi(y)$,
versus $y/(\pi R)$. For each boundary condition, labelled by
($V_0,V_\pi,V_\beta$), the eigenfunction corresponding to the
lightest non-vanishing mode is displayed.
\label{evenf}}
\end{figure}
%
%
%
%
\begin{figure}[!t]
\centerline{
\psfig{figure=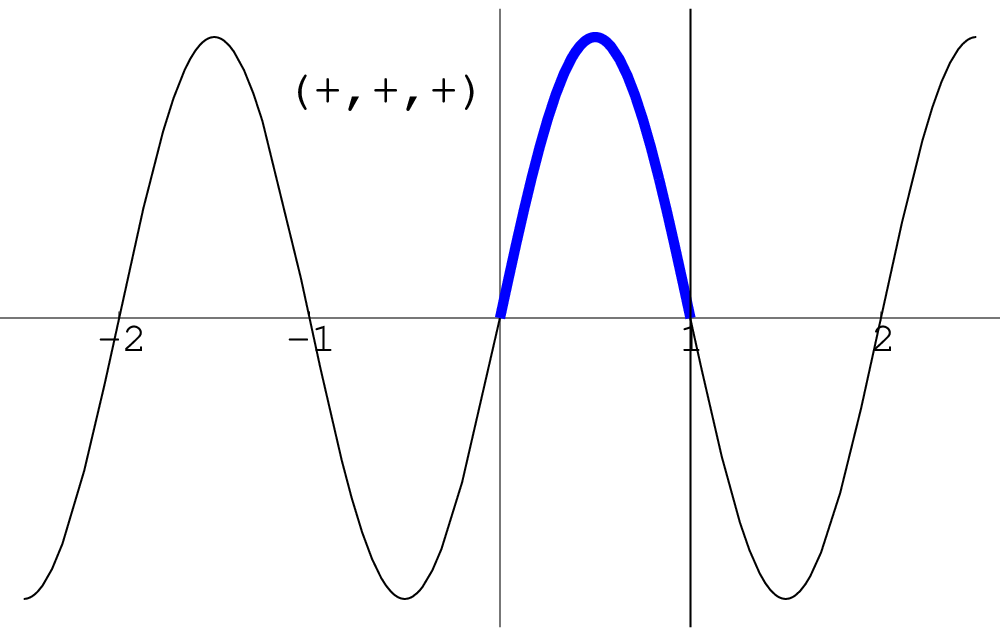,height=3cm,width=6cm}
\psfig{figure=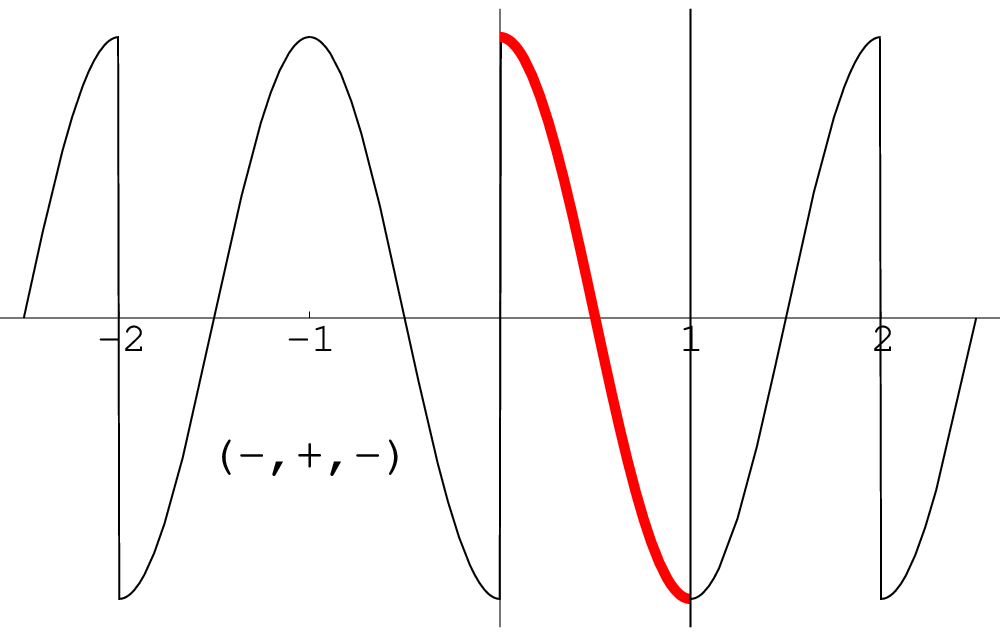,height=3cm,width=6cm}}
\vspace{0.3cm}
\centerline{
\psfig{figure=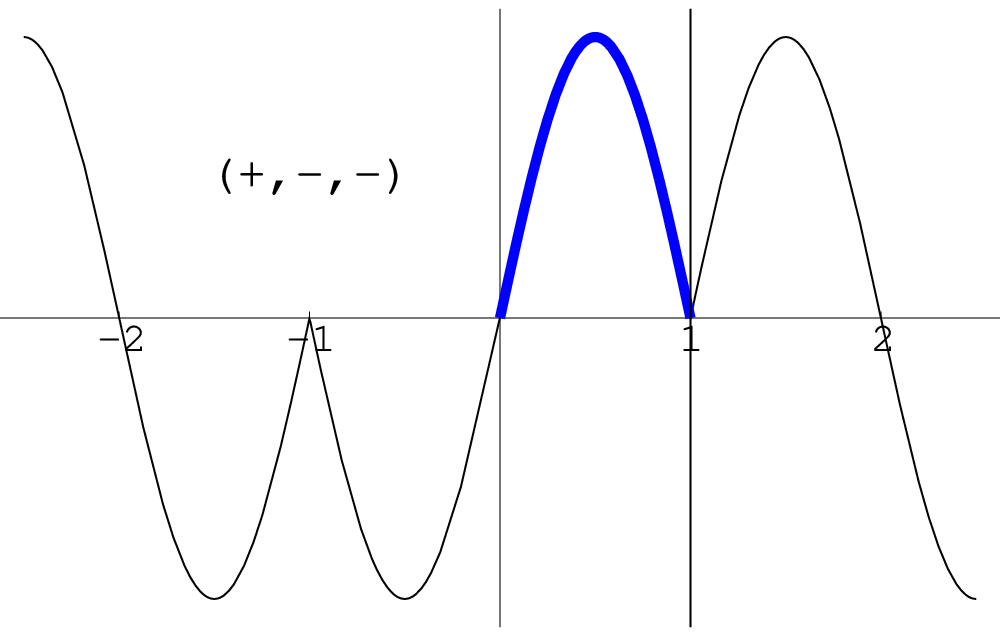,height=3cm,width=6cm}
\psfig{figure=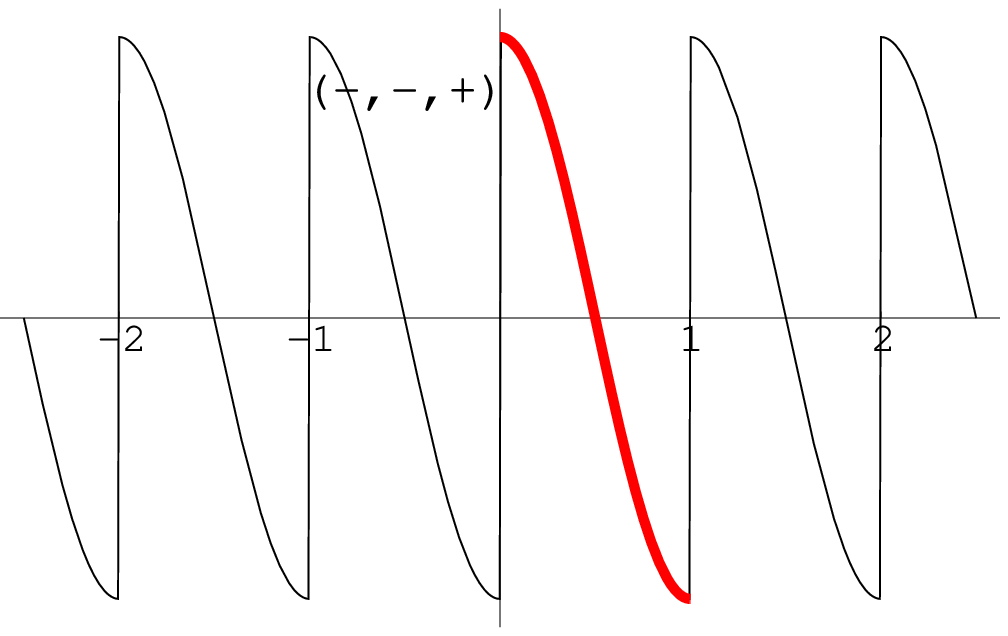,height=3cm,width=6cm}}
\vspace{0.3cm}
\centerline{
\psfig{figure=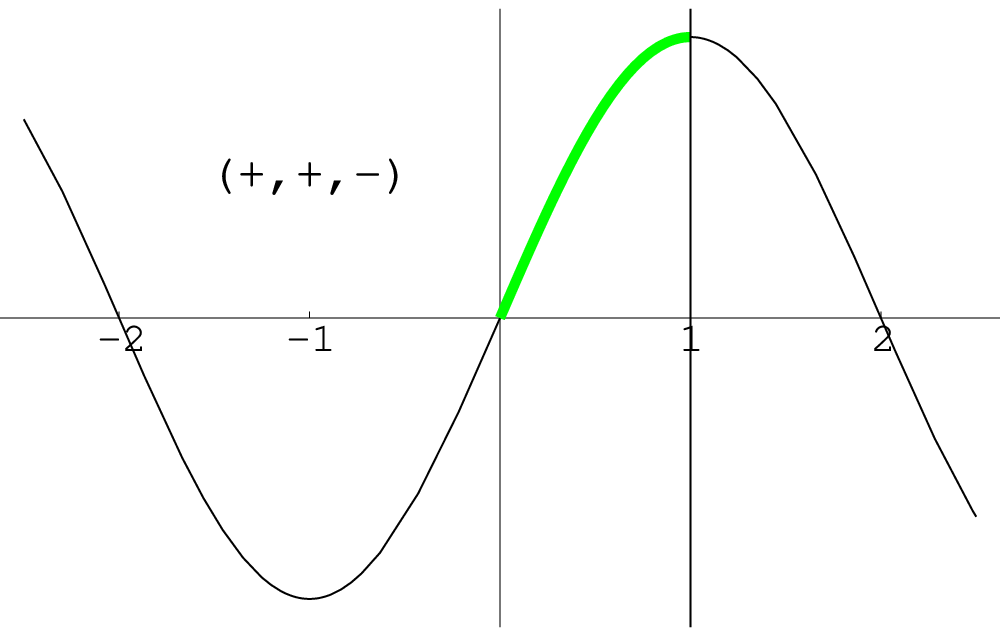,height=3cm,width=6cm}
\psfig{figure=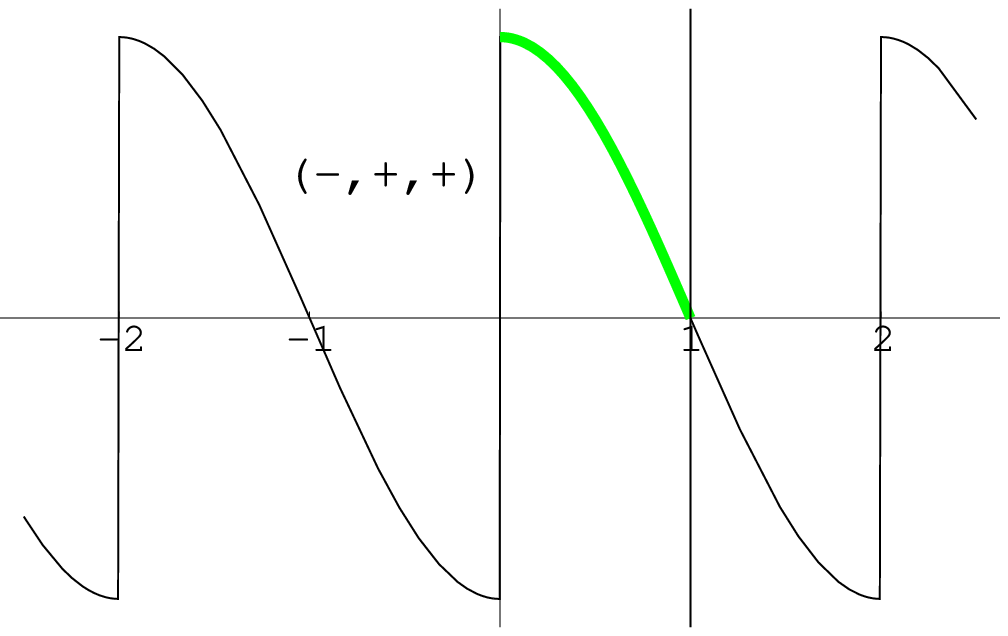,height=3cm,width=6cm}}
\vspace{0.3cm}
\centerline{
\psfig{figure=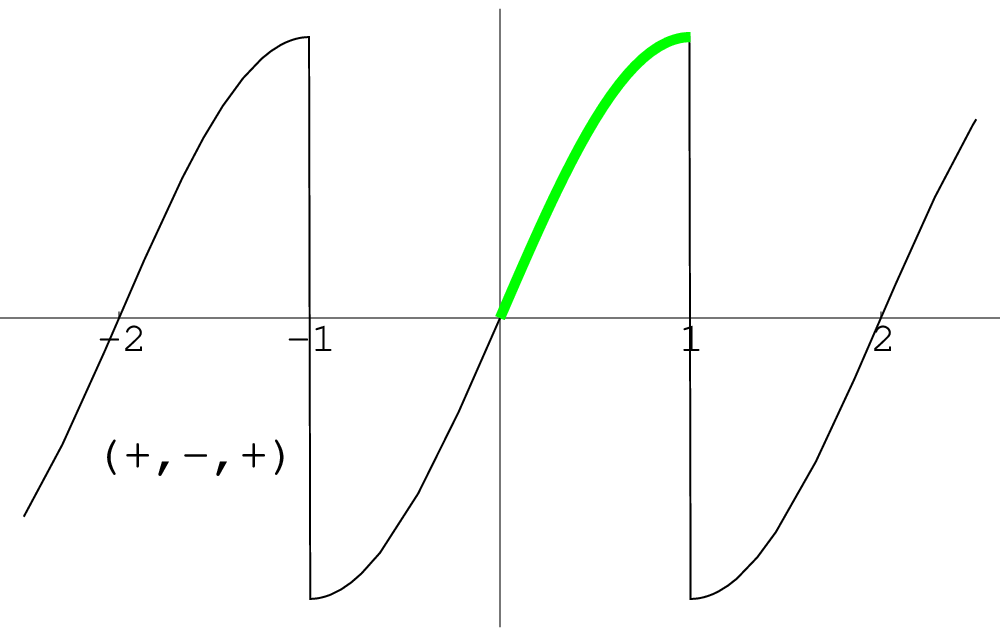,height=3cm,width=6cm}
\psfig{figure=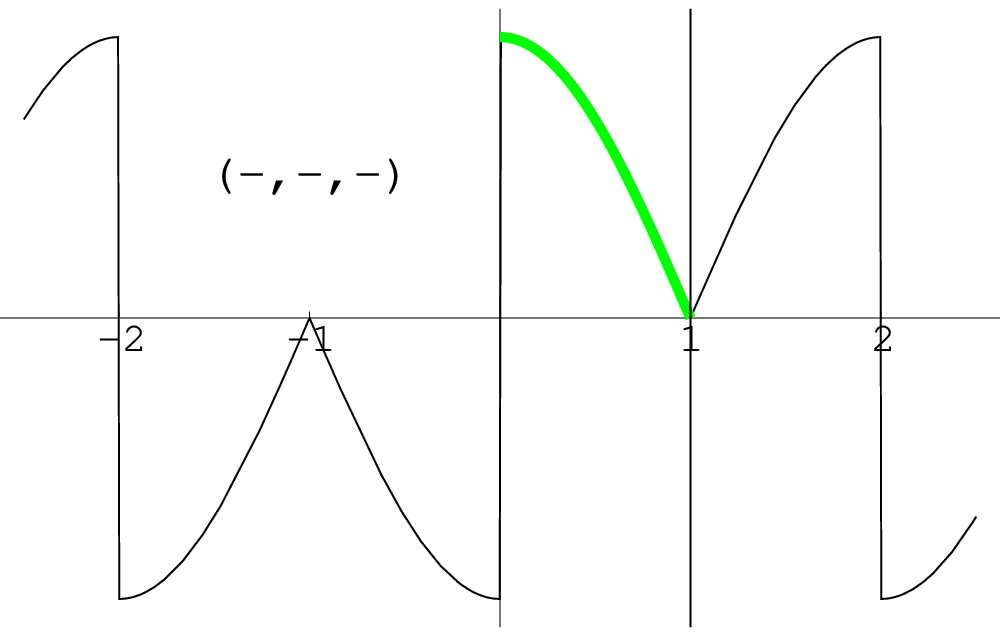,height=3cm,width=6cm}}
\caption{Eigenfunctions of $-\partial^2_y$, for a real odd field $\varphi(y)$,
versus $y/(\pi R)$. For each boundary condition, labelled by
($V_0,V_\pi,V_\beta$), the eigenfunction corresponding to the
lightest non-vanishing mode is displayed.
\label{oddf}}
\end{figure}
%
%
%
%
\begin{table}[!t]
\label{tab1}
\begin{center}
\begin{tabular}{|c|c|c|c|c|}
\hline
& & & &\\
$(V_0,V_\pi,V_\beta)$ & $m~ R$ & {\tt eigenfunctions} & $\varphi(0)$ &
$\varphi(\pi R)$\\
& & & &\\
\hline
& & & &\\
$(+,+,+)$ &  $n\ge 0$ & $\varphi_1(y)=\cos m y$ & $\ne 0$ & $\ne 0$\\
 &  $n> 0$ & $\varphi_5(y)=\sin m y$ & $0$ & $0$\\
& & & &\\
$(-,+,-)$ &  $n> 0$ & $\varphi_2(y)=\epsilon(y/2)~ \sin m y$ &
$0$ & $0$\\
&  $n\ge 0$ & $\varphi_6(y)=\epsilon(y/2)~ \cos m y$ &
${\tt jump}$ & $\ne 0$\\
& & & &\\
$(+,-,-)$ &  $n\ge 0$ & $\varphi_3(y)=\epsilon(y/2+\pi R/2)\cos m y$ &
$\ne 0$ & {\tt jump}\\
&  $n> 0$ & $\varphi_7(y)=\epsilon(y/2+\pi R/2)\sin m y$ &
$0$ & 0\\
& & & &\\
$(-,-,+)$ &  $n>0$ & $\varphi_4(y)=\epsilon(y)~ \sin m y$ &
$0$ & $0$\\
&  $n\ge 0$ & $\varphi_8(y)=\epsilon(y)~ \cos m y$ &
{\tt jump} &{\tt jump} \\
& & & &\\
$(+,+,-)$ &  $n+1/2$ & $\varphi_1(y)=\cos m y$ & $\ne 0$ & $0$\\
&  $n+1/2$ & $\varphi_5(y)=\sin m y$ & $0$ & $\ne 0$\\
& & & &\\
$(-,+,+)$ &  $n+1/2 $ & $\varphi_2(y)=\epsilon(y/2)~ \sin m y$ &
$0$ & $\ne 0$\\
&  $n+1/2 $ & $\varphi_6(y)=\epsilon(y/2)~ \cos m y$ &
{\tt jump}& $0$\\
& & & &\\
$(+,-,+)$ &  $n+1/2$ & $\varphi_3(y)=\epsilon(y/2+\pi R/2)\cos m y$ &
$\ne 0$ & 0\\
&  $n+1/2$ & $\varphi_7(y)=\epsilon(y/2+\pi R/2)\sin m y$ &
$0$ & {\tt jump}\\
& & & &\\
$(-,-,-)$ &  $n+1/2$ & $\varphi_4(y)=\epsilon(y)~ \sin m y$ &
$0$ & ${\tt jump}$\\
&  $n+1/2$ & $\varphi_8(y)=\epsilon(y)~ \cos m y$ &
${\tt jump}$& 0\\
& & & &\\
\hline
\end{tabular}
\end{center}
\caption{Eigenvalues and eigenfunctions of $-\partial^2_y$, for
even fields $\varphi_k$ $(k=1,2,3,4)$ and odd fields $\varphi_k$
$(k=5,6,7,8)$. The spectrum is given in terms of the non-negative
integer $n$.}
\end{table}
%
%
The eigenfunctions~(\ref{cos}) and~(\ref{epsin}) for $m=1/(2R)$ are
compared in the third row of fig.~\ref{evenf}.  We observe that
$V_0=+{\mathbf 1}$ implies $\partial_y\varphi=0$ at $y=0$, that is
Neumann boundary conditions.  Instead, if we take $V_0=-{\mathbf 1}$,
the even field $\varphi$ should vanish at $y=0$ as for a Dirichelet
boundary condition, and this produces a cusp at $y=0$. Despite this
difference, the two eigenfunctions are closely related.  If compared
in the region $0<y<\pi R$, they look the same, up to an exchange
between the two walls at $y=0$ and $y=\pi R$.  They both vanish at one
of the two walls and they have the same non-vanishing value at the
other wall, with the same profile in between.  Indeed, the two cases
are related by a coordinate transformation and a field redefinition:
\be
\varphi_2(y)=\epsilon(y/2)~\varphi_1(y+\pi R)~.
\label{bos-redef}
\ee
If $\varphi_1(y)$ is even, continuous and anti-periodic,
it is easy to see that the function $\varphi_2(y)$ defined
in~(\ref{bos-redef}) is even, periodic and has a cusp in $y=y_{2q}$,
where it vanishes. The equations of motion are not affected by the
translation $y\to y+\pi R$, which simply exchanges the boundary
conditions at $y=y_{2q}$ and $y=y_{2q+1}$. Moreover, the physical
properties of a quantum mechanical system are invariant under
a local field redefinition. Therefore the two systems related
by eq.~(\ref{bos-redef}) are equivalent.

In table 2.1 we collect spectrum and eigenvalues for all
possible cases that are allowed by an even or odd field $\varphi$.  We
have found it useful to express the solutions in terms of the sign
function, which specifies the singularities of the system. Indeed,
$\epsilon(y/2)$ is singular in $y=y_{2q}$, $\epsilon(y/2+\pi R/2)$ in
$y=y_{2q+1}$ and $\epsilon(y)$ in all $y=y_q$. The correct parity of
the solutions is guaranteed by the properties of $\epsilon(y)$.  Also
the periodicity can be easily determined from the fact that
$\epsilon(y)$ is periodic, whereas $\epsilon(y/2)$ and
$\epsilon(y/2+\pi R/2)$ are anti-periodic.

There are three types of spectra: first, the ordinary KK tower $n/R$
that includes a zero mode; second, an identical spectrum with the
absence of the zero mode and, finally, the KK tower shifted by
$1/2R$. All systems that possess the same spectrum can be related by
field redefinitions that can be easily derived from table 2.1. The
only non-trivial transformation, applying only to the case of
semi-integer spectrum, is the one in~(\ref{bos-redef}). For
semi-integer spectrum, all $\varphi_k$ $(k=1,...8)$ are related. For
integer and non-negative spectrum, we have maps among $\varphi_1$,
$\varphi_3$, $\varphi_6$ and $\varphi_8$. For strictly positive
integer spectrum, $\varphi_2$, $\varphi_4$, $\varphi_5$ and
$\varphi_7$ are related.  Thanks to these relations, we can always go
from a description in terms of discontinuous field variables to a
descriptions by the smooth fields $\varphi_1(y)$ or $\varphi_5(y)$.
Also, as can be seen from figs.~\ref{evenf} and~\ref{oddf}, the
behavior in the vicinity of the fixed points is the same for all the
eigenfunctions representing the same type of spectrum, up to a
possible exchange between the two fixed points.  In the presence of a
single real field $\varphi$ the parity $Z$ does not seem to have an
absolute physical meaning. We find that there are equivalent physical
systems with opposite $Z$ parities for $\varphi$.

In conclusion, there are less physically inequivalent systems than
independent boundary conditions. There are different boundary
conditions that lead to the same spectrum and the corresponding
systems are related by field redefinitions. The parameter $V_0\cdot
V_\pi\cdot V_\beta$ is equal in equivalent systems.  When $V_0\cdot
V_\pi\cdot V_\beta=+{\mathbf 1}$, $mR$ is integer whereas for
$V_0\cdot V_\pi\cdot V_\beta=-{\mathbf 1}$, $mR$ is semi-integer.

\subsection{More Scalar Fields}
\label{GBCbos-more}
Several scalar fields lead to the possibility of exploiting continuous
global symmetries to characterize boundary conditions.  As an example,
we consider here the case of a 5D complex scalar field
$\varphi(y)\equiv (\varphi_1(y) + i~ \varphi_2(y))/\sqrt{2}$.  Its
equation of motion:
\be
-\partial^2_y \varphi = m^2 \varphi~,
\ee
is invariant under global O(2) transformations, acting on
$(\varphi_1,\varphi_2)$.

We first discuss the case where $Z=diag(+1,-1)$ in the basis
$(\varphi_1,\varphi_2)$. In this case we can take:
\be
U_\gamma=
\left(
\begin{array}{cc}
\cos\theta_\gamma & \sin\theta_\gamma\\
-\sin\theta_\gamma & \cos\theta_\gamma
\end{array}
\right)~,
\label{rotation}
\ee
which is a symmetry of the theory and satisfies~(\ref{cons2}).  In
general, we can choose three independent angles
$\theta_\gamma=(\beta,\delta_0,\delta_\pi)$ for the twist and the two
jumps at $y=0,\pi R$ respectively.  The solution of the equation of
motion subjected to these boundary conditions can be obtained by the
same method used in section~\ref{GBCbos-one}. We find:
\be
\varphi(y)=e^{\dd i (m y - \alpha(y))}~~~,
\label{sol2}
\ee where \be m=\frac{n}{R} - \frac{(\beta-\delta_0-\delta_\pi)}{2 \pi
R}~~~ ~~~~~~~(n\in {\mathbb Z})~, 
\ee 
and $\alpha(y)$ is the function introduced in
section~\ref{GBCferm-one}.  The presence of this function explains why
the shift of the spectrum with respect to the KK levels is given by
$\beta-\delta_0-\delta_\pi$ and not by $\beta$ as in the conventional
SS mechanism.

When $\beta-\delta_0-\delta_\pi=0~~~({\rm mod}~ 2\pi)$, the masses are
$n/R$ and we can order all massive modes in pairs. Indeed each
physical non-vanishing mass $\vert m\vert=\vert n\vert/R$ $(n\ne 0)$
corresponds to two independent eigenfunctions.  For instance, when
$\beta=\delta_0=\delta_\pi=0$, we have $\varphi_{\pm}^n=\exp(\pm i n
y/R)$.  This infinite series of degenerate 4D doublets can be
interpreted as a consequence of the $O(2)$ symmetry, which is
unbroken.  A non-vanishing shift of the KK levels induces an explicit
breaking of the $O(2)$ symmetry. The order parameter is
$\beta-\delta_0-\delta_\pi$ $({\rm mod}~ 2\pi)$.  When
$\beta-\delta_0-\delta_\pi$ is non-vanishing (and not a multiple of $2
\pi$) the eigenfunctions of the massive modes are no longer paired,
each of them corresponding now to a different physical mass.  As for
the case of a single real field, different boundary conditions may
lead to the same spectrum. For instance, it is possible that $O(2)$
remains unbroken, despite the existence of non-trivial boundary
conditions, if twist and jumps are such that the combination
$\beta-\delta_0-\delta_\pi$ vanishes mod $2\pi$. Moreover, starting
from a generic system with both twist $\beta$ and jumps
$\delta_{0,\pi}$ different from zero, we can always move to an
equivalent `smooth' theory where the jumps vanish and the twist
$\beta^c$ of the new scalar field $\varphi^c(y)$ is given by
$\beta-\delta_0-\delta_\pi$. The map between the two systems is given
by
\be
\varphi^c(y)=e^{\dd{i\alpha(y)}}\varphi(y)~.
\label{map}
\ee
The multiplicative factor $e^{\dd{i\alpha(y)}}$ removes the
discontinuities from $\varphi(y)$ and add a twist $-\delta_0
-\delta_\pi$ to the wave function.

Another interesting case is that of $Z$ proportional to the identity.
If we assign the same $Z$ parity to the real components
$\varphi_{1,2}(y)$, then $U_\gamma$ commutes with $Z$ and the
condition~(\ref{cons2}) implies that its eigenvalues are $\pm 1$. If
also $[U_0,U_\pi]=0$, then it is not restrictive to go to a field
basis where all $U_\gamma$ are diagonal, with elements $\pm 1$.  This
would lead to a discussion qualitatively close to that of
section~\ref{GBCbos-one}, where twist and jumps were quantized. A new
feature occurs if $[U_0,U_\pi]\ne 0$. Consider as an example
$Z=U_\beta=diag(+1,+1)$ in the basis $(\varphi_1,\varphi_2)$. A
consistent choice for $U_0$ and $U_\pi$ is:
\be
U_0=
\left(
\begin{array}{cc}
\cos\delta_0 & -\sin\delta_0\\
-\sin\delta_0 & -\cos\delta_0
\end{array}
\right)~,~~~~~~~~~
U_\pi=
\left(
\begin{array}{cc}
1 & 0\\
0& -1
\end{array}
\right)~.
\label{ncjumps}
\ee
Notice that the $O(2)$ matrices $U_0$ and $U_\pi$ square to 1, as
required by the condition~(\ref{cons2}).  The solutions of the
equations of motion are:
\bea
\label{sol3}
\phi_1(y)&=&\cos(my-\alpha(y))\nn\\
\phi_2(y)&=&\epsilon(y)\sin(my-\alpha(y))~,
\eea
where
\be
m=\frac{n}{R} + \frac{\delta_0}{2 \pi R}~~~
~~~~~~~(n\in \mathbb{Z})
\ee
and in $\alpha(y)$ we have to set $\delta_{\pi}=0$.  It is interesting
to note that this choice of boundary conditions leads to a theory that
is physically equivalent to that studied at the beginning of this
section, where the fields $\varphi_1$ and $\varphi_2$ had opposite
parity.  We can go back to that system and consider the case of
periodic fields with a jump at $y=0$: $Z=diag(+1,-1)$,
$U_\pi=U_\beta=1$ and $U_0$ as in~(\ref{rotation}) with
$\theta_0=\delta_0$. If we now perform the field redefinition:
\be
\varphi_1(y)\to\varphi_1(y)~,~~~~~~~~~~~
\varphi_2(y)\to\epsilon(y)~\varphi_2(y)~,
\ee
the new field variables are both even and periodic and their jumps are
those given in~(\ref{ncjumps}).  It is easy to see that also the
solutions~(\ref{sol2}) are mapped into~(\ref{sol3}). Moreover, it will
be now possible to describe the theory defined by the
jumps~(\ref{ncjumps}) in terms of smooth field variables,
characterized by a certain twist.

This correspondence provides another example of equivalent systems,
despite a different assignment of the orbifold parity.  The presence
of discontinuous fields is a generic feature of field theories on
orbifolds. The present discussion suggests that at least in some cases
these discontinuities may not have any physical significance, being
only related to a particular and not compelling choice of field
variables.

\subsection{Brane Action for Bosonic System}
\label{GBCbos-mass}
In the previous sections we showed the equivalence between bosonic
systems characterized by discontinuous fields and `smooth' systems in
which fields are continuous but twisted.  For each pair of systems
characterized by the same mass spectrum we were able to find a local
field redefinition, plus a possible discrete translation, mapping the
mass eigenfunctions of one system into those of the other system.
Here we would like to further explore the relation between smooth and
discontinuous systems by showing that the field discontinuities are
strictly related to lagrangian terms localized at the fixed points.

We begin by discussing the case of one real scalar field.  To fix the
ideas we focus on the equivalence between the cases $(+,+,-)$ and
$(+,-,+)$ with $Z=1$ of table 2.1. The other cases can be discussed
along similar lines. We denote by $\varphi^c$ the continuous field
$(+,+,-)$ with twist $U_\beta=-1$ and by $\varphi$ the periodic field
$(+,-,+)$ that has a jump $U_\pi=-1$. If we start from the lagrangian
$\mathcal{L}$ for the boson $\varphi^{c}$
\be
\mathcal{L}(\varphi^{c},\partial\varphi^{c})=-\frac{1}{2}
\partial_{M}\varphi^{c} \partial^{M}\varphi^{c}~,
\ee
and we perform the field redefinition:
\be
\varphi^{c}(y) = \chi(y) \varphi(y)~~~~~~~~
\chi(y)\equiv\epsilon(y/2+\pi R/2)~,
\ee
we obtain an expression in terms of discontinuous fields and their
derivatives, from which it is difficult to derive the correct equation
of motion for the system. Indeed the new lagrangian is highly singular
and the naive use of the variational principle would lead to
inconsistent results.  In order to avoid these problems we regularize
$\chi(y)$ by means of a smooth function $\chi_\lambda(y)$
($\lambda>0$) which reproduces $\chi(y)$ in the limit $\lambda\to 0$.
By performing the substitution:
\be
\varphi^{c}(y) = \chi_{\lambda}(y)~\varphi(y)~,
\ee
we obtain
\be
\label{regl}
\mathcal{L}(\varphi^{c},\partial\varphi^{c}) =
-\frac{1}{2}\chi_{\lambda}^{2} \partial_{M}\varphi \partial^{M}\varphi -
\chi_{\lambda}\chi_{\lambda}'\varphi\partial_{y}\varphi -
\frac{1}{2}\chi_{\lambda}'^{2} \varphi^{2}~.
\ee
Since the field $\varphi(y)$ is periodic, we can work in the interval
$0\le y \le 2\pi R$. In the limit $\lambda\to 0$, we find:
\be
\mathcal{L}(\varphi^{c},\partial\varphi^{c}) =
-\frac{1}{2} \chi^{2} \partial_{M}\varphi \partial^{M}\varphi
+ 2\chi~ \delta(y-\pi R)~ \varphi\partial_{y}\varphi -
 2 \delta^2(y-\pi R)~ \varphi^{2}~.
\ee
The action contains quadratic terms for the field $\varphi(y)$ that
are localized at $y=\pi R$.  However these terms are quite singular
and, strictly speaking, are mathematically ill-defined even as
distributions.  For this reason we derive the equation of motion for
$\varphi(y)$ using the regularized action, eq.~(\ref{regl}), from
which we get:
\be
\chi_\lambda\left(
 \chi_{\lambda}\partial_{y}^{2}\varphi +
 2~\chi_{\lambda}'\partial_{y}\varphi +
 \chi_{\lambda}''\varphi +
 \chi_{\lambda} m^{2} \varphi\right) = 0~,
\ee
where we identified $\partial_{\mu}\partial^{\mu}\varphi$ with
$m^{2}\varphi$.  The term in brackets should vanish everywhere, since
it is continuous and we can choose $\chi_\lambda(y)$ different from
zero everywhere except at one point between 0 and $2\pi R$.  If we
finally take the limit $\lambda\to 0$ we obtain the equation of motion
for the discontinuous fields:
\be
\label{discf}
 \chi \partial_{y}^{2}\varphi
 - 4 \delta(y-\pi R) \partial_{y}\varphi
 - 2 \delta'(y-\pi R) \varphi +
 \chi~ m^{2} \varphi = 0~.
\ee
Away from the point $y=\pi R$ this equation reduces to the equation of
motion for continuous fields: terms with delta functions disappear and
we can divide by $\chi(y)$. We obtain:
\be
 \partial_{y}^{2}\varphi +
 m^{2} \varphi = 0~.
\ee
Moreover, by integrating eq.~(\ref{discf}) and its primitive around $y
= \pi R$, we find:
\be
 \begin{array}{l}
  \varphi (\pi R + \xi ) = - \varphi (\pi R - \xi )\\
  \varphi '(\pi R + \xi ) = - \varphi '(\pi R - \xi )~,
 \end{array}
\ee
which are just the expected jumps.

There is another possibility to derive the correct equation of motion
from a singular action, beyond that of adopting a convenient
regularization.  We illustrate this procedure in the case of one
complex scalar field $\varphi(y)$.  The basic idea is to use a set of
field variables such that their infinitesimal variations, implied by
the action principle, are continuous functions of $y$. The action
principle requires that the variation of the action $S$, assumed to be
a smooth functional of $\varphi$ and $\partial\varphi$, vanishes for
infinitesimal variations of the fields from the classical trajectory:
\be
\delta S=\int d^4x~ dy~ \frac{\delta\mathcal{L}}{\delta\varphi}
~\delta\varphi=0~.
\ee
If the system is described by discontinuous fields, in general we
cannot demonstrate that $\delta\mathcal{L}/\delta\varphi$ vanishes at
the singular points, since multiplication/division by discontinuous
functions like $\delta\varphi$ is known to produce inequivalent
equalities.  An exception is the case of fields whose generic
variation $\delta\varphi$ is a continuous function, despite the
discontinuities of $\varphi$. In this case the action principle leads
directly to the usual equation of motion.

We consider as an example the case discussed at the beginning of
section~\ref{GBCbos-more}.  Real and imaginary components of $\varphi$
are respectively even and odd functions of $y$ and we have boundary
conditions specified by the matrices $U_\gamma$ in
eq.~(\ref{rotation}).  In particular, the discontinuities of
$\varphi_i$ $(i=1,2)$ and its $y$-derivative across 0 and $\pi R$ are
given by:
\be
\left(
\begin{array}{c}
\varphi_{1(2)}\\
\partial_y\varphi_{1(2)}
\end{array}
\right)(\gamma^+)
-
\left(
\begin{array}{c}
\varphi_{1(2)}\\
\partial_y\varphi_{1(2)}
\end{array}
\right)(\gamma^-)
=
(-)2
\tan\frac{\delta_\gamma}{2}
\left(
\begin{array}{c}
\varphi_{2(1)}\\
\partial_y\varphi_{2(1)}
\end{array}
\right)(\gamma)~,
\label{disc1}
\ee
where $\gamma$ stands for $0$ or $\pi R$, $\gamma^{+(-)}$ denotes
$0^{+(-)}$ or $\pi R^{+(-)}$ and
\be
\left(
\begin{array}{c}
\varphi_{1(2)}\\
\partial_y\varphi_{1(2)}
\end{array}
\right)(\gamma)
\equiv
\frac{1}{2}
\left[
\left(
\begin{array}{c}
\varphi_{1(2)}\\
\partial_y\varphi_{1(2)}
\end{array}
\right)(\gamma^+)
+
\left(
\begin{array}{c}
\varphi_{1(2)}\\
\partial_y\varphi_{1(2)}
\end{array}
\right)(\gamma^-)
\right]~.
\ee
From this we see that a generic variation of $\varphi_2$ is
discontinuous. The jump of $\delta\varphi_2$ across $0$ or $\pi R$ is
proportional to the value of $\delta\varphi_1$ at that point, which in
general is not zero.  However we can move to a new set of real fields
$\theta$ and $\rho$:
\be
\varphi =\rho~ e^{i \theta}~,
\ee
whose discontinuities from eq.~(\ref{disc1}) read:
\be
\left(
\begin{array}{c}
\rho\\
\partial_y\rho
\end{array}
\right)(\gamma^+)
=
\left(
\begin{array}{c}
\rho\\
\partial_y\rho
\end{array}
\right)(\gamma^-)~~~~~~~~~~
\left(
\begin{array}{c}
\theta\\
\partial_y\theta
\end{array}
\right)(\gamma^+)
-
\left(
\begin{array}{c}
\theta\\
\partial_y\theta
\end{array}
\right)(\gamma^-)
=
\left(
\begin{array}{c}
\delta_\gamma\\
0
\end{array}
\right)~.
\label{disc2}
\ee
The discontinuity of $\theta$ at each fixed point is a constant,
independent from the value of $\varphi$ at that point. As a
consequence, the infinitesimal variation $\delta\theta$ relevant to
the action principle is continuous everywhere, including the points
$y=0$ and $y=\pi R$.  We can derive the action for $(\rho,\theta)$, by
starting from the lagrangian expressed in terms of
$\varphi^c\equiv\rho\exp[i(\theta+\alpha)]$, where the function
$\alpha$ has been defined in eq.~(\ref{alpha}):
\be
\mathcal{L}(\varphi^c,\partial\varphi^c) =
-\partial_{M}\varphi^{c\dag} \partial^{M}\varphi^c~.
\ee
In terms of $\rho$ and $\theta$ we have:
\be
\mathcal{L}(\rho,\partial\rho,\theta,\partial\theta) =
-\partial_{M}\rho\ \partial^{M}\rho -
\rho^{2}\partial_{M}(\theta +\alpha)\ \partial^{M}(\theta +\alpha)~.
\ee
The lagrangian now contains singular terms, localized at the fixed
points.  The equations of motions, derived from the variational
principle, read:
\be
\left\{\begin{array}{l}
\partial_{M}\partial^{M}\rho -
\rho\ \partial_{M}(\theta +\alpha)\ \partial^{M}(\theta +\alpha)=0\\
\partial_{M} [\rho^{2}\partial^{M}(\theta +\alpha)]=0
\end{array}\right. ~.
\label{ch}
\ee
In the bulk $\alpha$ is constant and drops from the previous
equations, which then become identical to the equations for the
continuous field $\varphi^c$, in polar coordinates.  Moreover, by
integrating eq.~(\ref{ch}) around the fixed points and by recalling
the properties of the function $\alpha$, we reproduce precisely the
jumps of eq.~(\ref{disc2}).  The same results can be obtained by
introducing a regularization for $\alpha$.

To summarize, when going from a smooth to a discontinuous description
of the same physical system, singular terms are generated in the
lagrangian. In our examples we have quadratic terms localized at the
orbifold fixed point which, despite their highly singular behaviour,
are necessary for a consistent description of the system. Indeed they
encode the discontinuities of the adopted field variables which can be
reproduced via the classical equation of motion after appropriate
regularization or through a careful application of the standard
variational principle. Conversely, when localized terms for bulk
fields are present in the 5D lagrangian, as for many phenomenological
models currently discussed in the literature, the field variables are
affected by discontinuities. These can be derived by analyzing the
regularized equation of motion and can be crucial to discuss important
physical properties of the system, such as its mass spectrum. In some
case we can find a field redefinition that eliminate the
discontinuities and provide a smooth description of the system. It
would be nice to know precisely when we can perform the redefinition
that completely eliminate the localized terms and when they cannot be
removed. In the next section we will study this issue.


\section[Equivalent Effective Lagrangians]{Equivalent Effective Lagrangians}
\label{bfwz}
In previous sections we showed that different effective lagrangians
can describe the same physical system. This is due to the equivalence
between the SS twist and the generalized boundary conditions described
in detail in sections~\ref{GBCferm-bc} and~\ref{GBCbos-bc}, which are
associated to localized mass terms.  In this section we find out the
class of equivalence of the lagrangians associated to the same
spectrum and we classify the conditions the 5D mass terms should
satisfy in order to be ascribed to a SS twist. We illustrate this in
detail only for fermions; analogous considerations also apply to
bosons.

\subsection{Generation of 5D Mass Terms for Periodic Fields}
\label{bfwz-mass}
We consider the lagrangian of eq.~(\ref{lagrBFWZ}) with twisted
periodicity conditions~(\ref{twistBFWZ})-(\ref{betaBFWZ}). The
solution to the equation of motion with these boundary conditions is
displayed in eqs.~(\ref{ferm-massesBFWZ})-(\ref{gammaBFWZ}). We now
move to a class of equivalent descriptions of this system exploiting
the fact that S-matrix elements do not change if we perform a local
and non-singular field redefinition. We replace the twisted fields
$\Psi(y)$ by periodic ones $\widetilde{\Psi}(y)$:
\be
\label{redefBFWZ}
\Psi(y)  =  V(y) \, \widetilde{\Psi}(y) \, ,
\;\;\;\;\;\;\;\;
\widetilde{\Psi} (y+2\pi R)  =
\widetilde{\Psi} (y) \, ,
\ee
where $V(y)$ must then be a $2\times 2$ matrix satisfying
\begin{subequations}
\label{reqall}
\be
V(y+2 \pi R)=U_{\vec{\beta}} \, V(y) \, ,
\label{req1}
\ee
\end{subequations}
as can be immediately checked from eqs.~(\ref{twistBFWZ})
and~(\ref{redefBFWZ}). Besides condition~(\ref{req1}), we will impose
for our convenience two additional constraints on the matrix
$V(y)$. One is
\addtocounter{equation}{-1}
\begin{subequations}
\addtocounter{equation}{1}
\be
V(y)\in SU(2) \, ,
\label{req2}
\ee
\end{subequations}
which guarantees that the redefinition is non-singular, and that the
kinetic terms for $\widetilde{\Psi}(y)$ remain canonical.  We also
require that the new fields $\widetilde{\psi}_1(y)$ and
$\widetilde{\psi}_2(y)$ have the same parities as the original ones
${\psi}_1 (y)$ and ${\psi}_2 (y)$:
\addtocounter{equation}{-1}
\begin{subequations}
\addtocounter{equation}{2}
\be
\left\{
\begin{array}{rclcc}
V_{ij}(-y)&=&+V_{ij}(y)  & \phantom{bla} &(ij=11,22)\\
V_{ij}(-y)&=&-V_{ij}(y)  & \phantom{bla} &(ij=12,21)
\end{array} \right. \, .
\label{req3}
\ee
\end{subequations}
Notice that eq.~(\ref{req3}) implies $V(0)=\exp \, ( \, i \, \theta \,
\widehat{\sigma}^3 \, )$, with $\theta \in {\mathbb R}$.

Before exploring the effects of the field redefinition of
eq.~(\ref{redefBFWZ}), we observe that the solution to the
conditions~(\ref{reqall}) is by no means unique. Starting from any
given solution $V(y)$, a new set of solutions $V'(y)$ can be generated
via matrix multiplication:
\be
V'(y) = W_L(y) \, V(y) \, W_R(y) \, ,
\label{newsol}
\ee
provided that the following conditions are satisfied:
\begin{subequations}
\be
W_L(y+2\pi R) \, U_{\vec{\beta}} = U_{\vec{\beta}} \, W_L(y)
\, , \;\;\;\;\; W_R(y+2\pi R) = W_R(y) \, ,
\label{wy1}
\ee
\be
W_{L,R}(y)\in SU(2) \, ,
\label{wy2}
\ee
\be
\left\{
\begin{array}{rclcc}
(W_{L,R})_{ij}(-y)&=&+(W_{L,R})_{ij}(y)& \phantom{bla} &(ij=11,22)\\
(W_{L,R})_{ij}(-y)&=&-(W_{L,R})_{ij}(y)& \phantom{bla} &(ij=12,21)
\end{array} \right. \, .
\label{wy3}
\ee
\end{subequations}

We are now ready to explore the effects of the field redefinition of
eq.~(\ref{redefBFWZ}). The lagrangian $\cL$, expressed in terms of the
periodic field $\widetilde{\Psi}(y)$, describes exactly the same
physics as before, but its form is now different:
\bea
\label{result}
\cL(\Psi,\partial\Psi) & = &
\cL(\widetilde{\Psi},\partial\widetilde{\Psi}) + \left\{
-\frac{i}{2}\left[m_1(y)+i \, m_2(y)\right] \widetilde{\psi}_1
\widetilde{\psi}_1 \right. \nn \\ & + & \left.
\frac{i}{2}\left[m_1(y)-i \, m_2(y)\right]
\widetilde{\psi}_2 \widetilde{\psi}_2 + i \, m_3(y) \,
\widetilde{\psi}_1 \widetilde{\psi}_2 + {\rm h.c.} \right\} \, ,
\eea
where the mass terms $m_a(y)$ $(a=1,2,3)$ are the coefficients
of the Maurer-Cartan form
\be
m(y) \equiv m_a(y) \, \widehat{\sigma}^a =
- i \, V^\dagger(y) \partial_y V(y) \, ,
\label{mcform}
\ee
and satisfy:
\begin{subequations}
\label{mmmm}
\be
m_a(y+2\pi R) = m_a(y) \, ,
\label{I}
\ee
\be
m_a(y) \in {\mathbb R} \, ,
\label{II}
\ee
\be
m_{1,2}(-y) = + m_{1,2}(y) \, ,
\;\;\;\;\;
m_{3}(-y) = -m_{3}(y) \, ,
\label{III}
\ee
$$
U_{\vec{\beta}}  =   V(0) \,
P \left[ \exp \left( {\dd i \,
\int_0^y dy' m(y')}
\right) \right] \,
P_< \left[ \exp \left( {\dd i \,
\int_y^{y+2\pi R} dy' m(y')}
\right) \right] \, \times \phantom{blablabla}
$$
\be
P\,' \left[
\exp \left( {\dd - i \, \int_0^y
dy' m(y')}\right)\right]
V^\dagger(0) \, ,
\;\;\;\;\;
\left\{
\begin{array}{cccc}
P=P_< &  P\,'=P_> & {\rm for} & y>0  \\
P=P_> &  P\,'=P_< & {\rm for} & y<0
\end{array}
\right. \; .
\label{IV}
\ee
\end{subequations}

Properties~(\ref{I})-(\ref{III}) are in one-to-one correspondence with
conditions~(\ref{req1})-(\ref{req3}) on $V(y)$. Eq.~(\ref{IV}) is
related with eq.~(\ref{req1}), and prescribes how the information on
the twist of the original fields $\Psi(y)$ is encoded in the new
lagrangian. The symbols $P_<$ and $P_>$ denote inequivalent
definitions of path-ordering, specified in appendix~\ref{path-ord}
with some useful properties and the proof of eq.~(\ref{IV}). Notice
that, by taking the trace of both members in eq.~(\ref{IV}), we obtain
a relation between the Wilson loop and the twist parameter $\beta$:
\be
\cos \beta =
{1 \over 2} {\rm tr} \, U_{\vec{\beta}} =
{1 \over 2}{\rm tr} \, P_< \left[ \exp \left( {\dd i \,
\int_y^{y+2\pi R} dy' m(y')} \right) \right] \, .
\label{wloop}
\ee
Notice also that, because of the freedom of performing global $SU(2)$
transformations with the constant matrix $V(0)$, which are invariances
of the lagrangian, different values of the twist $\vec{\beta}$ with
the same value of $\beta$ correspond to physically equivalent
descriptions.

The mass terms in eq.~(\ref{result}) are of three different types,
associated with the three possible bilinears $\widetilde{\psi}_1
\widetilde{\psi}_1$, $\widetilde{\psi}_2 \widetilde{\psi}_2$ and
$\widetilde{\psi}_1 \widetilde{\psi}_2$. Because of eqs.~(\ref{mmmm}),
they do not correspond to the most general set of $y$-dependent mass
terms allowed by 4D Lorentz invariance, which would be characterized
by three independent complex functions. The r\^ole of the
conditions~(\ref{mmmm}) is to guarantee the equivalence between the
descriptions on the two sides of eq.~(\ref{result}).

Also the converse is true. Given a lagrangian such as the one on the
right-hand side of eq.~(\ref{result}), expressed in terms of periodic
fields $\widetilde{\psi}_i(y)$ $(i=1,2)$ and with mass terms
satisfying eqs.~(\ref{mmmm}), we can move to the equivalent lagrangian
of eq.~(\ref{lagrBFWZ}), where all mass terms have been removed and
the fields satisfy the periodicity conditions of
eq.~(\ref{twistBFWZ}), by performing the field redefinition of
eq.~(\ref{redefBFWZ}). As shown in appendix~\ref{path-ord}, $V(y)$ is
given by:
\be
V(y) = V(0) \, P \left[ \exp \left( {\dd i
\, \int_0^y dy' m(y') } \right) \right] \, ,
\;\;\;\;\;
P = \left\{
\begin{array}{ccc}
P_< & {\rm for} & y > 0 \\
P_> & {\rm for} & y < 0
\end{array}
\right.
\, .
\label{converse}
\ee
For any $V(0) = \exp \, ( \, i \, \theta \, \widehat{\sigma}^3 \, )$
($\theta \in {\mathbb R}$), conditions~(\ref{reqall}) are satisfied
with $U_{\vec{\beta}}$ given by eq.~(\ref{IV}).  The arbitrariness in
$V(0)$ reflects the fact that physically distinct theories are
characterized by $\beta$, not by $\vec{\beta}$.

Up to now we have considered only free theories. What happens if we
turn on interactions in the lagrangian~(\ref{lagrBFWZ})? First of all
we ask that the interaction terms are invariant under $SU(2)$. If
these terms do not contains $y$-derivatives, they remain unchanged
after the redefinition~(\ref{redefBFWZ}), so eq.~(\ref{result}) still
holds true. If instead terms involving $\partial_y \Psi$ are present,
they will generate, after the field redefinition, additional but
controllable contributions to the right hand side of
eq.~(\ref{result}). Anyway, since the equivalence between two
lagrangians related by a local field redefinition holds irrespectively
of the explicit form assumed by the interaction terms, the two
descriptions of the system will still be equivalent.

\subsection{Examples and Localization of 5D Mass Terms}
\label{bfwz-ex}
From the discussion of the previous section, it is clear that mass
`profiles' $m_a(y)$ for periodic fields, of the type specified in
eq.~(\ref{mmmm}), do not have an absolute physical meaning. They can
be eliminated from the lagrangian and replaced by a twist, the two
descriptions being completely equivalent. Moreover, all lagrangians
with the same interaction terms and mass profiles corresponding to the
same twist $\vec{\beta}$, as computed from eq.~(\ref{IV}), are just
different equivalent descriptions of the same physics. Indeed, suppose
that $\cL^1$ and $\cL^2$ are two such lagrangians, and call $V^I(y)$
$(I=1,2)$ the local redefinitions mapping $\cL^I$ into the lagrangian
$\cL$ for the twisted, massless 5D fields $\Psi(y)$. Then $\cL^1$ and
$\cL^2$ are related by the local non-singular field redefinition $V(y)
= {V^2}^\dagger(y) V^1(y)$. This shows that, in the class of
interacting models under consideration, what matters is the twist
$\vec{\beta}$ and not the specific form of the mass terms $m_a(y)$
enjoying the properties~(\ref{mmmm})~\footnote{Actually, in view of
the observations after eqs.~(\ref{wloop}) and~(\ref{converse}), what
really matters is $\beta$.}.  We can make use of this freedom to show
that $m_{1,2}(y)$ can be localized at the fixed points $y=0$ and/or
$y=\pi R$, without affecting the physical properties of the theory.

As an example, we consider the simple case in which the twist parameter
is just
\be
\vec{\beta}= (0,\beta,0) \, .
\label{twist2}
\ee
Then a frequently used solution to eq.~(\ref{reqall}), for
the twist specified by eq.~(\ref{twist2}), is
\be
V^O(y) = \exp \left( i \, \beta \widehat{\sigma}^2
\frac{y}{2\pi R} \right) \, ,
\label{vss}
\ee
the symbol `$O$' standing for {\em `ordinary'}. Starting from the
lagrangian $\cL$ for the periodic fields $\widetilde{\Psi}(y)$,
defined by $V^O(y)$ via the redefinition of eq.~(\ref{redefBFWZ}), and
performing the standard Fourier decomposition of the 5D fields into 4D
modes, we can immediately check that the 4D mass eigenvalues and
eigenfunctions are indeed given by
eqs.~(\ref{ferm-massesBFWZ})-(\ref{gammaBFWZ}), with
$\beta_1=0$. Applying eq.~(\ref{mcform}) to $V^O(y)$, we find the
constant mass profile:
\be
m_1^O(y)=m_3^O(y)=0 \, ,
\;\;\;\;\;
m_2^O(y)=\frac{\beta}{2 \pi R} \, ,
\label{mo}
\ee
and we can check that, in agreement with eq.~(\ref{wloop}):
\be
\beta = \int_y^{y+2\pi R}d y' m_2^O(y') \, .
\label{sametwist}
\ee

We now move to a more general solution of eqs.~(\ref{reqall})
and~(\ref{twist2}), where, in a basis of periodic fields, the system
is described by a different lagrangian $\cL^G$ (the symbol `$G$'
stands for {\em `generalized'}). $\cL^G$ is still of the general form
of eq.~(\ref{result}), including interaction terms, but now:
\be
m_1^G(y)=m_3^G(y)=0 \, ,
\;\;\;\;\;
m_2^G(y)\ne 0 \, ,
\label{mgen}
\ee
and $m_2^G(y)$ is an otherwise arbitrary real, periodic, even function
of $y$, with the property that
\be
\int_y^{y+2\pi R}d y' m_2^G(y') =
\int_y^{y+2\pi R}d y' m_2^O(y') =
\beta \, .
\ee
As long as the above properties are satisfied, the two lagrangians
$\cL^O$ and $\cL^G$ are physically equivalent. Two representative and
equivalent choices of $m_2^G(y)$ are illustrated in fig.~\ref{m2y}:
\begin{figure}[!t]
\begin{center}
\epsfig{figure=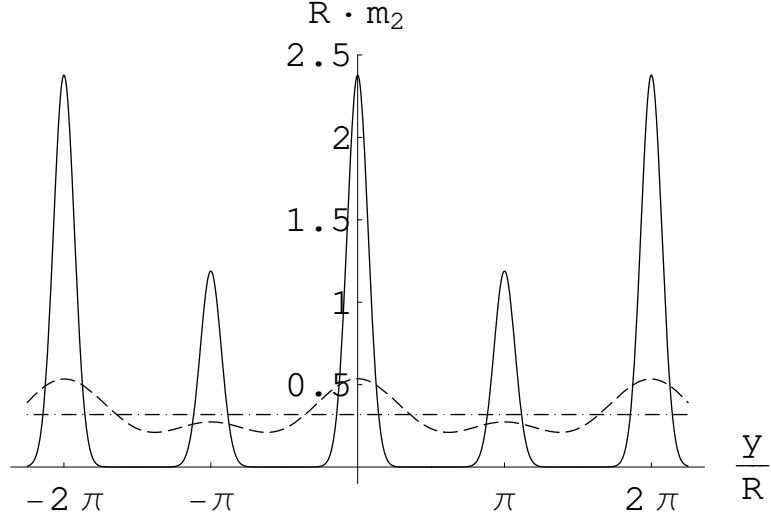,height=7.5cm}
\end{center}
\caption{Two representative and equivalent choices for $m_2^G(y)$,
corresponding to $\beta=2$. For reference, the dash-dotted line
shows the equivalent constant profile $m_2^O(y)= 1 / (\pi R)$.}
\label{m2y}
\end{figure}
\begin{figure}[!t]
\begin{center}
\epsfig{figure=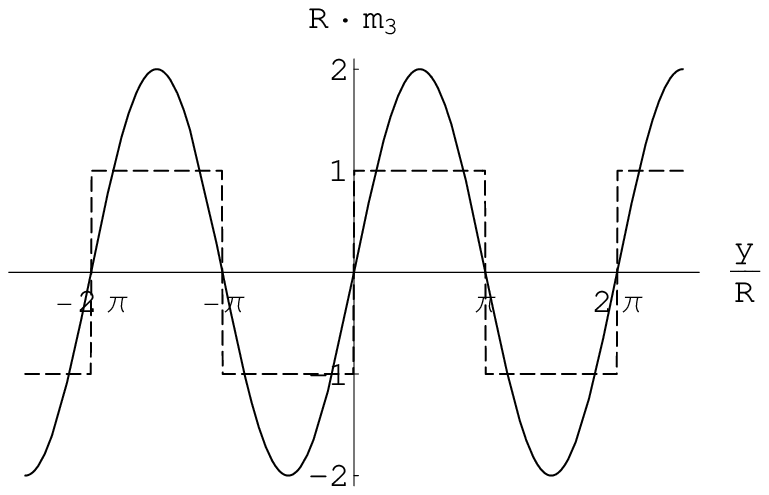,height=7.5cm}
\label{m3y}
\end{center}
\caption{Two representative and equivalent choices for $m_3(y)$: the
solid line corresponds to $m_3(y)=(2 \sin y)/R$, the dashed one to
$m_3(y)=\epsilon(y)/R$.}
\label{m3y}
\end{figure}
the dashed line shows a mild (gaussian) localization around the
orbifold fixed points, the solid line a strong localization. The
interactions between $\widetilde{\Psi}(y)$ and other fields are not
determined by the shapes of the fermion eigenmodes and, indirectly, by
the profile of $m_2(y)$. Neither the mass spectrum, nor the
interactions depend on shapes, which are an artifact of the choice of
field variables. As long as the twist is kept fixed, shapes can be
arbitrarily deformed along $y$, without changing the physics.

A possible special choice for $m_2^G(y)$ is the singular limit:
\be
m_2^G(y)=\sum_{q=-\infty}^{+\infty}
\left[\delta_0 \, \delta(y-2 q \pi R)+
\delta_\pi \, \delta(y-(2 q+1) \pi R)\right] \, ,
\;\;\;\;\;
\delta_0+\delta_\pi=\beta \, ,
\label{mg}
\ee
where what we actually mean is a suitably regularized version of the
distribution in eq.~(\ref{mg}). This description is apparently quite
remote from the `ordinary' one. The mass terms vanish everywhere but
at the orbifold fixed points, where there are localized contributions
to $m_2(y)$. The redefinitions bringing from the massive periodic
fields of $\cL^O$ and $\cL^G$ to the corresponding massless twisted 5D
fields are:
\be
\Psi(y)=V^{O,G}(y)~ \widetilde{\Psi}^{O,G}(y) \, ,
\label{redefog}
\ee
with $V^O(y)$ given by eq.~(\ref{vss}) and
\be
V^G(y)=\exp \left[i \, \alpha(y) \widehat{\sigma}^2\right] \, .
\label{redefg}
\ee
Here $\alpha(y)$ is the step function introduced in
section~\ref{GBCferm-one}.
The local field redefinition that relates the two lagrangians $\cL^O$
and $\cL^G$ is:
\be
\widetilde{\Psi}^G(y)={V^G}^\dagger (y) V^O(y)~ \widetilde{\Psi}^O(y)
\,.
\label{redefgo}
\ee
Notice that the periodic fields $\widetilde{\Psi}^G(y)$ are not smooth
but only piecewise smooth. This can be checked either by integrating
the equations of motion for $\widetilde{\Psi}^G(y)$, derived from
$\cL^G$, in a small region around the fixed points, or by making use
of the field redefinition in eq.~(\ref{redefgo}), recalling that
$\widetilde{\Psi}^O(y)$ and $V^O(y)$ are smooth while $V^G(y)$ is not.
We find that the fields $\widetilde{\Psi}^G(y)$ have cusps and
discontinuities described by:
\be
\left\{
\begin{array}{l}
\widetilde{\Psi}^G(2 q \pi R+\xi) =
e^{\dd{\, i \, \delta_0 \sigma^2}}
\widetilde{\Psi}^G(2 q \pi R-\xi)
\\
\widetilde{\Psi}^G[(2 q+1) \pi R+\xi] =
e^{\dd{\, i \, \delta_\pi\sigma^2}} \,
\widetilde{\Psi}^G[(2 q+1) \pi R-\xi]
\end{array}
\right.
\, ,
\;\;\;\;\;
(0 < \xi \ll 1 \, ,
\, q \in {\mathbb Z} )
\, ,
\label{jumps}
\ee 
where the jumps of the field variables are parametrized by
$\delta_{0,\pi}$.

Another simple but instructive example corresponds to a lagrangian for
periodic fields of the form in eq.~(\ref{result}), where now
\be
m_1(y)=m_2(y)=0 \, ,
\;\;\;\;\;
m_3(y) \ne 0 \, ,
\label{oddm}
\ee
and $m_3(y)$ is an otherwise arbitrary real, odd, periodic function of
$y$, as prescribed by eqs.~(\ref{I})-(\ref{III}). Notice that, for
any such function, eq.~(\ref{wloop}) gives always $\beta=0$, since
\be
\int_y^{y+2 \pi R} dy' m_3(y') = 0  \, .
\ee
In other words, real, periodic, odd mass profiles can be completely
removed by a field redefinition without introducing a non-trivial
twist. Such a field redefinition corresponds to:
\be
V(y) = \exp \left[ i \int_0^y dy' m_3(y') \right] \, .
\ee
Some representative profiles for $m_3(y)$ are exhibited in
fig.~\ref{m3y}.  Notice that no constant $m_3(y) \ne 0$ is allowed by
eqs.~(\ref{I})-(\ref{III}), and also a $m_3(y) \ne 0$ completely
localized at $y = 2 q \pi R$ and/or $y = (2q + 1) \pi R$ is
forbidden. An allowed possibility is a piecewise constant $m_3(y)$,
for example:
\be
m_3(y) = \mu \, \epsilon(y) \, ,
\ee
where $\epsilon(y)$ is the periodic sign function and $\mu$ a real
constant with the dimension of a mass.


\section[Application to Gauge Symmetry Breaking]{Application to 
Gauge Symmetry Breaking}
\label{gauge}
The generalized boundary conditions discussed in previous sections can
be exploited to spontaneously break the gauge invariance of a 5D
system.  This is well-known as far as the twist is concerned, as
described in section~\ref{gauge-breaking}. A non-trivial twist induces
a shift in the KK levels which lifts the zero modes of the gauge
vector bosons: from a 4D point of view the gauge symmetry is thus
broken.  As we have seen in section~\ref{GBCbos}, also the
discontinuities of the fields and their first derivatives have a
similar effect on the spectrum and we may expect that, in the context
of a gauge theory, they lead to spontaneous breaking of the 4D gauge
invariance. To analyze these aspects, we focus on a 5D gauge theory
defined on our orbifold and based on the gauge group SU(2).

Not all $\mathbb{Z}_2$ parity assignments for the gauge fields
$A^a_M(x,y)$ $(a=1,2,3)$, $(M=\mu,5)$ are now possible. The gauge
invariance imposes several restrictions. First of all, the action of
$\mathbb{Z}_2$ on the 4D vector bosons $A^a_\mu$ should be compatible
with the algebra of gauge group. In other words, we should embed
$\mathbb{Z}_2$ into the automorphism algebra of the gauge
group~\cite{automorphism}. For SU(2) this leaves two possibilities:
either all $A^a_\mu$ are even, or two of them are odd and one is even.
Furthermore, a well-defined parity for the field strength implies that
the parity of $A^a_5$ should be opposite to that of $A^a_\mu$. In the
basis $(A^1_\mu,A^2_\mu,A^3_\mu)$, up to a re-labeling of the three
gauge fields, we can consider:
\bea
\label{parity} 
({\rm A})~~~~~~~~Z&=&diag(+1,+1,+1)\nn\\
({\rm B})~~~~~~~~Z&=&diag(-1,-1,+1)~. 
\eea
The boundary conditions on the $A^a_\mu$ are specified by $3\times3$
matrices $U_\gamma$ that satisfy the consistency
relation~(\ref{cons2}) and leave the SU(2) algebra invariant. This
last requirement can be fulfilled by requiring that $U_\gamma$ is an
SU(2) global transformation that acts on $(A^1_\mu,A^2_\mu,A^3_\mu)$
in the adjoint representation.  Finally, to preserve gauge invariance,
the boundary conditions on the scalar fields $A^a_5$ should be the
same as those on the corresponding 4D vector bosons. This can be seen
by asking that the various components of the field strength $F^a_{MN}$
possess well-defined boundary conditions.

For instance, in the case (A) where all fields $A^a_\mu$ have even
$\mathbb{Z}_2$ parity, a consistent assignment is:
\be 
\label{bcg1} 
U_\beta
= diag(-1,-1,+1)~,~~~~~~~~U_0=U_\pi=diag(+1,+1,+1)~.
\ee 
In the gauge $\partial^M A^a_M=0$, the 5D equations of motion read:
\be 
-\partial^2_y A^a_M = m^2 A^a_M~.
\label{eomg} 
\ee 
The solutions with the appropriate boundary
conditions are: 
\be 
A^a_\mu(x,y)=A^{a(n_a)}_\mu(x) \cos m_a y
~~~~~~~~m_{1,2} R = n_{1,2}+\frac{1}{2} ~~~~~~~~m_3 R= n_3 ~,
\label{spg1} 
\ee 
where $n_{1,2,3}$ are non-negative integers. The only zero mode of the
system is $A^{3(0)}_\mu(x)$ and, from a 4D point of view the original
gauge symmetry is broken down to the U(1) associated to this massless
gauge vector boson. The breaking of the 5D SU(2) gauge symmetry is
spontaneous and each mode in~(\ref{spg1}), but $A^{3(0)}_\mu(x)$,
becomes massive via a Higgs mechanism. The unphysical Goldstone bosons
are the modes of the fields $A^a_5(x,y)$, which are all absorbed by
the corresponding massive vector bosons. On the wall at $y=0$ all the
gauge fields and the parameters of the gauge transformations are
non-vanishing.  Here all the constraints coming from the full 5D gauge
invariance are effective. On the contrary on the wall at $y=\pi R$,
only $A^3_\mu(x,y)$ and the corresponding gauge parameter are
different from zero. Therefore the effective symmetry at the fixed
point $y=\pi R$ is the U(1) related to the 4D gauge boson
$A^3_\mu(x,\pi R)$.  This kind of setup where the gauge symmetry is
broken by twisted orbifold boundary conditions and the two fixed
points are characterized by two different effective 4D symmetries has
recently received lot of attention, for its successful application in
the context of GUTs (see sections~\ref{gauge-breaking}).

We have just shown how a gauge symmetry can be broken by means of
twisted fields.  It is interesting to note that the same physical
system can be described by using periodic field variables, with
discontinuities at the fixed points. This is achieved, for instance,
by means of the boundary conditions
\be 
U_\beta =
U_0=diag(+1,+1,+1)~,~~~~~~~~U_\pi=diag(-1,-1,+1)~.
\label{bcg2} 
\ee 
The solutions to the equations of motion~(\ref{eomg}) are:
\bea 
\label{spg2} 
A^a_\mu(x,y)&=&A^{a(n_a)}_\mu(x)~\epsilon(y/2+\pi R/2)\cos m_a y
~~~~~~m_a R = n_a+\frac{1}{2}~~~~~~~~(a=1,2)\nn\\
A^3_\mu(x,y)&=&A^{3(n_3)}_\mu(x) \cos m_3 y~~~~~~~~~~~~~~~~~~~~~~~~~~
m_3 R=n_3~, 
\eea 
where $n_{1,2,3}$ are non-negative integers. The new solutions
$A^{1,2}_\mu$ have cusps at $y=\pi R$ (mod $2\pi$), as the profiles
denoted by $(+,-,+)$ in fig.~\ref{evenf}. The two descriptions are
related by the field redefinition
\be
\label{gauge-redef}
A^a_\mu(x,y)~\to~ \epsilon(y/2+\pi R/2)~
A^a_\mu(x,y)~~~~~~~~(a=1,2) 
\ee
which is a local transformation. As already stressed, this assures
that the description in terms of smooth twisted fields is physically
equivalent to the one with periodic and discontinuous fields.

In the previous example the boundary conditions $U_\gamma$ commute
among themselves and with the parity $Z$. As a consequence the rank of
the gauge group SU(2) is conserved in the symmetry breaking. We can
lower the rank by assuming $[U_\gamma,Z]\ne 0$. As an example, we
consider the parity (B) of eq.~(\ref{parity}) and boundary conditions
described by:
\be
U_\gamma=e^{\dd \theta_\gamma T^2}~~~~~~~~ T^2= \left(
\begin{array}{ccc}
0& 0& 1\\
0& 0& 0\\
-1& 0& 0
\end{array}
\right)~, 
\label{bg3} 
\ee 
in the basis $(A^1_\mu,A^2_\mu,A^3_\mu)$. We allow, at the same time,
for a twist $\theta_\beta\equiv\beta$ and two jumps
$\theta_{0(\pi)}\equiv\delta_0(\delta_\pi)$. The matrices $U_\gamma$
are block diagonal and do not mix the index 2 with the indices
(1,3). Thus the boundary conditions are trivial for the odd field
$A^2_\mu$ and its derivative. Non-trivial boundary conditions involve
the fields $A^1_\mu$ and $A^3_\mu$. By solving the
equations~(\ref{eomg}), we obtain:
\bea
A^1_\mu(x,y)&=&A^{(n)}_\mu(x)~\sin(m y-\alpha(y))\nn\\
A^2_\mu(x,y)&=&A^{2(n_2)}_\mu(x)~\sin m_2 y~~~~~~~~~~~~~~~~~~m_2 R=n_2\\
A^3_\mu(x,y)&=&A^{(n)}_\mu(x)~\cos(m y-\alpha(y))~~~~~~~~~m
R = n - \frac{\beta-\delta_0-\delta_\pi}{2 \pi}~, \nn
\eea 
where $n\in \mathbb{Z}$, $n_2$ is a positive integer and the function
$\alpha(y)$ has been defined in section~\ref{GBCferm-one}. If
$\beta-\delta_0-\delta_\pi=0$ (mod $2\pi$), we have a zero mode
$A^{(0)}_\mu(x)$ and the gauge symmetry is spontaneously broken down
to U(1), as in the previously discussed examples. When
$\beta-\delta_0-\delta_\pi\ne0$ (mod $2\pi$), there are no zero modes
and SU(2) is completely broken. We can go continuously from this phase
to the phase where a U(1) survives, by changing the twist and/or the
jump parameters. We may thus have a situation where U(1) is broken by
a very small amount, compared to the scale $1/R$ that characterizes
the SU(2) breaking. The U(1) breaking order parameter is the
combination $\beta-\delta_0-\delta_\pi$.  The same physical system is
described by a double infinity of boundary conditions, those that
reproduce the same order parameter. All these descriptions are
equivalent and are related by field redefinitions. In the class of all
equivalent theories one of them is described by continuous fields
$A^{a\,c}_\mu(x,y)$.  We go from the generic theory described in terms
of $(\beta,\delta_0,\delta_\pi)$ to that characterized by
$(\beta^c\equiv\beta-\delta_0-\delta_\pi,\delta^c_0\equiv
0,\delta^c_\pi\equiv 0)$, via the field transformation:
\be \left(
\begin{array}{c}
A^{1\,c}_\mu\\
A^{3\,c}_\mu
\end{array}
\right) = \left(
\begin{array}{cc}
\cos\alpha(y)&-\sin\alpha(y)\\
\sin\alpha(y)&\cos\alpha(y)\\
\end{array}
\right) \left(
\begin{array}{c}
A^{1}_\mu\\
A^{3}_\mu
\end{array}
\right)~. 
\ee

In section~\ref{GBCbos-mass} we learned that discontinuous fields are
associated to localized lagrangian terms. We would like to derive
these terms also in this case with gauge fields. The main difference
with respect to the free theory examples discussed in
section~\ref{GBCbos-mass} is that now the lagrangian contains
derivative interactions among the gauge bosons. In principle these
could lead to localized interaction terms, that would provide a
non-trivial extension of the framework considered up to now. To
investigate this point we start from the 5D SU(2) Yang-Mills theory
defined by the parity (A) of eq.~(\ref{parity}) and by the boundary
conditions of eq.~(\ref{bcg1}).  No jumps are present and the overall
lagrangian is given only by the `bulk' term:
\be 
\label{lagg} 
\mathcal{L} =
-\frac{1}{4} F^{a\,c}_{MN} F^{a\,c\, MN}~. 
\ee
It is particularly convenient to discuss the physics in the unitary
gauge, where all the would-be Goldstone bosons, eaten up by the
massive KK modes, vanish:
\be 
\label{ugauge} 
A_5^{a\,c}(x,y)\equiv0~~~~~~~(a=1,2,3)~. 
\ee
In this gauge $F^{a\,c}_{5\mu}\equiv\partial_y A^{a\,c}_\mu$ and the
lagrangian~(\ref{lagg}) reads:
\be 
\label{laggug} 
\mathcal{L} = -\frac{1}{4}
F^{a\,c}_{\mu\nu} F^{a\,c\,\mu\nu} -\frac{1}{2}\partial_y A^{a\,c}_\mu
\partial_y A^{a\,c\,\mu}~. 
\ee 
To discuss the case of discontinuous gauge vector bosons, such as
those associated to boundary conditions~(\ref{bcg2}), we can simply
perform the field redefinition:
\be
A^{\hat{a}}_M(x,y)~=~ \chi_\lambda(y)~
A^{\hat{a} c}_M(x,y)~~~~~~~~({\hat{a}}=1,2)~, 
\label{red6} 
\ee 
where the function $\chi_\lambda(y)$ represents a regularized version
of $\epsilon(y/2+\pi R/2)$. Such redefinition maps the twisted, smooth
fields obeying~(\ref{bcg1}) into periodic variables, discontinuous at
$y=\pi R$, as specified in~(\ref{bcg2}). Notice that this redefinition
does not change the gauge condition~(\ref{ugauge}). If we plug the
transformation~(\ref{red6}) into the lagrangian~(\ref{laggug}), we
obtain the lagrangian for the system characterized by discontinuous
fields. We stress that, since this field redefinition is local, the
physics remains the same: the two systems are completely
equivalent. The S-matrix elements computed with the two lagrangians
are identical and, of course, this equivalence includes the
non-trivial non-abelian interactions. Our aim here is only to
understand how the physics, in particular the non-abelian
interactions, are described by the new, discontinuous variables.  From
eq.~(\ref{laggug}) we can already conclude that no localized
non-abelian interaction terms arise from the field
redefinition~(\ref{red6}), in the unitary gauge. Indeed, the only term
containing a $y$ derivative is quadratic, and, after the
substitution~(\ref{red6}), we will obtain terms analogous to those
discussed in~(\ref{regl}) for the case of a single scalar field.  We
find:
\begin{eqnarray}
\label{last-lagrangian} 
\mathcal{L}& = & -\frac{1}{4}
\tilde{F}^{3}_{\mu\nu} \tilde{F}^{3 \mu\nu} -\frac{1}{2}\partial_y
A^3_\mu \partial_y A^{3 \mu} +
\nonumber \\[0.15cm] 
&+&\frac{1}{2} \chi_\lambda^{2} f^{3\hat{b}\hat{c}}
A^{\hat{b}}_{\mu} A^{\hat{c}}_{\nu}
   \tilde{F}^{3 \mu\nu} -\frac{1}{4} \chi_\lambda^{4} 
f^{3\hat{b}\hat{c}} f^{3\hat{d}\hat{e}}
   A^{\hat{b}}_{\mu} A^{\hat{c}}_{\nu} A^{\hat{d}\mu} A^{\hat{e}\nu} -
\nonumber \\[0.15cm] 
&-&\frac{1}{4} \chi_\lambda^{2} F^{\hat{a}}_{\mu\nu}
F^{\hat{a}\mu\nu} -\frac{1}{2}\chi_\lambda^{2} \partial_y
A^{\hat{a}}_\mu \partial_y A^{{\hat{a}} \mu} -
\nonumber \\[0.15cm] 
& -&\frac{1}{2} ~\chi_\lambda'^{2}~
   A^{\hat{a}}_{\mu} A^{\hat{a}\mu}
- \chi_\lambda \chi_\lambda'~A^{\hat{a}\mu}~ \partial_{5}
A^{\hat{a}}_{\mu} ~,
\end{eqnarray}
where $\tilde{F}^{3}_{\mu\nu} = \partial_{\mu} A^{3}_{\nu}
-\partial_{\nu} A^{3}_{\mu}$, $f^{abc}$ are the structure constants of
SU(2) and the indices $\hat{a},\hat{b},...$ run over 1,2. In the limit
$\lambda\to 0$, the non-abelian interactions are formally unchanged,
whereas the last line represents a set of localized quadratic
terms. As discussed in the case of a real scalar field, such terms
guarantee, via the equations of motion, that the fields obey the new
boundary conditions~(\ref{bcg2}).  In a general gauge, interaction
terms localized at $y=\pi R$ are present, but they involve non-physical
would-be Goldstone bosons.\\


\begin{table}[!t]
\label{kawa-tab}
\centering{
\begin{tabular}{|c||c|c||c|c||c|}
\hline
& & & & & \\
Fields & $(Z,U_\beta)$ & $\psi_{K}(y)$
       & $(Z,U_\pi)$   & $\psi_{our}(y)$ & $m$ \\
& & & & & \\
\hline
& & & & & \\
$A^{a}_{\mu}, \lambda_{L}^{2a}, H^{D}_{u}, H^{D}_{d}$ &
$(+,+)$  & $cos(my)$ & $(+,+)$  & $cos(my)$ & $\frac{2n}{R}$ \\
& & & & & \\
\hline
& & & & & \\
$A^{\hat{a}}_{\mu}, \lambda_{L}^{2\hat{a}}, H^{T}_{u}, H^{T}_{d}$
&
$(+,-)$  & $cos(my)$ & $(+,-)$  
& $\epsilon (\frac{y}{2}+\frac{\pi R}{2})cos(my)$ & $\frac{2n+1}{R}$ \\
& & & & & \\
\hline
& & & & & \\
$A^{\hat{a}}_{5}, \Sigma^{\hat{a}},\lambda_{L}^{1\hat{a}},
\widehat{H}^{T}_{u}, \widehat{H}^{T}_{d}$ &
$(-,-)$  & $sin(my)$ & $(-,-)$  
& $\epsilon (\frac{y}{2}+\frac{\pi R}{2})sin(my)$ & $\frac{2n+1}{R}$ \\
& & & & & \\
\hline
& & & & & \\
$A^{a}_{5}, \Sigma^{a},\lambda_{L}^{1a}, \widehat{H}^{D}_{u},
\widehat{H}^{D}_{d}$ &
$(-,+)$  & $sin(my)$ & $(-,+)$  & $sin(my)$ & $\frac{2n+2}{R}$ \\
& & & & & \\
\hline
\end{tabular}}
\caption{Boundary conditions, eigenfunctions and spectra for
fields in the Kawamura model in the traditional scheme ($2^{nd}$,
$3^{rd}$ and $6^{th}$ columns) and in our framework ($4^{th}$,
$5^{th}$ and $6^{th}$ columns).} 
\end{table}

After having studied the case of $SU(2)$, we now briefly apply our
considerations to a realistic model. We consider the GUT proposed by
Kawamura in ref.~\cite{kawa} and described in
section~\ref{gauge-breaking} and we re-interpret this model with
generalized boundary conditions. We choose parity assignments
identical to the original ones, but, instead of a twist, we require
that some fields jump in $y=\pi R$. Among gauge fields only the vector
bosons of the coset $SU(5)~/~SU(3)\times SU(2)\times U(1)$ jump. In
table 2.2 boundary conditions, eigenfunctions and eigenvalues for both
the original model and for our framework are shown. We can observe
that the spectra are the same for all fields in both cases, while
eigenfunctions are identical for continuous fields but different for
jumping fields. The relation between the two sets of eigenfunctions is
analogous to the one between eq.~(\ref{spg1}) and eq.~(\ref{spg2}) and
they are related by a local field redefinition analogous to
eq.~(\ref{gauge-redef}).
 
Also in this case we can perform the field redefinition at the level
of the lagrangian in order to find the localized terms related to
jumps in this model. For simplicity we consider only the Yang-Mills
term of the lagrangian, neglecting both the supersymmetric part and
the Higgs terms:
\begin{equation}
\label{lagr_SU(5)} 
\mathcal{L} = -\frac{1}{4}
F^{\alpha}_{MN} F^{\alpha MN}~.
\end{equation}
Applying the usual field redefinition and choosing the unitary gauge,
this becomes:
\begin{eqnarray}
\label{lagr_SU(5)-jumps} 
\mathcal{L}& = & -\frac{1}{4}
\tilde{F}^{a}_{\mu\nu} \tilde{F}^{a \mu\nu} -\frac{1}{2}\partial_y
A^a_\mu \partial_y A^{a \mu} +
\nonumber \\[0.15cm] 
&+&\frac{1}{2} \epsilon^{2} f^{a\hat{b}\hat{c}}
A^{\hat{b}}_{\mu} A^{\hat{c}}_{\nu}
   \tilde{F}^{a \mu\nu} 
-\frac{1}{4} \epsilon^{4} f^{a\hat{b}\hat{c}} f^{a\hat{d}\hat{e}}
   A^{\hat{b}}_{\mu} A^{\hat{c}}_{\nu} A^{\hat{d}\mu} A^{\hat{e}\nu} -
\nonumber \\[0.15cm] 
&-&\frac{1}{4} \epsilon^{2} F^{\hat{a}}_{\mu\nu}F^{\hat{a}\mu\nu} 
-\frac{1}{2}\epsilon^{2} \partial_y
A^{\hat{a}}_\mu \partial_y A^{{\hat{a}} \mu} -
\nonumber \\[0.15cm] 
& -& 2 ~\delta^2(y-\pi R)~
   A^{\hat{a}}_{\mu} A^{\hat{a}\mu}
+2 \epsilon \delta(y-\pi R)~A^{\hat{a}\mu}~ \partial_{5}
A^{\hat{a}}_{\mu} ~,
\end{eqnarray}
where $\tilde{F}^{a}_{\mu\nu} = \partial_{\mu} A^{a}_{\nu}
-\partial_{\nu} A^{a}_{\mu} - f^{abc} A^{b}_{\mu} A^{c}_{\nu}$. Also
in this example we obtain mass terms that are localized where the
fields have discontinuities.


\section[Application to Supersymmetry Breaking]{Application to 
Supersymmetry Breaking}
\label{susy}

%
%
In this section we exploit the demonstrated equivalence between
``twist and jumps'' to show how brane-induced SUSY breaking in 5D,
which reproduces the main features of gaugino condensation in M-theory,
is equivalent to SS SUSY breaking.

We consider pure 5D Poincar\'e supergravity in its on-shell
formulation~\cite{sugra}. The supergravity multiplet contains the
f\"unfbein $e_M^{\;A}$, the gravitino $\Psi_M$ and the graviphoton
$B_M$. The 5D bulk Lagrangian is:
\bea
\label{BFZ2-lbulk}
\kappa {\cal L}_{bulk}
& = & - {1 \over 2 \kappa^2} e_5 R_5  - {1 \over 4} e_5 F_{MN} F^{MN}
- {\kappa \over 6 \sqrt{6}} \epsilon^{MNOPQ} F_{MN} F_{OP} B_Q + \nn \\
&   & + i  \epsilon^{MNOPQ} \ov{\Psi}_M \Sigma_{NO} D_P \Psi_Q - i
\sqrt{3 \over 2} {\kappa \over 2} e_5 F_{MN} \ov{\Psi}^M \Psi^N + \nn \\
& & + i \sqrt{3 \over 2} {\kappa \over 4} \epsilon^{MNOPQ} F_{MN}
\ov{\Psi}_O \Gamma_P \Psi_Q + {\rm 4\!\!-\!\!fermion~terms}~.
\eea
Here $\kappa=M_5^{-1}=(\pi R /M_{Pl}^2)^{\frac{1}{3}}$ is the inverse
reduced 5D Planck mass ($R$ is the compactification radius), $R_5$ is
the 5D scalar curvature, $e_5 = \det e_M^{\;A}$, $e_4 = \det
e_m^{\;a}$ (where the latter are the components of the f\"unfbein with
4D indices), $\epsilon^{MNOPQ} = e_5 e_A^{\;M} e_B^{\;N} e_C^{\;O}
e_D^{\;P} e_E^{\;Q} \epsilon^{ABCDE}$, $\epsilon^{mnop} = e_4
e_a^{\;m} e_b^{\;n} e_c^{\;o} e_d^{\;p} \epsilon^{abcd}$ and
$\epsilon^{\hat{0} \hat{1} \hat{2} \hat{3} \hat{5}} =
\epsilon^{\hat{0} \hat{1} \hat{2} \hat{3}} = + 1$. This lagrangian is
invariant under appropriate SUSY transformations (see eq.~(2.2) of
ref.~\cite{bfz2}).

We work on the orbifold $S^1/\mathbb{Z}_2$ and we assume that our
fields are fluctuations of the background
\be
\label{BFZ2-bg}
\langle g_{MN} \rangle =
\left(
\begin{array}{cc}
\eta_{mn} & 0 \\
0 & r^2
\end{array}
\right)~,
\ee
where $r^2=(R M_5)^2=(M_{Pl}^2 R^2/\pi)^{2/3}$ and all other
background fields are assumed to vanish.  We define the action of the
orbifold symmetry in such a way that the action, the transformation
laws and the background are all invariant. Writing the spinors in the
notation (A) of eq.~(\ref{ferm-notations}), we assign even
$\mathbb{Z}_2$-parity to
\be
\label{BFZ2-evenfields}
e_m^{\;a}~~~~~
e_{5 \hat{5}}~~~~~
B_5~~~~~
\psi_m^1~~~~~
\psi_5^2~~~~~
\eta^1~,
\ee
and odd $\mathbb{Z}_2$-parity to
\be
\label{BFZ2-oddfields}
e_5^{\;a}~~~~~
e_{m \hat{5}}~~~~~
B_m~~~~~
\psi_m^2~~~~~
\psi_5^1~~~~~
\eta^2~.
\ee
From a 4D point of view, the physical spectrum contains one massless
$N=1$ gravitational multiplet, with spins $(2,3/2)$, built from the
zero modes of $e_m^{\;a}$ and $\psi_m^1$; one massless $N=1$ chiral
multiplet, with spins $(1/2,0)$, composed of the zero modes of
$\psi_5^2$, $e_{5 \hat{5}}$ and $B_5$; and an infinite series of
massive multiplets of $N=2$ supergravity, with spins $(2,3/2,3/2,1)$
and squared masses
\be
\label{BFZ2-kkmass}
M_n^2 =  {n^2 \over R^2}~,~~~~~
(n=1,2,\ldots)~.
\ee
The KK tower gains mass through an infinite series of Higgs and
super-Higgs effects, each occurring at its own mass level. The KK
gravitons and graviphotons gain mass by eating the Fourier modes of
the fields $g_{m5}$, $g_{55}$ and $B_5$, while the massive gravitinos
eat the Fourier modes of the field $\Psi_5$. This is consistent with
the fact that the parameter of 5D SUSY, $\eta(x^M)$, has an infinite
number of Fourier modes.  Each of the modes generates a SUSY; in the
absence of matter, all but one are spontaneously broken. The broken
SUSYs implement an infinite series of super-Higgs effects for the
massive gravitinos.  The remaining SUSY is the $N=1$ of the 4D
low-energy effective action.

We now introduce the brane action. Since we are not interested in the
brane dynamics, we imagine that the brane fields are integrated out,
leaving a constant superpotential vev on each brane. The action is:
\be
\label{BFZ2-sbrane}
S_{brane} = {\kappa^2 \over 2} \,
\int d^4 x \int_{- \pi \kappa}^{+ \pi \kappa}
dy \;  e_4 \left[ \delta(y)
\ov{P_0} + \delta(y - \pi \kappa) \ov{P_{\pi}}  \right]
\, \psi_a^1 \sigma^{ab} \psi_b^1 +
{\rm \; h.c.}~,
\ee
where $P_0$ and $P_{\pi}$ are complex constants with the dimension of
$(mass)^3$ which parametrize the vevs of the superpotentials. With a
simple modification of the bulk SUSY transformations the whole action
$S_{bulk}+S_{brane}$ is invariant (for details see ref.~\cite{bfz2}).

The presence of these superpotential vevs induces SUSY breaking. In
section 4 of ref.~\cite{bfz2} the symmetry breaking mechanism as well
as the super-Higgs effect are studied in details. Here we report only
the fundamental results.  When $\ov{P_0}\ne-\ov{P_{\pi}}$ SUSY is
broken spontaneously: the fields $\psi^1_{5,\rho}(x^{\mu})$,
$\psi^2_{5,0}(x^{\mu})$ and $\psi^2_{5,\rho}(x^{\mu})$ ($\rho>0$ is
the KK index) are the goldstinos absorbed by the gravitinos
$\psi^1_{m,0}(x^{\mu})$, $\psi^1_{m,\rho}(x^{\mu})$ and
$\psi^2_{m,\rho}(x^{\mu})$. If we move to the unitary gauge the
goldstinos are eliminated from the lagrangian. On the contrary, when
$\ov{P_0}=-\ov{P_{\pi}}$ SUSY is preserved and we are left with a
massless gravitino that indicates that $N=1$ SUSY is left
unbroken. The gravitino mass spectrum is the following:
\be
\label{BFZ2-spectrum}
{\cal M}_{3/2}^{(\rho)} = {\rho \over R} + 
{\delta_0 + \delta_\pi \over 2 \pi R}~, ~~~~~~~~
(\rho\in \mathbb{Z})~,
\ee 
where
\be
\label{BFZ2-deltas}
\delta_{0 \, (\pi)} = 2 \,
\arctan {\kappa^3 P_{0\,(\pi)} \over 2} \, .
\ee
We can directly observe that when $\ov{P_0}\ne-\ov{P_{\pi}}$ the
gravitino masses are shifted with respect to their SUSY partners and
the lightest gravitino has a non-vanishing mass. These facts show that
SUSY is indeed spontaneously broken.

In previous paragraphs we showed that localized superpotential vevs
can induce SUSY breaking. Here we would like to reproduce the same
results by means of a SS twist, working in the opposite direction with
respect to what done in section~\ref{gauge} for the Kawamura model.

We start from the lagrangian~(\ref{BFZ2-lbulk}) for smooth fields
$\Psi^c$. Since it is invariant under the transformations of a global
$SU(2)_R$ symmetry, we can twist the periodicity conditions for the
gravitino (now written as in notation (B) of
eq.~(\ref{ferm-notations})):
\be
\Phi^c_M(y+2 \pi\kappa)
=
{e^{\dd - i \beta^c \widehat{\sigma}^2}} 
\Phi^c_M(y)~.
\label{BFZ2-twist}
\ee
The spectrum corresponding to these boundary conditions turns out to be:
\be
\label{BFZ2-spectwist}
{\cal M}_{3/2}^{(\rho)} = {\rho \over R} - {\beta^c \over 2 \pi R}~,~~~~~~~~
(\rho\in \mathbb{Z})~,
\ee
that is the classical SS spectrum. We observe that if we choose
$\beta^c=-\delta_0-\delta_{\pi}$ it coincides with the gravitino
spectrum of eq.~(\ref{BFZ2-spectrum}). Are the two descriptions really
equivalent? To prove this we perform the following field redefinition:
\be
\label{BFZ2-redef}
\Phi^c_M (y) = e^{\dd - i \alpha(y) \widehat{\sigma}^2}
\Phi_M (y)~, 
\ee
where $\alpha(y)$ is the step function defined in eq.~(\ref{alpha}).
It is possible to check that the fields $\Phi_M(y)$ obey the
following jump conditions at orbifold fixed points:
\be
\label{BFZ2-jumpcond}
\Phi_M (+\xi) =  e^{\dd i \delta_0 \widehat{\sigma}^2}
\Phi_M (- \xi)~,~~~~~~~~
\Phi_M (\pi \kappa + \xi) =  
e^{\dd i \delta_\pi \widehat{\sigma}^2}
\Phi_M (\pi \kappa - \xi)~.
\ee
They also have a twist $\beta = \beta^c + \delta_0 + \delta_\pi$. If
we choose once again $\beta^c = - \delta_0 - \delta_{\pi}$ they become
periodic, as in the framework of brane-induced SUSY breaking. 

We perform now the field redefinition~(\ref{BFZ2-redef}) in the bulk
action and we obtain:
\be
\label{BFZ2-finalaction}
S_{bulk}(\Psi^c,\partial\Psi^c) = 
S_{bulk}(\Psi,\partial\Psi) + S_{brane}(\Psi,\partial\Psi)
\ee
where $S_{brane}(\Psi,\partial\Psi)$ coincides with
eq.~(\ref{BFZ2-sbrane}).  Starting from a bulk action for continuous
and twisted fields we arrived, simply performing a local field
redefinition, to a bulk-plus-brane action for periodic fields. The
localized terms can be interpreted as superpotential vevs, remnants of
some brane dynamics, which are responsible of SUSY breaking. Since the
two systems are equivalent, we can assert that brane-induced SUSY
breaking is equivalent to SS SUSY breaking.

In the previous discussion we have performed a particular field
redefinition, in order to prove the equivalence of the SS
compactification to the mechanism of ref.~\cite{bfz2}. In
section~\ref{bfwz} we showed that there is an infinity of equivalent
lagrangians, corresponding to the general field redefinition
\be
\label{2.5-redef}
\Psi^c_M (y)  =  V(y) \, \widetilde{\Psi}_M(y)~,
\ee
where $\widetilde{\Psi}_M(y)$ are periodic fields and $V(y)$ satisfies
the conditions of eqs.~(\ref{reqall}). If we perform this field
redefinition in the bulk lagrangian of eq.~(\ref{BFZ2-lbulk}) we
obtain:
\bea
\label{2.5-result}
\cL(\Psi_M,\partial\Psi_M) & = &
\cL(\widetilde{\Psi}_M,\partial\widetilde{\Psi}_M)
+ \left\{
-\frac{i}{2}\left[m_1(y)+i \, m_2(y)\right]
\widetilde{\psi}_{1m} \sigma^{mn} \widetilde{\psi}_{1n}
\right. \\ 
& + & \left.
\frac{i}{2}\left[m_1(y)-i \, m_2(y)\right]
\widetilde{\psi}_{2m} \sigma^{mn} \widetilde{\psi}_{2n}
+ i \, m_3(y) \, \widetilde{\psi}_{1m} \sigma_{mn}
\widetilde{\psi}_{2n} + {\rm h.c.} \right\}~,\nn
\eea
with $m_i$ defined in section~\ref{bfwz}. These infinite lagrangians
describe all the same physics of the brane-induced SUSY breaking
scheme of ref.~\cite{bfz2}.

\clearemptydoublepage


\chapter[Flavour Symmetry Breaking]{Flavour Symmetry Breaking} 
\label{chap3}


\section[Flavour Physics in Four and More Dimensions]{Flavour Physics in Four and More Dimensions}
\label{3.1}

Since its foundation in the 1960's, the SM had passed all experimental
tests and had been confirmed even in precision measurements, first in
the sector of gauge interactions and more recently in the quark
sector. Recent experiments on neutrino physics have clarified the
picture in the lepton sector, suggesting that neutrinos should be
massive and that a mixing should exist also for leptons. In spite of
its success, also guided form hints coming from neutrino physics, many
physicists believe that the SM does not represent the final theory,
but rather that it has to be considered as an effective theory
originating from a more fundamental one. For instance there is a
little understanding, within the SM, about the intrinsic physics of
ESB, the hierarchy of charged fermion mass spectra, the smallness of
neutrino masses and the origin of flavour mixing and CP
violation. Moreover fermion replica are introduced on the base of the
experience, but the theory is not able to explain why there are
precisely three families.

The investigation of fermion masses and flavour mixing problems can be
traced back to the early 1970's. Since then many approaches have been
developed, but our understanding of flavour physics still remains
unsatisfactory: in many cases the problem seems to be transferred from
one place to another, instead of being solved.  The main idea
developed in most of the models is that of a flavour symmetry holding
at some energy scale. Certainly at our energies this symmetry must be
broken, but we do not know at which scale it is restored. If at this
scale a conventional 4D picture still holds, we can analyze the
flavour problem in the context of a local quantum field theory in four
space-time dimensions.  Here the most powerful tool that we have to
decipher the observed hierarchy among the different masses and mixing
angles is that of spontaneously broken flavour symmetries~\cite{fn}.
In the idealized limit of exact symmetry, only the heaviest fermions
are massive: the top quark and, maybe, the whole third family. The
lightest fermions and the small mixing angles originate from breaking
effects. This beautiful idea has been widely explored in many possible
versions, with discrete or continuous symmetries, global or local
ones.  A realistic description of fermion masses in this framework
typically requires either a large number of parameters or a high
degree of complexity and we are probably unable to select the best
model among the many existing ones. Moreover, in 4D we have little
hopes to understand why there are exactly three generations.

It might be the case that at the energy scale characterizing flavour
physics a 4D description breaks down.  For instance this happens in
superstring theories where the space-time is ten or
eleven-dimensional.  In the 10D heterotic string six dimensions can be
compactified on a Calabi-Yau manifold~\cite{can} or on orbifolds and
the flavour properties are strictly related to the features of the
compact space. In Calabi-Yau compactifications the number of chiral
generations is proportional to the Euler characteristics of the
manifold. In orbifold compactifications, matter in the twisted sector
is localized around the orbifold fixed points and the Yukawa
couplings, arising from world-sheet instantons, have a natural
geometrical interpretation~\cite{orb}.  Recently string realizations
where the light matter fields of the SM arises from intersecting
branes have been proposed.  Also in this context the flavour dynamics
is controlled by topological properties of the geometrical
construction~\cite{int}, having no counterpart in 4D field theories.

Perhaps in the future the flavour mystery will be unraveled by string
theory, but in the meantime it would be interesting to explore, in a
pure field theoretical construction, the new possibilities offered by
extra space-like dimensions.  For instance in orbifold
compactifications light 4D fermions may be either localized at the
orbifold fixed points or they may arise as zero modes of
higher-dimensional spinors, with a wave function suppressed by the
square root of the volume of the compact space. This led to several
interesting proposals.  It has been suggested that the smallness of
neutrino masses could be reproduced if the left-handed active
neutrinos sit at a fixed point and the right-handed sterile partners
live in the bulk of a large fifth dimension~\cite{nuex}.  In 5D GUTs
the heaviness of the third generation can be explained by localizing
the corresponding fields on a fixed point, whereas the relative
lightness of the first two generations as well as the breaking of the
unwanted mass relations can be obtained by using bulk
fields~\cite{hn2,gutex}.  Unfortunately in most models fermions are
put ``by hands'' in the bulk or on the branes, while we would like to
have a criterion suggesting their location.

A dynamical localization of chiral fermions is possible when a
higher-dimensional spinor interacts with a non-trivial background of
solitonic type.  It has been known for a long time that this provides
a mechanism to obtain massless 4D chiral fermions~\cite{sol}.  As an
example we can consider a theory with one infinite extra dimension and
we can couple the fermion field $\psi$ to a scalar background
$\phi$. The lagrangian of the system is:
\be
\cL = i\bar{\psi}\Gamma^A\partial_A\psi + g \phi \bar{\psi} \psi~,
\ee
where $\psi = (\psi_L~\psi_R)^T$, with $\psi_{L,R}$ 4D chiral fermions
depending also on $x_5$, $\phi$ is an $x_5$-dependent scalar field and
$A=\mu,5$. The 4D zero modes are the solution of
\be
i\Gamma^5\partial_5\psi + g \phi \psi = 0
\ee
and they are precisely:
\be
\psi_{L,R}(x,x_5) \propto e^{\dd{\mp g \int_{x_5^0}^{x_5} du~\phi (u)}} \psi_{L,R}(x)~. 
\label{pippo} 
\ee
We immediately observe that $\phi$ cannot be a constant since in a non
compact space $\psi_{L,R}$ would not be normalizable. On the contrary
if $\phi$ is soliton-like we can obtain one normalizable chiral zero
mode. For example if $\phi (x_5)=\epsilon (x_5-x_5^0)$, where
$\epsilon (x)$ is the sign function, and $g>0~(g<0)$, only the
left(right) mode will survive and moreover it will be localized around
$x_5^0$ that is the core of the topological defect:
\be
\psi_{L,R}(x,x_5) \propto e^{\dd{\mp g |x_5 - x_5^0|}}~ \psi_{L,R}(x)~.
\ee 
Coupling a higher-dimensional spinor to a solitonic background is then
a method to get chiral fermions without introducing any orbifold
compactification. Moreover fermions are localized and we can decide
the localization point simply by adding a mass term to the lagrangian:
if we add $M\bar{\psi}\psi$ the fermion will be concentrated around
the point where $\phi (x_5) = -M$.

\begin{figure}[!t]
\begin{center}
\epsfig{figure=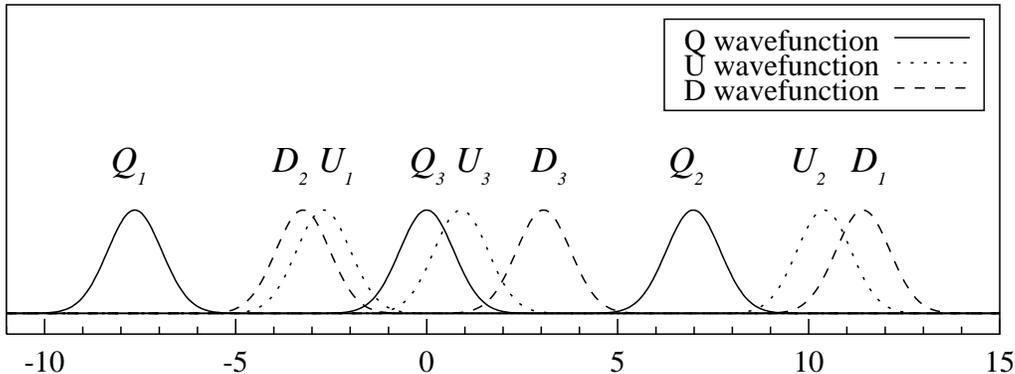,width=0.9\textwidth}
\end{center}
\vspace{-0.5cm}
\caption{Locations and profiles of quark wave functions in the fifth
dimension. $Q_i$ are the quark doublets, $D_i$ are $d_R$, $s_R$, $b_R$
and $U_i$ are $u_R$, $c_R$, $t_R$.}
\label{ms}
\end{figure}

The dynamical localization of zero modes can be used to explain the
observed hierarchy in the fermion spectrum and the intergenerational
mixing. This has been suggested first in ref.~\cite{ark} and then has
been extensively used in many following
proposals~\cite{mir-schm,dv-shif,ark2}.  In this kind of models mass
terms arise from the overlap among fermion and Higgs wave
functions. If we consider the simplest case of constant Higgs vev, an
ad hoc localization of fermions in the extra dimensions such as the
one proposed in ref.~\cite{mir-schm} (see fig.~\ref{ms}) can lead to
the observed mass spectrum.
\begin{figure}
\begin{center}
\epsfig{figure=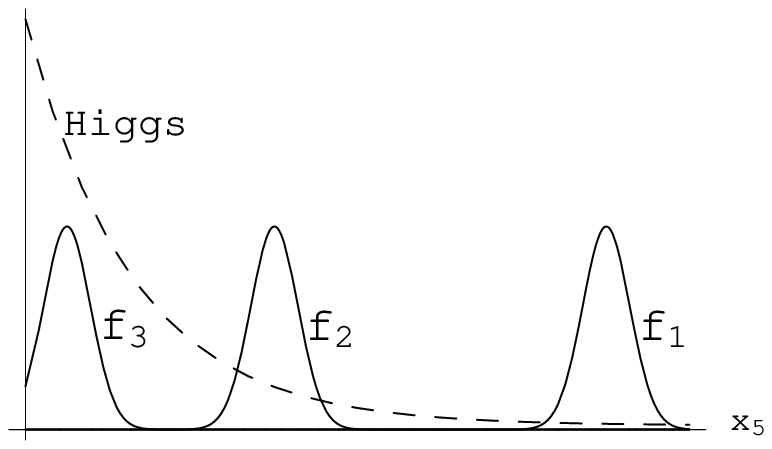,width=0.6\textwidth}
\end{center}
\vspace{-0.5cm}
\caption{Locations and profiles of fermion wave functions in the fifth
dimension with an exponentially decaying Higgs; here only one fermion
per family is displayed.}
\label{ds}
\end{figure}
In fact we see that the left and right wave functions of heavy quarks
have a greater overlap than the ones of light quarks, the latter being
located farer from each others in the extra dimension.  If we relax
the hypothesis of constant Higgs we can also obtain the correct mass
spectrum with only three localization points, one per generation.  For
instance this can be achieved with an exponentially decaying Higgs
(see fig.~\ref{ds})~\cite{dv-shif}.  In both cases there is an
exponential mapping between the parameters of the higher-dimensional
theory and the 4D masses and mixing angles, so that even with
parameters of order one large hierarchies are created. The number of
parameters we need is not lowered, but now they are all of the same
magnitude.

In orbifold compactifications, solitons are simulated by scalar fields
with a non-trivial parity assignment that forbids constant
non-vanishing vevs.  Under certain conditions, the energy is minimized
by field configurations with a non-trivial dependence upon the compact
coordinates~\cite{ggh}. As an example we consider the orbifold
$M^4\times S^1/\mathbb{Z}_2$ and we choose $\psi_L(\psi_R)$ even(odd)
under $\mathbb{Z}_2$ and $\phi$ odd, from which it follows that $\int
du\,\phi(u)$ is even. If then we calculate the zero modes we see from
eq.~(\ref{pippo}) that only $\psi_L$ is acceptable: $\psi_R$ would be
even, contrary to the initial assignment. In particular if we choose
$\phi(x_5)=\epsilon(x_5)$, where $\epsilon(x_5)$ is now the periodic
sign function, we obtain:
\be
\psi_L(x,x_5) \propto e^{\dd{-g|x_5|_{per}}}~\psi_L(x)~.  
\ee 
$\psi_L$ is localized at $x_5=0$ or at $x_5=\pi R$ depending on the
sign of $g$. Also in this case we have obtained a chiral localized
zero mode, but while with an infinite extra dimension it was the
normalization making the theory chiral, here it is a consequence of
orbifold projection.

In models like the ones described above, from one higher-dimensional
spinor we get one 4D chiral zero mode. Is it possible to obtain as
many chiral fermions as we want starting from one single
higher-dimensional spinor?  The answer to this question was given
already in the 80's (see refs.~\cite{sol}) when it had been shown that
coupling a fermion to a topologically non-trivial background could
produce a number of zero modes related to the topology of the
background itself.  Recently these results have been re-derived in a
more phenomenological context. For instance, in the model studied in
ref.~\cite{russi}, the authors consider a topological defect in two
infinite extra dimensions whose core corresponds to our 4D
world. Chiral fermion zero modes are trapped in the core by specific
interactions with the vortex of winding number $k$ which builds up the
defect. In this case an index theorem guarantees that the number of
chiral zero modes is equal to the topological charge $k$. Starting
from one single 6D spinor and choosing $k=3$ they obtain three 4D
fermion species with identical gauge and global quantum number, which
differ among themselves only in the dependence on the extra
coordinates. If we trap in the core the Higgs field by coupling it to
the vortex, his overlap with different fermion functions generates the
observed hierarchy of masses. To obtain the correct mixing angles the
model has to be complicated a little by introducing another scalar
field which, combined with the first one, builds up a new
defect. Although this model is not very simple, it gives an answer
both to the flavour problem and to the question of the number of
families.  Recently it has been extended to the case of two extra
dimensions compactified on a sphere~\cite{russi2}. A compact extra
space has the advantage of localizing the gauge fields in a finite
region.

Among the previously discussed models only the latter try to give an
explanation of the origin of fermion replica. Then one may think that
only with the introduction of a topological defect it is possible to
address both the issues. This is not true. It is well known in fact
that a spinor in $d$ dimensions have $2^{[d/2]}$ complex components,
that means it can contain many 4D replica. In 5D a fermion contains
two 4D components with opposite chirality; after projecting out the
wrong chirality, for example with an orbifold projection, we are left
with a single generation. One extra dimension is then not enough to
our purpose. In 6D fermions can be chiral and a 6D chiral fermion has
the same content of a 5D fermion. If we consider a 6D vector-like
fermion and we project out through orbifolding the unwanted chirality
we are left with two replica with the same chirality and the same
quantum numbers. In order to have a realistic model we need three
families, so also two extra dimensions are not enough. But in order to
understand if this kind of models can account both for the mass
hierarchy and the flavour mixing and how this can work, we can begin
to explore this simplified case. This has been done in
ref.~\cite{bfmpv} where the authors proposed a 6D toy model that we
describe in the following section.


\section[A Toy Model for Two Generations]{A Toy Model for Two Generations}
\label{3.2}

\subsection{Localized Fermions}
\label{3.2.1}

\begin{figure}[!b]
\centerline{
\psfig{figure=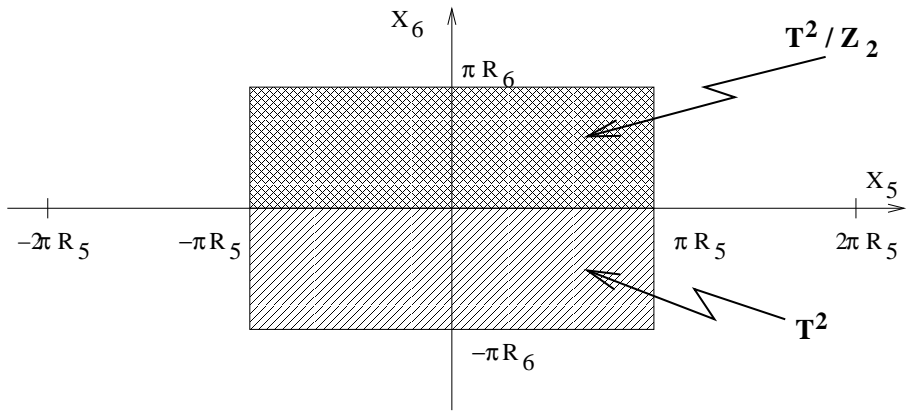,width=0.8\textwidth}}
\caption{Fundamental region of the orbifold $T^2/\mathbb{Z}_2$.}
\label{orb}
\end{figure}

We consider two extra spatial dimensions compactified on the orbifold
$T^2/\mathbb{Z}_2$, where $T^2$ is the torus defined by $x_i \to x_i+2
\pi R_i$ $(i=5,6)$ and $\mathbb{Z}_2$ is the parity symmetry
$(x_5,x_6)\to (-x_5,-x_6)$. As fundamental region of the orbifold we
can take, for instance, the rectangle $(\vert x_5 \vert \le \pi R_5)$,
$(0\le x_6 \le \pi R_6)$ (see fig.~\ref{orb}).
There are four inequivalent fixed points under $\mathbb{Z}_2$. In the
chosen fundamental region they can be identified with
$(x_5,x_6)=(0,0)$, $(\pi R_5,0)$, $(0,\pi R_6)$, $(\pi R_5,\pi R_6)$.

Since we want to reproduce the SM, we ask invariance under the gauge
group SU(3)$\otimes$SU(2)$\otimes$U(1). As discussed in the previous
section, we want to end with two fermion replica, so that we need to
start with vector-like fermions. To justify this choice we also ask
invariance under 6D parity.  As a consequence, the lagrangian has 6D
vector-like fermions $\Psi^{(\alpha)}$ $(\alpha=1,...5)$, one for each
irreducible representation of the SM, as summarized in
table~\ref{tab-bfmpv}.  With this set of fermion fields, our model is
automatically free from 6D gauge anomalies.  As we will see later on,
requiring exact 6D parity symmetry is too strong an assumption to
obtain a `realistic' fermion spectrum. Although eventually we will
relax this assumption, for the time being we carry on our construction
by enforcing 6D parity invariance.
%
\begin{table}[!t]
\begin{center}
\begin{tabular}{|c|c|c|c|}   
\hline
& & & \\                         
{\tt field} & SU(3) & SU(2) & U(1) \\ 
& & & \\
\hline
& & & \\                         
$\Psi^{(1)}$& 3& 2& +1/6\\ 
& & & \\
\hline
& & & \\                         
$\Psi^{(2)}$& 3& 1& +2/3\\ 
& & & \\
\hline
& & & \\                         
$\Psi^{(3)}$& 3& 1& -1/3\\ 
& & & \\
\hline
& & & \\                         
$\Psi^{(4)}$& 1& 2& -1/2\\ 
& & & \\
\hline
& & & \\                         
$\Psi^{(5)}$& 1& 1& -1\\ 
& & & \\
\hline
\end{tabular}
\caption{Vector-like 6D fermions and their gauge quantum numbers.}
\label{tab-bfmpv}
\end{center}
\end{table}
We have:
\be
{\cal L}_g={\cal L}_{gauge}+i\sum_{\alpha=1}^5 \overline{\Psi^{(\alpha)}}
\Gamma^M D_M \Psi^{(\alpha)}~,
\label{lg}
\ee
where ${\cal L}_{gauge}$ stands for the 6D kinetic term for the gauge
vector bosons $A_M$ $(M=\mu ;i =0,...3;5,6)$ of
SU(3)$\otimes$SU(2)$\otimes$U(1), $\Gamma^M$ are the 6D gamma matrices
and $D_M \Psi^{(\alpha)}$ denotes the appropriate fermion covariant
derivative. We recall that, up to the $(x_5,x_6)$ dependence, a 6D
vector-like spinor is equivalent to a pair of 4D Dirac spinors:
$\Psi=(\eta,\chi)^T$. Moreover each 6D fermion can be split into two
chiralities $\Psi=\Psi_{+}+\Psi_{-}$, eigenstates of $\Gamma_7$:
$\Psi_{\pm}=(1\pm\Gamma_7)/2~~ \Psi$.  We choose a representation for
the Dirac matrices in 6D where $\Gamma_7=\gamma_5\otimes\sigma_3$ (see
appendix~\ref{gamma-mat}), where $\sigma_3$ is the third Pauli matrix,
so that in terms of 4D chiralities we have:
$\Psi_{+}=(\eta_R,\chi_L)^T$ and $\Psi_{-}=(\eta_L,\chi_R)^T$. Each
component $\eta_{L,R}$, $\chi_{L,R}$ transforms in the same way under
the gauge group.  All fields are assumed to be periodic in $x_5$ and
$x_6$. By inspecting the kinetic terms, we see that consistency with
the orbifold projection requires a non-trivial assignment of the
$\mathbb{Z}_2$ parity. We take $A_\mu$ even under $\mathbb{Z}_2$ and
$A_i$ $\mathbb{Z}_2$-odd.  In the fermion sector, $\eta_{R(L)}$ and
$\chi_{R(L)}$ should have the same $\mathbb{Z}_2$ parity, which should
be opposite for $\eta_{R(L)}$ and $\chi_{L(R)}$. We choose
$\mathbb{Z}_2(\eta^{(\alpha)}_R,\chi^{(\alpha)}_L,\eta^{(\alpha)}_L,\chi^{(\alpha)}_R)$
equal to $(-1,+1,+1,-1)$ for $\alpha=1,4$, and $(+1,-1,-1,+1)$ for
$\alpha=2,3,5$.  At this level the zero modes are the gauge vector
bosons of the SM and two independent chiral fermions for each
irreducible representation of the SM, describing two massless
generations.  There are no gauge anomalies in our model.  Bulk
anomalies are absent because the 6D fermions are vector-like.  There
could be gauge 4D anomalies localized at the four orbifold fixed
points~\cite{orban,asa}. In our model based on $T^2/\mathbb{Z}_2$, the
anomalies are the same at each fixed point and they actually vanish
with the quantum number assignments of
table~\ref{tab-bfmpv}\footnote{We have explicitly checked this by
adapting the analysis described in ref.~\cite{asa}.}. Indeed they are
proportional to the anomalies of the 4D zero modes, which form two
complete fermion generations, thus providing full 4D anomaly
cancellation.  Fermion masses in 6D and Yukawa couplings do not modify
this conclusion.

In the absence of additional interactions, each zero mode is constant
with respect to $x_5$ and $x_6$. Even by introducing a 6D (parity
invariant) Yukawa interaction between fermions and a Higgs electroweak
doublet, we do not break the 4D flavor symmetry, which is maximal.
The first step to distinguish the two fermion generations is to
localize them in different regions of the compact space.  In our model
this can be done in a very simple way, by introducing a 6D fermion
mass term
\bea
{\cal L}_m&=&\sum_{\alpha=1}^5 m_{(\alpha)} \overline{\Psi^{(\alpha)}}
\Psi^{(\alpha)}\nn\\
&=&\sum_{\alpha=1}^5 m_{(\alpha)} \overline{\Psi^{(\alpha)}}
\left(\frac{1-\Gamma_7}{2}\right)\Psi^{(\alpha)}+h.c.~,
\label{lm}
\eea
where 6D parity requires $m_{(\alpha)}$ to be real.  This term is
gauge invariant and relates left and right 4D chiralities. Therefore
the mass parameters $m_{(\alpha)}$ are required to be
$\mathbb{Z}_2$-odd and cannot be constant in the whole $(x_5,x_6)$
plane.  The simplest possible choice for $m_{(\alpha)}$ is a constant
in the orbifold fundamental region\footnote{Of course there is not a
unique way of choosing the fundamental region and this leads to
several possible choices for $m_{(\alpha)}$. Although we are now
regarding $\mu_{(\alpha)}$ as real parameters, in the next section we
will also need results for complex $\mu_{(\alpha)}$. For this reason
we carry out our analysis directly in the complex case.}:
\be
m_{(\alpha)}(x_5,x_6)=\mu_{(\alpha)} \epsilon(x_6)~,
\ee 
where $\epsilon(x_6)$ denotes the (periodic) sign function.  This
function can be regarded as a background field.  In a more fundamental
theory it could arise dynamically from the vev of a gauge singlet
scalar field, periodic and $\mathbb{Z}_2$-odd~\cite{ggh}.  Then the parameters
$\mu_{(\alpha)}$ would essentially represent Yukawa couplings.  In our
toy model we regard $\epsilon(x_6)$ as an external fixed background
and neglect its dynamics.

The properties of the 4D light fermions are now described by the zero
modes of the 4D Dirac operator in the background proportional to
$\epsilon(x_6)$. These zero modes are the normalized solutions to the
differential equations:
\bea
(\partial_5+i\partial_6) \chi^{(\alpha)}_L+ \mu_{(\alpha)} \epsilon(x_6) 
\eta^{(\alpha)}_L &=&0\nn\\
(\partial_5-i\partial_6) \eta^{(\alpha)}_L+ \mu_{(\alpha)}^*\epsilon(x_6) 
\chi^{(\alpha)}_L &=&0\nn\\
-(\partial_5+i\partial_6) \chi^{(\alpha)}_R+\mu_{(\alpha)}^*\epsilon(x_6) 
\eta^{(\alpha)}_R &=&0\nn\\
-(\partial_5-i\partial_6) \eta^{(\alpha)}_R+ \mu_{(\alpha)}\epsilon(x_6) 
\chi^{(\alpha)}_R &=&0~,
\label{fstorder}
\eea 
with periodic boundary conditions for all fields and with the $\mathbb{Z}_2$
parities defined above. Combining eqs.~(\ref{fstorder}) together we
obtain the following second order partial differential equations,
holding in the whole $(x_5,x_6)$ plane:
\bea (\partial_5^2 +
\partial_6^2) \ \chi_L^{\alpha} - |\mu_{(\alpha)}|^2 \ \epsilon^2(x_6)
\ \chi_L^{\alpha} - 2 \ i \ \mu_{(\alpha)} \ (-1)^k \ \delta_k(x_6) \
\eta_L^{\alpha} &=& 0 \nonumber \\ (\partial_5^2 + \partial_6^2) \
\eta_L^{\alpha} - |\mu_{(\alpha)}|^2 \ \epsilon^2(x_6) \
\eta_L^{\alpha} + 2 \ i \ \mu_{(\alpha)}^* \ (-1)^k \ \delta_k(x_6) \
\chi_L^{\alpha} &=& 0 \nonumber \\ (\partial_5^2 + \partial_6^2) \
\chi_R^{\alpha} - |\mu_{(\alpha)}|^2 \ \epsilon^2(x_6) \
\chi_R^{\alpha} + 2 \ i \ \mu_{(\alpha)}^* \ (-1)^k \ \delta_k(x_6) \
\eta_R^{\alpha} &=& 0 \nonumber \\ (\partial_5^2 + \partial_6^2) \
\eta_R^{\alpha} - |\mu_{(\alpha)}|^2 \ \epsilon^2(x_6) \
\eta_R^{\alpha} - 2 \ i \ \mu_{(\alpha)} \ (-1)^k \ \delta_k(x_6) \
\chi_R^{\alpha} &=& 0
\label{app-eom}
\eea
where $k$ is an integer, $\delta_k(x_6) \equiv \delta(x_6 - k \pi
R_6)$ and the sum over $k$ is understood. The advantage of working
with eqs.~(\ref{app-eom}) is that in the bulk these equations are
decoupled and identical for all fields. Away from the lines $x_6 = k
\pi R_6$, $k\in \mathbb{Z}$, they read:
\be (\partial_5^2 + \partial_6^2) \ \phi - |\mu_{(\alpha)}|^2 \ \phi =
0
\label{app-eombulk}
\ee
with appropriate boundary conditions. Here $\phi$ stands for
$\chi_L^{\alpha}$, $\eta_L^{\alpha}$, $\chi_R^{\alpha}$,
$\eta_R^{\alpha}$.  In each strip $k \pi R_6 < x_6 < (k+1) \pi R_6$,
the general solution to this equation can be written in the form:
\be \phi^{(k)}(x,x_5,x_6) = \sum_{n\in \mathbb{Z}} \ \left[ \ C^{(k)}_n(x) \
e^{\dd\alpha_n \frac{x_6}{R_5}} + C'^{(k)}_n(x) \ e^{\dd-\alpha_n
\frac{x_6}{R_5}}\ \right] \ e^{\dd i n \frac{x_5}{R_5}}
\label{app-sol}
\ee
with $\alpha_n = \sqrt{n^2 + |\mu_{\alpha}|^2 R_5^2}$.  These
solutions can be glued together by imposing periodicity along $x_6$,
$\mathbb{Z}_2$ parity and the appropriate discontinuity across the lines $x_6 =
k \pi R_6$. This last requirement can be directly derived from
eqs.~(\ref{fstorder}) and eqs.~(\ref{app-eom}).  The fields $\phi$
should be continuous everywhere, whereas their first derivatives have
discontinuities $\Delta^{(k)}(\partial_6\phi)$ given by:
\be
\begin{array}{ll}
\Delta^{(2k)}(\partial_6\phi) = -2\, i\, s(\phi ')\, \phi '(2k\pi R_6)
 \quad & \textrm{at}\quad x_6=2k\pi R_6 \\
 \Delta^{(2k+1)}(\partial_6\phi) = 2\, i\, s(\phi ')\, \phi
 '((2k+1)\pi R_6) \quad & \textrm{at}\quad x_6=(2k+1)\pi R_6
\end{array}
\label{app-jumps}
\ee
where $(\phi,\phi ') = (\chi_L^{\alpha},\eta_L^{\alpha}),\,
                       (\eta_L^{\alpha},\chi_L^{\alpha}),\,
                       (\chi_R^{\alpha},\eta_R^{\alpha}),\,
                       (\eta_R^{\alpha},\chi_R^{\alpha})$ 
and
\be
s(\phi ') = \left\{
\begin{array}{ll}
-\mu_{\alpha}\ & \textrm{if}\ \phi '=\eta_L^{\alpha},\,
\chi_R^{\alpha}\\ \mu_{\alpha}^*\ & \textrm{if}\ \phi
'=\eta_R^{\alpha},\, \chi_L^{\alpha}
\end{array}\right. ~.
\label{app-esse}
\ee
These requirements have a non-trivial solution only for $n=0$, that
means the zero modes are independent of $x_5$. We obtain:
\begin{itemize}
\item{$\alpha=1,4$}
\end{itemize}
\bea
\label{zm14}
\left(
\begin{array}{c}
\eta^{(\alpha)}_R\\
\chi^{(\alpha)}_R
\end{array}
\right)&=&0\\
\left(
\begin{array}{c}
\eta^{(\alpha)}_L\\
\chi^{(\alpha)}_L
\end{array}
\right)&=&
f^{(\alpha)}_1(x)
\left(
\begin{array}{c}
1\\
i \dd\frac{\mu_{(\alpha)}}{\vert\mu_{(\alpha)}\vert}
\end{array}
\right)
\xi^{(\alpha)}_1(x_5,x_6)
+
f^{(\alpha)}_2(x)
\left(
\begin{array}{c}
1\\
- i \dd\frac{\mu_{(\alpha)}}{\vert\mu_{(\alpha)}\vert}
\end{array}
\right)
\xi^{(\alpha)}_2(x_5,x_6)~,\nn
\eea
\begin{itemize}
\item{$\alpha=2,3,5$}
\end{itemize}
\bea
\left(
\begin{array}{c}
\eta^{(\alpha)}_R\\
\chi^{(\alpha)}_R
\end{array}
\right)&=&
f^{(\alpha)}_1(x)
\left(
\begin{array}{c}
1\\
-i \dd\frac{\mu_{(\alpha)}^{~*}}{\vert\mu_{(\alpha)}\vert}
\end{array}
\right)
\xi^{(\alpha)}_1(x_5,x_6)
+
f^{(\alpha)}_2(x)
\left(
\begin{array}{c}
1\\
i \dd\frac{\mu_{(\alpha)}^{~*}}{\vert\mu_{(\alpha)}\vert}
\end{array}
\right)
\xi^{(\alpha)}_2(x_5,x_6)\nn\\
\left(
\begin{array}{c}
\eta^{(\alpha)}_L\\
\chi^{(\alpha)}_L
\end{array}
\right)&=&0~,
\label{zm235}
\eea
where $f^{(\alpha)}_{1,2}(x)$ are 4D chiral spinors:
\be
\begin{array}{ll}
f^{(1)}_1=
\left(
\begin{array}{c}
u_L\\
d_L
\end{array}
\right)&~~~~~
f^{(1)}_2=
\left(
\begin{array}{c}
c_L\\
s_L
\end{array}
\right)\nn\\
f^{(2)}_1=u_R&~~~~~ f^{(2)}_2=c_R\nn\\
f^{(3)}_1=d_R&~~~~~ f^{(3)}_2=s_R\nn\\
f^{(4)}_1=
\left(
\begin{array}{c}
\nu_{e L}\\
e_L
\end{array}
\right)&~~~~~
f^{(4)}_2=
\left(
\begin{array}{c}
\nu_{\mu L}\\
\mu_L
\end{array}
\right)\nn\\
f^{(5)}_1=e_R&~~~~~ f^{(5)}_2=\mu_R
\end{array}~,
\ee
whereas $\xi^{(\alpha)}_{1,2}(x_5,x_6)$ are functions describing the
localization of the zero modes in the compact space:
\bea
\xi^{(\alpha)}_1(x_5,x_6)&=&
\frac{e^{\dd{-\pi \vert\mu_{(\alpha)}\vert R_6}}}{\sqrt{2\pi R_5}}~
\frac{\sqrt{\vert\mu_{(\alpha)}\vert}}{\sqrt{1-e^{\dd{-2\pi\vert\mu_{(\alpha)}\vert R_6}}}}~~
e^{~\dd{\vert \mu_{(\alpha)} x_6\vert_{per}}}\nn\\
\xi^{(\alpha)}_2(x_5,x_6)&=&
\frac{1}{\sqrt{2\pi R_5}}~
\frac{\sqrt{\vert\mu_{(\alpha)}\vert}}{\sqrt{1-e^{\dd{-2\pi\vert\mu_{(\alpha)}\vert R_6}}}}~~
e^{\dd{-\vert \mu_{(\alpha)} 
x_6\vert_{per}}}~.
\label{xi}
\eea
In the above equations $\vert x_6\vert_{per}$ denotes a periodic
function, coinciding with the ordinary $\vert x_6\vert$ in the
interval $[-\pi R_6,\pi R_6]$. As in the case $m_{(\alpha)}=0$, for
each 6D spinor we have two independent chiral zero modes, whose 4D
dependence is described by $f^{(\alpha)}_{1,2}$.  They are still
constant in $x_5$, but not in $x_6$.
\begin{figure}[!h]
\centerline{
\psfig{figure=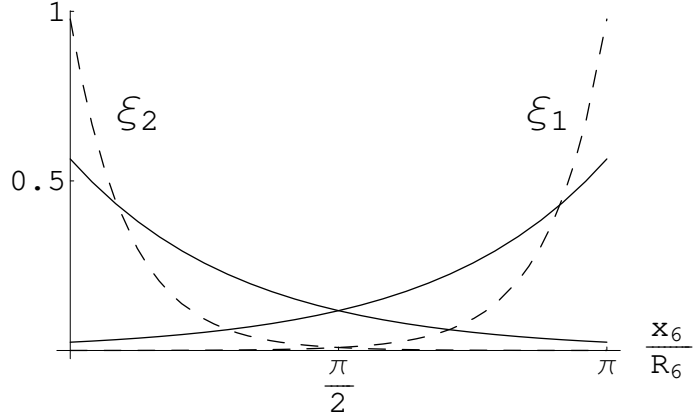,width=0.7\textwidth}}
\caption{Wave functions $\xi^{(\alpha)}_{1,2}(x_5,x_6)$, 
in units of $\sqrt{R_5 R_6}$. 
They have been obtained by choosing $\vert\mu_{(\alpha)}\vert R_6=1(3)$
for continuous(dashed) lines.}
\label{zeromodes}
\end{figure}
Indeed, the zero mode proportional to $f^{(\alpha)}_2$ is localized at
$x_6=0$ (mod $2\pi R_6$), whereas that proportional to
$f^{(\alpha)}_1$ is peaked around $x_6=\pi R_6$ (mod $2\pi R_6$) (see
fig.~\ref{zeromodes}).  The two zero modes with well-defined
localization properties in the compact space have non-trivial
components both along $\eta$ and along $\chi$ and they are orthogonal
to each other. The constant factors in eqs.~(\ref{xi}) normalize the
zero modes to 1. In our toy model, the number of zero modes is not
related to a non-trivial topological property of the background
$\epsilon(x_6)$. The two zero modes are determined by the orbifold
projection. The presence of the background only induces a separation
of the corresponding wave functions in the compact space. Actually we
can go smoothly from localized to constant wave functions, by turning
off the constants $\mu_{(\alpha)}$, as apparent from eqs.~(\ref{xi}).

With the introduction of the background, we now have two fermion
generations, one sat at $x_6=0$ and the other at $x_6=\pi R_6$. From
the point of view of the 4D observer, who cannot resolve distances in
the extra space, there is still a maximal flavour symmetry and,
indeed, all fermions are still massless at this level.  Fermions can
acquire masses in the usual way, by breaking the electroweak symmetry
via the non-vanishing vev of a Higgs doublet $H$.  If such a vev were
a constant in $x_6$, then we would obtain equal masses for the two
fermion generations.  Thus, to break the 4D flavour symmetry we need a
non-trivial dependence of the Higgs vev upon $x_6$. There are several
ways to achieve this. For instance, we might assume that $H$ is a bulk
field. Under certain conditions it may happen that the minimum of the
energy is no longer $x_6$-constant.  Examples of this kind are
well-known in the literature~\cite{ncvev}.  If $H$ interacts with a
suitable $x_6$-dependent background, there is a competition between
the kinetic energy term, which prefers constant configurations, and
the potential energy term, which may favour a $x_6$-varying vev. In
non-vanishing portions of the parameter space the minimum of the
energy can depend non-trivially on $x_6$.  In the minimal version of
our toy model we will simulate this dependence in the simplest
possible way, by introducing a Higgs doublet $H$ with hypercharge +1/2
localized along the line $x_6=0$\footnote{Alternatively, we could
assume that $H$ is localized at the orbifold fix point
$(x_5,x_6)=(0,0)$. From the point of view of fermion masses and mixing
angles, the two choices are equivalent. To avoid singular terms in the
action, we could also consider a mild localization, described by some
smooth limit of the Dirac delta functions involved in the present
treatment. Our results would not be qualitatively affected.}.  The
most general Yukawa interaction term invariant under $\mathbb{Z}_2$,
6D parity and SU(3) $\otimes$ SU(2) $\otimes$ U(1) reads:
\be
{\cal L}_Y=
\left[
y_u~ \tilde{H}^\dagger~
\overline{\Psi^{(2)}}\Psi^{(1)}
+
y_d~ H^\dagger~
\overline{\Psi^{(3)}}\Psi^{(1)}
+
y_e~ H^\dagger~
\overline{\Psi^{(5)}}\Psi^{(4)}
+h.c.\right]
\delta(x_6)~,
\label{lY}
\ee
where $\tilde{H}=i\sigma^2 H^*$. Notice that $H$ has dimension +3/2
and $y$ has dimension -3/2, in mass units.  In the next section we
will see how a realistic pattern of masses and mixing angles arises
from these Yukawa interactions.

Summarizing, our model is described by the lagrangian:
\be
{\cal L}={\cal L}_g+{\cal L}_m+{\cal L}_Y+{\cal L}_H~,
\ee
where ${\cal L}_g$, ${\cal L}_m$, ${\cal L}_Y$ are given in
eqs.~(\ref{lg}),~(\ref{lm}) and~(\ref{lY}), respectively, while ${\cal
L}_H$, localized at $x_6=0$, contains the kinetic term for the Higgs
doublet and the scalar potential that breaks spontaneously
SU(2)$\otimes$U(1).  The complex phases in $y_i$ can be completely
eliminated via field redefinitions: in the limit of exact 6D parity
symmetry all parameters are real.

\subsection{Masses and Mixing Angles}
\label{masses}

The fermion mass terms arise from ${\cal L}_Y$ after ESB, here
described by $\langle H\rangle=(0~ v/\sqrt{2})^T$.  To evaluate the
fermion mass matrices we should expand the 6D fermion fields in 4D
modes and then perform the $x_5$ and $x_6$ integrations. In practice,
if we focus on the lightest sector, we can keep only the zero modes in
the expansion. We obtain:
\bea
m_u&=&
\frac{y_u}{\sqrt{2}}~ v~ 
\frac{\sqrt{\vert \mu_{(1)} \mu_{(2)} \vert}}{2\sqrt{(1-\lambda_1^2)
(1-\lambda_2^2)}}
\left(
\begin{array}{cc}
c_{u-}~ \lambda_1 \lambda_2 & c_{u+}~ \lambda_2\\
c_{u+}~ \lambda_1 & c_{u-}
\end{array}
\right)\nn\\
m_d&=&
\frac{y_d}{\sqrt{2}}~ v~ 
\frac{\sqrt{\vert \mu_{(1)} \mu_{(3)} \vert}}{2\sqrt{(1-\lambda_1^2)
(1-\lambda_3^2)}}
\left(
\begin{array}{cc}
c_{d-}~ \lambda_1 \lambda_3 & c_{d+}~ \lambda_3\\
c_{d+}~ \lambda_1 & c_{d-}
\end{array}
\right)\nn\\
m_e&=&
\frac{y_e}{\sqrt{2}}~ v~ 
\frac{\sqrt{\vert \mu_{(4)} \mu_{(5)} \vert}}{2\sqrt{(1-\lambda_4^2)
(1-\lambda_5^2)}}
\left(
\begin{array}{cc}
c_{e-}~ \lambda_4 \lambda_5 & c_{e+}~ \lambda_5\\
c_{e+}~ \lambda_4 &c_{e-} 
\end{array}
\right)~,
\eea
where 
\be
c_{u\pm}=1\pm\frac{\mu_{(1)}\mu_{(2)}}{\vert
\mu_{(1)}\mu_{(2)}\vert}~,~~~~~~
c_{d\pm}=1\pm\frac{\mu_{(1)}\mu_{(3)}}{\vert
\mu_{(1)}\mu_{(3)}\vert}~,~~~~~~
c_{e\pm}=1\pm\frac{\mu_{(4)}\mu_{(5)}}{\vert
\mu_{(4)}\mu_{(5)}\vert}~,
\ee
and
\be
\lambda_\alpha=e^{\dd{-\pi\vert\mu_{(\alpha)}\vert R_6}}~.
\ee
These mass matrices, here given in the convention $\overline{f_R} m_f
f_L$, are not hermitian. It is interesting to see that, for generic
order-one values of the dimensionless combinations $c_{f\pm}$ and
$\mu_{(\alpha)} R_6$, the mass matrices display a clear hierarchical
pattern.  Fermion masses of the first generation are suppressed by
$\lambda_{(\alpha)}\lambda_{(\beta)}$ compared to those of the second
generation and mixing angles are of order $\lambda_{(\alpha)}$ or
$\lambda_{(\beta)}$. This is quite similar to what obtained in 4D
models with a spontaneously broken flavour symmetry.  Here the r\^ole
of small expansion parameters is played by the quantities
$\lambda_\alpha$.  However in our parity invariant model, the
parameters $\mu_{(\alpha)}$ are real and the coefficients $c_{f\pm}$
are `quantized'. Either $c_{f+}$ or $c_{f-}$ should vanish and this
implies no mixing.  Indeed when 6D parity is conserved, we have only
two possible orientations of the fermion zero modes in the
$(\eta,\chi)$ space: either $(1,i)$ or $(1,-i)$, as apparent from
eqs.~(\ref{zm14}) and~(\ref{zm235}). Thus the scalar product between
two zero modes in the $(\eta,\chi)$ space is either maximal or zero.
Modulo a relabeling among first and second generations, this gives
rise to a perfect alignment of mass matrices and a vanishing overall
mixing.  To overcome this problem, we should relax the assumption of
exact 6D parity symmetry\footnote{There are other possibilities that
lead to a non-vanishing mixing. For instance we could introduce
several independent backgrounds and couple them selectively to the
different fermion fields.  In our view, the solution discussed in the
text is the simplest one.}.  We will assume that 6D parity is broken
`softly', by the fermion-background interaction described by ${\cal
L}_m$.  This can be achieved by taking complex values for the mass
coefficients $\mu_{(\alpha)}$\footnote{All previous equations remain
unchanged, but the first equality in eq.~(\ref{lm}). Only the second
one is correct.}.  In a fundamental theory such a breaking could be
spontaneous: if $m_{(\alpha)}$ were complex fields, then the
lagrangian would still be invariant under 6D parity acting as
$m_{(\alpha)} \leftrightarrow m_{(\alpha)}^\dagger$.  It might occur
that the dynamics of the fields $m_{(\alpha)}$ led to complex vevs for
$m_{(\alpha)}$, thus spontaneously breaking parity. In our toy model
we will simply assume the existence of such a complex background.  All
the relations that we have derived hold true for the complex case as
well and we have now hierarchical mass matrices with a non-trivial
intergenerational mixing.  By expanding the results at leading order
in $\lambda_\alpha$ we find:
\bea
\label{mom}
m_c=\vert y_u \vert v \frac{\sqrt{\vert \mu_{(1)}\mu_{(2)}
\vert}}{2\sqrt{2}} \vert c_{u-} \vert ~~~~~~~~&
{\dd \frac{m_u}{m_c}=\frac{\vert c_{u+}^2 - c_{u-}^2 \vert}{\vert c_{u-}
\vert^2} \lambda_1 \lambda_2} \nn \\
m_s=\vert y_d \vert v \frac{\sqrt{\vert \mu_{(1)}\mu_{(3)}
\vert}}{2\sqrt{2}} \vert c_{d-} \vert ~~~~~~~~&
{\dd \frac{m_d}{m_s}=\frac{\vert c_{d+}^2 - c_{d-}^2 \vert}{\vert c_{d-}
\vert^2} \lambda_1 \lambda_3} \nn \\
m_\mu=\vert y_e \vert v \frac{\sqrt{\vert \mu_{(4)}\mu_{(5)}
\vert}}{2\sqrt{2}} \vert c_{e-} \vert ~~~~~~~~&~~
{\dd \frac{m_e}{m_\mu}=\frac{\vert c_{e+}^2 - c_{e-}^2 \vert}{\vert c_{e-}
\vert^2} \lambda_4 \lambda_5}~.
\eea
Finally, after absorbing residual phases in the definition
of the $s$ and $c$ 4D fields, the matrices $m_u^\dagger m_u$ and 
$m_d^\dagger m_d$ are diagonalized by orthogonal transformations
characterized by mixing angles $\theta_{u,d}$:
\be
\theta_{u,d}= \left\vert\frac{c_{u,d+}}{c_{u,d-}}\right\vert
\lambda_1~,
\label{ma}
\ee
still at leading order in $\lambda_\alpha$.
Therefore the Cabibbo angle is given by:
\be
\theta_C=\left(\left\vert\frac{c_{d+}}{c_{d-}}\right\vert-
\left\vert\frac{c_{u+}}{c_{u-}}\right\vert\right)\lambda_1~.
\label{ca}
\ee
Barring accidental cancellations in the relevant combinations of the
coefficients $c_{f\pm}$, the Cabibbo angle is of order $\lambda_1$.
Then, by assuming $\lambda_3\approx\lambda_1$ and
$\lambda_2\approx\lambda_1^3$ we reproduce the correct order of
magnitude of mass ratios in the quark sector.  These are small numbers
in the 4D theory, but can be obtained quite naturally from the 6D
point of view: $\mu_{(1)} R_6 \approx \mu_{(3)} R_6\approx 0.5$ and
$\mu_{(2)} R_6\approx 1.3$.  Similarly, by taking
$\lambda_4\lambda_5\approx\lambda_1^2$ we can naturally fit the lepton
mass ratio.

It can be useful to comment about the way flavour symmetry is broken
in this toy model. Before the introduction of the Yukawa interactions
and modulo U(1) anomalies, the flavour symmetry group is
U(2)$^5$. After turning the Yukawa couplings on, we can consider
several limits. When $R_6\to \infty$, the quantities
$\lambda_{(\alpha)}$ vanish and the flavour symmetry is broken down to
U(1)$^5$, acting non-trivially on the lightest sector.  If $R_6$ is
finite and non-vanishing, U(1)$^5$ is in turn completely broken down
by $\lambda_{(\alpha)}\ne 0$.  Nevertheless, contrary to what happens
in models with abelian flavour symmetries, the coefficients of order
one that multiply the symmetry breaking parameters
$\lambda_{(\alpha)}$ are now related one to each other. This can be
appreciated by taking the limit $R_6\to 0$. We have
$\lambda_{(\alpha)}=1$ and the residual flavour symmetry is a
permutation symmetry, separately for the lepton and the quark sectors:
$S_2\otimes S_2$.

Let us now briefly comment about neutrino masses and mixings in this
set-up.  The most straightforward way to produce neutrino masses is to
add a gauge singlet 6D fermion field, $\Psi^{(6)}$, with $\mathbb{Z}_2$
assignments $(+1,-1,-1,+1)$. As for the case of charged fermions, by
introducing a mass term for $\Psi^{(6)}$ as in eq.~(\ref{lm}) and a
Yukawa interaction with $\Psi^{(4)}$ and $\tilde H$ as in
eq.~(\ref{lY}), we obtain a Dirac neutrino mass term
\be   
m_\nu = \frac{y_\nu}{\sqrt{2}} v  
\frac{\sqrt{\vert\mu_{(4)} \mu_{(6)}\vert }}{2\sqrt{ (1- \lambda_4^2 ) (1- \lambda_6^2)}} 
\left( \matrix{  c_{\nu -} \lambda_4 \lambda_6 & c_{\nu +} \lambda_6 \cr 
c_{\nu +} \lambda_4 & c_{\nu -}}  \right)~.
\ee
A large mixing angle in the lepton sector, $\theta_L$, is obtained
for $\lambda_4 = O(1)$, in which case the neutrino mass hierarchy,
\be
\frac{m_{\nu_1}}{m_{\nu_2}} = \frac{|c_{\nu +}^2 - c_{\nu -}^2 |}{|c_{\nu -}|^2} \lambda_4 \lambda_6 ~,
\ee
is controlled by $\lambda_6$.  At leading order, the left mixings in
$m_e^\dagger m_e$ and $m_\nu^\dagger m_\nu$ correspond to
\be
\tan  2\theta_{e,\nu} = 
\frac{2~ c_{e,\nu +}~c_{e,\nu -}~\lambda_4}{(c_{e,\nu -}^2-c_{e,\nu +}^2~\lambda_4^2)}~,
\ee
so that $\theta_L \equiv \theta_\nu-\theta_e$ is naturally large.  As
in 4D, the smallness of these Dirac neutrino masses with respect to
the electroweak scale has to be imposed by an ad hoc suppression of
the Yukawa coupling $y_\nu$.  A natural suppression could be achieved
by considering also Majorana masses. A Majorana mass term for
right-handed neutrinos can be introduced in 6D as follows:
\be
\label{majo}
\cL_{M} = M_R\, \overline{\Psi^{(6)c}}\,\Psi^{(6)} =
          M_R\, \Psi^{(6)T}C\,\Psi^{(6)}~,
\ee
where $C$ is the 6D charge conjugation matrix defined in the second
part of appendix~\ref{gamma-mat}.  If we substitute the expression for
$\Psi$ in terms of 4D spinors we obtain:
\bea
\label{majo4D}
\cL_M &=& 2\, M_R\, \eta_R^{(6)T}\, \gamma^0\, \gamma^2\, \chi_R^{(6)} -
          2\, M_R\, \eta_L^{(6)T}\, \gamma^0\, \gamma^2\, \chi_L^{(6)} =
\nn\\[0.2cm]
      &=& 2\,i\, M_R\, \bigg(\, \overline{\eta_R^{(6)c}}\, \chi_R^{(6)} -
                       \overline{\eta_L^{(6)c}}\, \chi_L^{(6)}\, \bigg)~.
\eea
At this point we could work out the spectrum in the neutrino
sector. However we cannot proceed exactly as in the case of charged
fermions. There we calculated the spectrum in two steps: firstly we
considered only the lagrangian $\cL_g+\cL_m$ and calculated the zero
modes and secondly we introduced the Higgs sector and the Yukawa
couplings and we calculated the new masses as little perturbations of
KK levels. This was possible since the Higgs vev is assumed to be
small compared to the KK scale $1/R$ ($R\sim R_5\sim R_6$). If we now
introduce Majorana masses for neutrinos in order to realize a see-saw
mechanism, $M_R$ could be of the same order of magnitude of $1/R$ or
even larger. This implies that our two steps method is no more
justified, since we cannot neglect the Majorana mass term in the first
step of our calculation. How can we proceed?  One way could be to
integrate out right-handed neutrinos, leading to higher-dimensional
operators in the lagrangian. Starting from this new lagrangian we
could now proceed in two steps as before, finding again the mass
spectrum for all fermions. These calculations have not yet been
performed and they will be a subject for future investigation.

\subsection{Which Scale for Flavour Physics?}
\label{scale}

Our 6D toy model is non renormalizable. It is characterized by some
typical mass scale $\Lambda$. At energies larger than this typical
scale, the description offered by the model is not accurate enough and
some other theory should replace it.  Up to now we have not specified
$\Lambda$.  We could have in mind a traditional picture where
$\Lambda$ is very large, perhaps close to the 4D Planck scale, where
presumably all particle interactions, including the gravitational one,
are unified in a fundamental theory.  In this scenario we have the
usual hierarchy problem.  Clearly our simple model cannot explain why
$v<<\Lambda^{3/2}$ and we should rely on some additional mechanism to
render the ESB scale much smaller compared to $\Lambda$.  A SUSY or
warped version of our toy model could alleviate the technical aspect
of the hierarchy problem.  Alternatively, we could ask how small could
$\Lambda$ be without producing a conflict with experimental data.  For
simplicity we assume that the two radii $R_5$, $R_6$ are approximately
of the same order $R$. Due to the different dimension between 6D and
4D fields, coupling constants of the effective 4D theory are
suppressed by volume factors and we require $\Lambda R\ge1$ to work in
a weakly coupled regime.  Therefore, lower bounds on $1/R$ are also
lower bounds for $\Lambda$. Lower bounds on $1/R$ come from the search
of the first KK modes at the existing colliders or from indirect
effects induced by the additional heavy modes.  These last effects
lead to departures from the SM predictions in electroweak
observables. From the precision tests of the electroweak sector, we
get a lower bound on $1/R$ in the $TeV$ range (for a brief review on
constraints on extra dimensions see~\cite{PDG} and references
therein).  However, the most dangerous indirect effects are those
leading to violations of universality in gauge interactions and those
contributing to flavour changing processes.  Indeed, whenever we have
a source of flavour symmetry breaking, we expect a violation of
universality at some level.  In the SM such violation comes through
loop effects from the Yukawa couplings and it is tiny. In our model,
as we will see, such effects can already arise at tree level and, to
respect the experimental bounds, a sufficiently large scale $1/R$ is
needed.

Since in each fermion sector the two generations are described by two
copies of the same wave function, differing only in their localization
along $x_6$, the universality of the gauge interactions will be
guaranteed if the gauge vector bosons have a wave function perfectly
constant in $x_6$.  This is the case only for massless gauge vector
bosons, such as the photon, but, as we will see now, not necessarily
for the massive gauge vector bosons like $W$ and $Z$.  Moreover, also
the higher KK modes of all gauge bosons have non-constant
wave functions and their interactions with split fermions are in
general non-universal.

We start by discussing the interactions between the lightest fermion
generations and the observed $W$ and $Z$ vector bosons.  Consider, for
simplicity, the limit of vanishing gauge coupling $g'$ for
$U(1)$. Then the free equation of motion for the gauge bosons $W_\mu$
of $SU(2)$ reads:
\be
\Box W_\mu + \frac{g^2}{2} h^2(x_6) W_\mu=0~,
\label{gvb}
\ee
where $h(x_6)$ denotes the $x_6$-dependent vev of the Higgs doublet
$H$. To avoid problems in dealing with singular, ill-defined
functions, here $h(x_6)$ is a smooth function, vev of a 6D bulk
field. From the eq.~(\ref{gvb}) we will see that, if $h(x_6)$ is not
constant, then the lightest mode for the gauge vector bosons is no
longer described by a constant wave function. Therefore the 4D gauge
interactions, resulting from the overlap of fermion and vector bosons
wave functions, can be different for the two generations.

In general we are not able to solve the above equation exactly, but we
can do this by a perturbative expansion in $g^2$, which we could
justify a posteriori.  At zeroth order the $W^3$ mass and the
corresponding wave function are given by:
\be
(m^{(0)}_W)^2=0~~~~~~~~~~~~~
W^{(0)}_\mu=\frac{1}{\sqrt{2 \pi^2 R_5 R_6}}~.
\ee
At first order we find:
\bea
m_W^2&=&\frac{g^2}{2\pi R_6} \int_{0}^{+\pi R_6} dx_6~ h^2(x_6)\nn\\
W_\mu&=&W^{(0)}_\mu(1+\delta W_\mu(x_6))\nn\\
\delta W_\mu(x_6)&=&\int_0^{x_6}du\int_0^u dz(\frac{g^2}{2} h^2(z)-m_W^2)~,
\label{wf}
\eea
modulo an arbitrary additive constant in $W_\mu$, that can be adjusted
by normalization. We see that when $h(x_6)$ is constant, the usual
result is reproduced: $m_W^2=g^2 h^2/2$ and the corresponding wave
function does not depend on $x_6$.  Eq.~(\ref{wf}) allows us to
compute the fractional difference $(g_1-g_2)/(g_1+g_2)$ between the
SU(2) couplings to the first and second fermion generation,
respectively.  Focusing on $W^3_\mu$, we obtain:
\be
\left\vert\frac{g_1-g_2}{g_1+g_2}\right\vert=
\frac{\int_0^{\pi R_6}dx_6~
\left(\vert\xi^{(\alpha)}_1\vert^2-\vert\xi^{(\alpha)}_2\vert^2\right)
\delta W^3_\mu}
{\int_0^{\pi R_6}dx_6~
\left(\vert\xi^{(\alpha)}_1\vert^2+\vert\xi^{(\alpha)}_2\vert^2\right)}~,
\label{rg}
\ee
where $\alpha=1,4$.  As expected, if $\delta W^3_\mu$ is
$x_6$-constant, then the gauge couplings are universal.  From the
precision tests of the SM performed in the last decade at LEP and SLC
we expect that such a difference should not exceed, say, the per-mill
level.  We have analyzed numerically eq.~(\ref{rg}) for several
choices of the parameters and for several possible profiles of the vev
$h(x_6)$. We found that universality is respected at the per-mill
level for $m_W^2 R_6^2<O(10^{-3})$ or $1/R_6>3~TeV$.

We now consider the interactions of the higher modes arising from the
KK decomposition of the gauge vector bosons:
\be
A_\mu=i \sum_{c,m,n} t^c A_\mu^{c (m,n)}(x) z_{mn}(x_5,x_6)~,
\label{kkg}
\ee
where $t^c$ are the generators of the gauge group factor, $A_\mu^{c
(m,n)}(x)$ the corresponding 4D vector bosons and $z_{mn}(x_5,x_6)$
the periodic, $\mathbb{Z}_2$-even wave functions:
\be
\label{zmn}
z_{mn}(x_5,x_6)=\frac{1}{\sqrt{\pi^2 R_5 R_6 2^{\delta_{m,0}\delta_{n,0}}}}
\cos(m\frac{x_5}{R_5}+n\frac{x_6}{R_6})~.
\ee
In eq.~(\ref{kkg})-(\ref{zmn}) $m$ and $n\ge 0$ are integers: $m$ runs
from $-\infty$ to $+\infty$ for positive $n$ and from $0$ to $+\infty$
for $n=0$. From eq.~(\ref{lg}) we obtain the 4D interaction term:
\be
-\frac{g}{\Lambda} \sum_{a,b}\sum_{m,n} c_{ab}^{mn}~ 
\overline{f^{(\alpha)}_a(x)}~\gamma^\mu t^c~ f^{(\alpha)}_b(x)~ 
A_\mu^{c (m,n)}(x)~,
\label{hint}
\ee
where $g$ denotes the gauge coupling constant of the relevant group
factor and the scale $\Lambda$ has been included to make $g$
dimensionless; $a,b=1,2$ are generation indices. The coefficients
$c_{ab}^{mn}$ $(a,b=1,2)$, resulting from the integration over $x_5$
and $x_6$, describe the overlap among the fermion and gauge-boson wave
functions. We obtain:
\bea
c_{ab}^{mn}&=&0~~~~~~~~~~m\ne 0\nn\\
c_{ab}^{0n}&=&0~~~~~~~~~~a\ne b\nn\\
c_{11}^{0n}&=&
\frac{1}{\sqrt{\pi^2 R_5 R_6 2^{\delta_{n,0}}}}
            \frac{4 \vert\mu_{\alpha}\vert^2 R_6^2}
{n^2 + 4 \vert\mu_{\alpha}\vert^2 R_6^2} 
            \frac{(1-(-1)^n e^{2|\mu_{\alpha}| \pi R_6})}
                 {(1- e^{2|\mu_{\alpha}| \pi R_6})}\nn\\
c_{22}^{0n}&=&(-1)^n c_{11}^{0n}~.
\label{hcoef}
\eea
For odd $n$, the interactions mediated by $A_\mu^{c (0,n)}$ are
non-universal. By asking that universality holds within the
experimental limits, we get a lower bound on $1/R$ similar to that
discussed before, of the order of some $TeV$.

Until now we have considered only bounds from violation of
universality, while stronger limits are obtained by flavour changing
processes. Starting from eqs.~(\ref{hint}) and~(\ref{hcoef}), after
ESB, we should account for the unitary transformations bringing
fermions from the interaction basis to the mass eigenstate basis.  The
terms involving $A_\mu^{c (0,2n+1)}$ are not invariant under such
transformations and flavour changing interactions are produced. To
show how this works in detail we focus on the term:
\be
-\frac{g}{\Lambda}\sum_{n}
\left(\bar{d}_L~\bar{s}_L\right) \gamma^{\mu} t^c 
\left( \matrix{ c_{11}^{0n} & 0 \cr 
                0 & c_{22}^{0n} }  \right)
\left( \begin{array}{c}
       d_L\\ s_L
       \end{array} \right)
A_{\mu}^{c(0n)}
+(L\rightarrow R)~.
\ee
When we move to the interaction basis it becomes:
\be
-\frac{g}{\Lambda}\sum_{n}
\left(\bar{d}_L~\bar{s}_L\right) \gamma^{\mu} t^c 
V_L^d
\left( \matrix{ c_{11}^{0n} & 0 \cr 
                0 & c_{22}^{0n} }  \right)
V_L^{d\,\dagger}
\left( \begin{array}{c}
       d_L\\ s_L
       \end{array} \right)
A_{\mu}^{c(0n)}
+(L\rightarrow R)~.
\ee
For even $n$ $V_L^d$ commutes with the matrix $\left( c_{ij}^{0n}
\right)$ and the interactions remain diagonal; but for odd $n$ these
matrices do not commute and flavour changing interactions are induced
by:
\be
U_L^{d,2n+1} = c_{11}^{0,2n+1} V_L^d 
\left( \matrix{ 1 & 0 \cr 
                0 & -1 }  \right) V_L^{d\,\dagger}~.
\ee
Now we can integrate out the heavy modes $A_\mu^{c (0,2n+1)}$ and we
will obtain an effective, low-energy description of flavour violation
in terms of four-fermion operators, suppressed by $(1/R_6)^2$.  The
most relevant effects of these operators have been discussed by
Delgado, Pomarol and Quiros in ref.~\cite{dpq} in a 5D framework with
split fermions. Starting from the effective $\Delta S =2$ lagrangian
for down-type quarks, they calculated $\Delta m_K$ and $\epsilon_K$ as
function of $M_c = 1/R$ and they derived a lower bound on $1/R$ of
$O(100~TeV)$ and of $O(1000~TeV)$, respectively, which at least as an
order of magnitude applies also to our model.


\section[Towards a Realistic Model]{Towards a Realistic Model}
\label{3gen}

The model introduced in the previous section is simply a toy model but
can be considered as a first step towards the construction of a
realistic theory of flavour in which the fermion families are
dynamically generated. In this section we briefly discuss vices and
virtues of this toy model and we suggest how we can overcome its
problems, first of all the one of the number of generations.

One of the most interesting features of our toy model is that, starting
from a single 6D spinor for every representation of the SM, we obtain
two fermion families simply through compactification. Then masses and
mixings are obtained in two steps, firstly with the localization of
the two generations in two different regions in the extra dimensions
and then through the introduction of a localized Higgs vev. Here the
parameters we introduce are all of order one and the hierarchy is
generated thanks to the different overlap of the fermionic wave
functions with the Higgs. 

At a first sight, from the above description, it seems that this model
solves the hierarchy problem and explains why there are more fermion
replica. But this is not completely true. First of all in our toy
model we are able to obtain only two generations and, within this
framework, it is impossible to get more families since a 6D
vector-like spinor contains only two left(right)-handed 4D
spinors. Moreover, even if we obtained the hierarchy dynamically, we
have anyway a large number of parameters so that our theory is not
predictive. These are essentially the main problems related with our
toy model. Then of course there are other less important defects: for
example we do not have analyzed yet the problem of neutrino masses,
even if the possibility of writing Majorana mass terms lets us suppose
that probably a see-saw mechanism will be possible.

In what follows we try to suggest how to build a realistic and
predictive model for flavour. First of all we would like to have a
model with three generations and to reach this scope there are two
possibilities. Either we adopt a topologically non trivial background
which changes the number of zero modes or we increase the number of
extra dimensions. Since the first direction has already been
explored~\cite{russi,russi2} and the model proposed is quite
complicated\footnote{Moreover in these models the number of families
is still imposed by hands since it coincides with the winding number
of the topological defect which is required to be three.}, we explore
the other possibility, i.\,e. we decide to work in $d$ extra
dimensions, with $d>2$. How many extra dimensions are necessary to
build a realistic model? As mentioned before such a theory should be
able to explain, among other issues, the smallness of neutrino masses
and the simplest way is through the see-saw mechanism. Then an
analysis of Majorana masses in extra dimensions is in order.

The Majorana mass term is defined as:
\be
\label{general-majo}
\overline{\Psi^c}\,\Psi = \Psi^T\,C\,\Psi
\ee
where the charge conjugation matrix $C$ in D dimensions is defined by
\be
\label{charge}
C_{\pm}\, \Gamma_M\, C_{\pm}^{-1}\, = \, \pm \Gamma_M^T~.
\ee
In even dimensions both $C_-$ and $C_+$ exist, while in odd dimensions
only one is possible~\cite{cdf}. To form Majorana masses the matrix
$C$ must be antisymmetric~\cite{wett} and this condition is verified
for D=2,3,4,5,6 mod\,8. It follows that, with $\textrm{D}>6$, the
minimum number of extra dimensions we have to consider in order to
have Majorana masses is six.

The analysis of Majorana masses suggests us to work in 10D. How are
fermions in this scheme? Are they suitable to obtain three
generations? In 10D we can have Dirac, Majorana, Weyl and
Majorana-Weyl spinors~\cite{cdf}. A 10D Dirac spinor is composed by 32
complex components, so it contains eight 4D Dirac spinors. Majorana
and Weyl conditions halve the number of components. If we start with a
Weyl spinor it contains eight 4D Weyl spinors, four with left
chirality and four with right. Of course this is enough for our scopes
and we can assert that 10D is the right framework in which
investigate.  

Unfortunately the 10D Majorana mass term mixes the two 10D
chiralities, so we cannot start from a Weyl spinor. The simplest way
to proceed, in strict analogy with what done in the toy model, is to
start with vector-like fermions. Obviously the number of independent
4D spinor is doubled and now we have to find a way to obtain three
left(right)-handed fermions starting from eight left-handed and eight
right-handed spinors.  To reduce the number of 4D spinors, i.\,e. to
obtain only a few fermions with zero mode, there are essentially two
ways: orbifolding and/or twisting. With six extra dimension a lot of
possibilities are in principle allowed, with different orbifolds such
as $\mathbb{Z}_2$ and $\mathbb{Z}_3$ and maybe with twist. Although a
large amount of work has still to be done in order to obtain a
realistic and predictive theory, our toy model can be considered a
good step in this direction.

\clearemptydoublepage

\addcontentsline{toc}{chapter}{Conclusions}
\pagestyle{plain} 

\chapter*{Conclusions}

In this thesis we have analyzed the problem of symmetry breaking in
theories with extra spatial dimensions compactified on
orbifolds. Orbifold compactifications can be exploited to break gauge
symmetries, supersymmetry and flavour symmetries with features that
have no counterpart in four-dimensional theories. In particular in
chapter~\ref{chap2} we considered scenarios where gauge symmetry and
supersymmetry breaking is induced by the generalized boundary
conditions. In contrast, in chapter~\ref{chap3} we concentrated on the
flavour problem and proposed a six-dimensional toy model.\\

In chapter~\ref{chap2} we studied the new features of the
Scherk-Schwarz mechanism when applied to an orbifold
compactification. As explained in section~\ref{scherk-schwarz}, in a
compact space one can twist the periodicity conditions on the fields
by a symmetry of the action. This results in a shift of the
Kaluza-Klein levels and, in particular, yields a massive zero mode and
therefore can be used to break four-dimensional symmetries. When the
Scherk-Schwarz mechanism is implemented on orbifolds we find that
certain consistency conditions between the operators defining the
twist and those defining the orbifold parity must hold (see
section~\ref{ss-orb}). However at variance with theories defined on
manifolds where fields must be smooth everywhere, in orbifold theory
fields can have discontinuities at the fixed points, provided the
physical properties of the system remain well defined. So the most
general boundary conditions on fields are specified not only by parity
and periodicity, but also by possible jumps at the fixed points. In
sections~\ref{GBCferm-bc} and~\ref{GBCbos-bc} we have discussed the
most general boundary conditions respectively for fermions and bosons
on the orbifold $S^1/\mathbb{Z}_2$. We found that these boundary
conditions are identical, except that in the case of bosons they must
be imposed on the $y$-derivative of fields as well. The most general
boundary conditions include parity, periodicity and jumps at the two
fixed points and are represented by unitary matrices satisfying some
consistency requirements.

After the assignment of the boundary conditions, we calculated the
corresponding spectra and eigenfunctions in the case of one fermion
field (section~\ref{GBCferm-one}), one scalar field
(section~\ref{GBCbos-one}) and more scalar fields
(section~\ref{GBCbos-more}). In each case we found that the spectrum
has the typical Scherk-Schwarz-like form, where the Kaluza-Klein
levels are shifted by a constant amount which now depends on both the
twist and jump parameters. The corresponding eigenfunctions can be
discontinuous or have cusps at the fixed points and may be periodic or
not, depending on the twist.

As the spectrum we found is like the one corresponding to usual
twisted boundary conditions, we realized that it is possible to
recover our spectrum through a Scherk-Schwarz mechanism with
appropriate twist. The eigenfunctions associated to this twist are now
continuous and the different systems are related by a local field
redefinition.  We know that physics is invariant under such a local
field redefinition and thus we concluded that these two systems are
physically equivalent.

Therefore we could state there is an entire class of different
boundary conditions corresponding to the same spectrum, i.\,e. to the
same physical properties, with eigenfunctions related by local field
redefinitions. In sections~\ref{GBCferm-mass} and~\ref{GBCbos-mass} we
performed this redefinition in the action and observed that, both for
fermions and bosons, generalized boundary conditions correspond to
$y$-dependent five-dimensional mass terms that can be localized at the
orbifold fixed points. Although these terms can be singular, which
requires appropriate regularization, we conclude that they are
essential for the consistency of the theory, since they encode the
behaviour of fields at boundaries.

We have stated that the same four-dimensional spectrum can correspond
to different mass terms. Thus we would like to classify the most
general set of five-dimensional mass terms that leads to a given mass
spectrum. We discussed this in section~\ref{bfwz-mass}, where we gave
the conditions that a five-dimensional mass term should satisfy in
order to be ascribed to a Scherk-Schwarz twist. Moreover we found a
relationship between the Scherk-Schwarz twist parameter and the Wilson
loop obtained by integrating the mass terms around the
extra-dimension. In section~\ref{bfwz-ex} we discussed some examples
of equivalent mass terms.

In sections~\ref{gauge} and~\ref{susy} we studied some
phenomenological applications of our generalized boundary conditions
specifically for gauge symmetry breaking and supersymmetry
breaking. We studied the breaking of an $SU(2)$ gauge symmetry in a
toy model before moving on to a realistic $SU(5)$ model and then we
considered supersymmetry breaking in a pure five-dimensional
supergravity model.\\

In chapter~\ref{chap3} we focused on flavour symmetry and we
constructed a six-dimensional toy model for flavour where the number
of generations dynamically arises as a consequence of the presence of
extra dimensions. In section~\ref{3.1} we briefly reviewed the
literature on the subject, in particular showing that methods for
localizing four-dimensional chiral fermions along the extra dimensions
exist and have already been proposed to solve the flavour
problem. However we observed that earlier models are far from being
realistic and moreover require that the number of fermion replica is
introduced by hand. Indeed the only models that attempt to
simultaneously explain both the flavour problem and the number of
generations require a non-trivial topological background and are quite
complicated. In contrast, for our toy model we exploited the fact that
a higher-dimensional spinor can be decomposed into many
four-dimensional spinors to address the flavour $plus$ fermion replica
problem simultaneously.

In section~\ref{3.2.1} we illustrated the main features of our toy
model.  Note that although our construction is insufficient to obtain
automatically three generations, we decided it was worthwhile to study
a six-dimensional two-family model since it exhibits interesting new
features that should also arise in a more realistic three-family
setup.  Indeed a six-dimensional Dirac spinor contains two left-handed
and two right-handed four-dimensional spinors. Moreover using a Dirac
spinor has the advantage that gauge and gravitational anomalies are
absent. We compactified the extra dimensions on the orbifold
$T^2/\mathbb{Z}_2$ and, with appropriate parity assignments, we showed
that it is possible to project out the unwanted chirality states to
obtain two four-dimensional spinors with the same chirality and the
same quantum numbers, i.\,e. two replica of the same fermion. We
started from six vector-like fermions with the quantum numbers of the
standard model and, after orbifolding, we got two four-dimensional
chiral zero modes for every spinor, which we identified with
($q_{1L},u_R,d_R,l_{1L},e_R,\nu_{eR}$) and
($q_{2L},c_R,s_R,l_{2L},\mu_R,\nu_{\mu R}$). We showed that, in the
absence of other interactions, these zero modes have constant profiles
along the extra dimensions, but the introduction of a six-dimensional
Dirac mass term for every fermion can localize the two zero modes in
two different regions of the extra space where the level of
localization depends on the absolute value of this Dirac mass. The
$\mathbb{Z}_2$-symmetry implies that the Dirac masses must have odd
parity and so we chose them proportional to the periodic sign function
along one of the two extra coordinates (and constant along the other),
with the proportionality constant different for each fermion.

After localizing the two families in two different regions of the
compact space, we introduced a Higgs vev confined on the brane around
which the second generation lives and calculated the subsequent
fermionic mass spectrum (section~\ref{masses}). We discovered that the
masses are naturally hierarchical, i.\,e. from order one parameters of
the fundamental theory we obtained a four-dimensional
hierarchy. Moreover the mixing appeared to be related to a soft
breaking of the six-dimensional parity. Exploiting the fact that
Majorana masses are allowed in six dimensions, we proposed that the
smallness of neutrino masses could be explained through an
extra-dimensional see-saw mechanism. In section~\ref{scale} we put a
lower limit on the scale of extra dimensions by calculating the order
of magnitude of flavour changing neutral currents processes in our
model.

Although we found very interesting results, our toy model contains too
many parameters and so its predictability is weak. Moreover it
contains only two fermion replica and in six dimensions it is not
possible to obtain three generations without introducing a
topologically non-trivial background.  In section~\ref{3gen} we
discussed how to extend our toy model to a more realistic case with
three generations. We suggested that the most promising framework is
realized within a ten-dimensional space-time, where Majorana masses
are allowed and fermions contain a sufficient number of
four-dimensional components. Although our model is not completely
realistic and requires further development, we think that this toy
model is an important step towards the construction of a realistic
model that explains both the flavour and the fermion replica problems
within the same framework.

\clearemptydoublepage
\pagestyle{headings}

\appendix
\addcontentsline{toc}{chapter}{Acknowledgements}
\pagestyle{plain}

\chapter*{Acknowledgements}

First of all I warmly thank Ferruccio Feruglio for having been the
advisor everyone would like to have, for his precise and clear
explanations, for his enthusiasm in doing physics and for the
enjoyable and fruitful collaboration over these years.

Then I would like to thank my other collaborators, Isabella Masina,
Manuel P\'erez-Victoria, Andrea Wulzer and Fabio Zwirner, for
interesting discussions and pleasant collaborations.

I also would like to thank David Rayner for having put my English in a
``more English'' form.

I also thank the other Ph.D. students, in particular Alessandro
Torrielli for interesting $on$-$train$ discussions and wonderful
explanations and my friends Sandra Moretto and Francesca Della Vedova
for their enjoyable company and support.

Finally I thank my family and everyone that from afar believed in me
and supported me during these difficult but wonderful years.

\clearemptydoublepage
\pagestyle{headings}

\chapter[$\Gamma$ Matrices in Five and Six Dimensions]
{$\Gamma$ Matrices in Five and Six Dimensions}

\label{gamma-mat}

\subsection*{FIVE DIMENSIONS}

In 5D we work with the metric \be \eta_{MN} = \textrm{diag}(-1, 1, 1,
1, 1) \ee where $M,N = \mu,5 = 0,1,2,3,5$. The representation of 5D
$\Gamma$-matrices we use in the text is the following:
\be
\Gamma^{\mu} = \left(\begin{array}{cc}
0 & \sigma^{\mu} \\
\bar{\sigma}^{\mu} & 0
\end{array}\right)~, ~~~~~~
\Gamma^5 = i \left(\begin{array}{cc}
-\mathbf{1} & 0\\
0 &\mathbf{1}
\end{array}\right)
\ee
with $\sigma^{\mu} = (\mathbf{1}, \sigma^i)$, $\bar{\sigma}^{\mu}
= (\mathbf{1}, -\sigma^i)$ and $\sigma^i$ are the Pauli
matrices.\\[0.5cm]

\subsection*{SIX DIMENSIONS}

In order to have a greater overlap with the literature and an easier
comparison of our results with the existing ones, in 6D we change our
notations and we work with the metric \be \eta_{MN} = \textrm{diag}(1,
-1, -1, -1, -1, -1) \ee where $M,N = \mu,5,6 = 0,1,2,3,5,6$. The
representation of 6D $\Gamma$-matrices we adopt is
\be \Gamma^{\mu} = \left(\begin{array}{cc}
\gamma^{\mu} & 0\\
0 & \gamma^{\mu}
\end{array}\right)~, ~~~~~~
\Gamma^5 = i\ \left(\begin{array}{cc}
0 & \gamma_5\\
\gamma_5 & 0
\end{array}\right)~, ~~~~~~
\Gamma^6 = i\ \left(\begin{array}{cc}
0 & i\ \gamma_5\\
-i\ \gamma_5 & 0
\end{array}\right)~.
\ee
Here $\gamma^{\mu}$, $\gamma_5$ are 4D $\gamma$-matrices given by
\be \gamma^0 = \left(\begin{array}{cc}
\mathbf{1} & 0 \\
0 & \mathbf{-1}
\end{array}\right)~, ~~~~~~
\gamma^i = \left(\begin{array}{cc}
0 & \sigma^i \\
-\sigma^i & 0
\end{array}\right)~, ~~~~~~
\gamma^5 = \left(\begin{array}{cc}
0 & \mathbf{1} \\
\mathbf{1} & 0
\end{array}\right)~,
\ee
where $\sigma^i$ are the Pauli matrices.

In 6D the analogous of $\gamma_5$, $\Gamma_7\ (=\Gamma^7)$, is defined by:
\be \Gamma_7 = \Gamma_0\Gamma_1\Gamma_2\Gamma_3\Gamma_5\Gamma_6 =
\left(\begin{array}{cc}
\gamma^5 & 0 \\
0 & -\gamma^5
\end{array}\right)~.
\ee

The 6D charge conjugation matrix, defined by
\be
C_+\, \Gamma_M\, C_+^{-1}\, = \, \Gamma_M^T~,
\ee
is
\be
C = \Gamma_0\Gamma_2\Gamma_5=
\left(\begin{array}{cc}
0 & \gamma^0\gamma^2\gamma^5\\
\gamma^0\gamma^2\gamma^5 & 0 
\end{array}\right)~.
\ee

\clearemptydoublepage

\chapter[Path-Ordered Products]{Path-Ordered Products}
\label{path-ord}

We collect here some useful formulae and results concerning
path-ordered products and show how they can be used to prove
eq.~(\ref{converse}) and eq.~(\ref{IV}). First, we distinguish the two
inequivalent definitions of path-ordering, introducing the symbols:
\bea 
\label{Pdef}
P_> [m(y_1) m(y_2)] & \equiv & 
m(y_1) m(y_2) \Theta(y_1 - y_2) + 
m(y_2) m(y_1) \Theta(y_2 - y_1) \, , 
\nn \\ 
P_< [m(y_1) m(y_2)] & \equiv & 
m(y_1) m(y_2) \Theta(y_2 - y_1) + 
m(y_2) m(y_1) \Theta(y_1 - y_2) \, ,
\eea
where $m(y)$ is a $y$-dependent matrix and
\be
\label{theta}
\Theta(y) = \left\{
\begin{array}{lcc}
1 & {\rm for} & y>0  \\
0 & {\rm for} & y<0
\end{array}
\right.
\ee
is the Heaviside step function. From the above definitions, and
assuming $y_1 < y_2 < y_3$, the following properties follow:
\be
\label{propa}
P_> \left[\exp\left({\dd i \, 
\int_{y_1}^{y_2} dy' m(y')}\right)\right] 
\, \cdot \,
P_<\left[\exp\left({\dd - i \, 
\int_{y_1}^{y_2} dy' m(y')}\right)\right] 
= {\mathbf 1} \, ,
\ee
\be
\label{propb}
P_> \left[\exp\left({\dd  i \!\! 
\int_{y_1}^{y_3} \!\! dy' m(y')}
\right)\right]
=
P_> \left[\exp\left({\dd  i \!\! 
\int_{y_2}^{y_3} \!\! dy' m(y')}
\right)\right]
\, \cdot \,
P_>\left[\exp\left({\dd  i \!\! 
\int_{y_1}^{y_2} \!\! dy' m(y')}
\right)\right]  
\, ,
\ee
\be
\label{propc}
P_< \left[\exp\left({\dd i \!\!
\int_{y_1}^{y_3} \!\! dy' m(y')}
\right)\right]
=
P_< \left[\exp\left({\dd i \!\! 
\int_{y_1}^{y_2} \!\! dy' m(y')}
\right)\right]
 \, \cdot \,
P_<\left[\exp\left({\dd i \!\! 
\int_{y_2}^{y_3} \!\! dy' m(y')}\right)\right]  
\, .
\ee
If the $y$-dependent matrix $V(y)$ satisfies the differential
equation:
\be
\label{dv}
\partial_y \, V(y) = i \, V(y) \, m(y) \, ,
\ee
then it is immediate to prove that
\be
\label{Pv}
V(y) = V(y_0) \, P \left[\exp\left({\dd i \, 
\int_{y_0}^y dy' m(y')}\right)\right] \, ,
\;\;\;\;\;
P = \left\{
\begin{array}{ccc}
P_< & {\rm for} & y_0 < y \\
P_> & {\rm for} & y_0 > y
\end{array}
\right.
\, .
\ee
Correspondingly, if $m(y)$ is hermitian, then $V^\dagger(y)$ obeys the
equation
\be
\label{dvd}
\partial_y \, V^\dagger(y) = - i \, m(y) \,  V^\dagger(y) \, ,
\ee
which is solved by
\be
\label{Pvd}
V^\dagger(y) = P \left[\exp \left( {\dd - i \, 
\int_{y_0}^y dy' m(y')} \right) \right] V^\dagger(y_0) 
\, ,
\;\;\;\;\;
P = \left\{
\begin{array}{ccc}
P_> & {\rm for} & y_0 < y \\
P_< & {\rm for} & y_0 > y
\end{array}
\right. \, .
\ee
Showing that eq.~(\ref{mcform}) implies eq.~(\ref{converse}) is now a
simple application of eqs.~(\ref{dv}) and (\ref{Pv}). To show instead
that eqs.~(\ref{req1}) and (\ref{mcform}) imply eq.~(\ref{IV}), it is
sufficient to solve eq.~(\ref{req1}) for $U_{\vec{\beta}}$,
\be
U_{\vec{\beta}} = V(y + 2 \pi R) V^\dagger (y) \, ,
\ee
and to insert the explicit form of the solutions of
eq.~(\ref{mcform}), namely eqs.~(\ref{Pv}) and (\ref{Pvd}).  It is
also easy to show that the second member of eq.~(\ref{IV}) is indeed
$y$-independent.

\clearemptydoublepage

\addcontentsline{toc}{chapter}{Bibliography}

\clearemptydoublepage

\end{document}